\documentclass[twoside,twocolumn,9pt]{article}
\usepackage{extsizes}
\usepackage[super,sort&compress,comma]{natbib} 
\usepackage[version=3]{mhchem}
\usepackage[left=1.5cm, right=1.5cm, top=1.785cm, bottom=2.0cm]{geometry}
\usepackage{balance}
\usepackage{mathptmx}
\usepackage{sectsty}
\usepackage{graphicx} 
\usepackage{lastpage}
\usepackage[format=plain,justification=justified,singlelinecheck=false,font={stretch=1.125,small,sf},labelfont=bf,labelsep=space]{caption}
\usepackage{float}
\usepackage{fancyhdr}
\usepackage{fnpos}
\usepackage[english]{babel}
\addto{\captionsenglish}{%
  \renewcommand{\refname}{Notes and references}
}
\usepackage{array}
\usepackage{droidsans}
\usepackage{charter}
\usepackage[T1]{fontenc}
\usepackage[usenames,dvipsnames]{xcolor}
\usepackage{setspace}
\usepackage[compact]{titlesec}

\usepackage{siunitx} 
\usepackage{svg} 
\usepackage{subcaption} 
\usepackage{placeins} 

\definecolor{cream}{RGB}{222,217,201}
\usepackage{hyperref} 

\begin{document}

\pagestyle{fancy}
\thispagestyle{plain}
\fancypagestyle{plain}{
\renewcommand{\headrulewidth}{0pt}
}

\makeFNbottom
\makeatletter
\renewcommand\LARGE{\@setfontsize\LARGE{15pt}{17}}
\renewcommand\Large{\@setfontsize\Large{12pt}{14}}
\renewcommand\large{\@setfontsize\large{10pt}{12}}
\renewcommand\footnotesize{\@setfontsize\footnotesize{7pt}{10}}
\makeatother

\renewcommand{\thefootnote}{\fnsymbol{footnote}}
\renewcommand\footnoterule{\vspace*{1pt}%
\color{cream}\hrule width 3.5in height 0.4pt \color{black}\vspace*{5pt}} 
\setcounter{secnumdepth}{5}

\makeatletter 
\renewcommand\@biblabel[1]{#1}            
\renewcommand\@makefntext[1]%
{\noindent\makebox[0pt][r]{\@thefnmark\,}#1}
\makeatother 
\renewcommand{\figurename}{\small{Fig.}~}
\sectionfont{\sffamily\Large}
\subsectionfont{\normalsize}
\subsubsectionfont{\bf}
\setstretch{1.125} 
\setlength{\skip\footins}{0.8cm}
\setlength{\footnotesep}{0.25cm}
\setlength{\jot}{10pt}
\titlespacing*{\section}{0pt}{4pt}{4pt}
\titlespacing*{\subsection}{0pt}{15pt}{1pt}

\fancyfoot{}
\fancyfoot[RO]{\footnotesize{\sffamily{1--\pageref{LastPage} ~\textbar  \hspace{2pt}\thepage}}}
\fancyfoot[LE]{\footnotesize{\sffamily{\thepage~\textbar\hspace{2pt} 1--\pageref{LastPage}}}}
\fancyhead{}
\renewcommand{\headrulewidth}{0pt} 
\renewcommand{\footrulewidth}{0pt}
\setlength{\arrayrulewidth}{1pt}
\setlength{\columnsep}{6.5mm}
\setlength\bibsep{1pt}

\makeatletter 
\newlength{\figrulesep} 
\setlength{\figrulesep}{0.5\textfloatsep} 

\newcommand{\topfigrule}{\vspace*{-1pt}%
\noindent{\color{cream}\rule[-\figrulesep]{\columnwidth}{1.5pt}} }

\newcommand{\botfigrule}{\vspace*{-2pt}%
\noindent{\color{cream}\rule[\figrulesep]{\columnwidth}{1.5pt}} }

\newcommand{\dblfigrule}{\vspace*{-1pt}%
\noindent{\color{cream}\rule[-\figrulesep]{\textwidth}{1.5pt}} }

\makeatother

\twocolumn[
  \begin{@twocolumnfalse}
\vspace{1em}
\sffamily
\begin{tabular}{m{0cm} p{17cm} }

 & \noindent\LARGE{\textbf{Novel fast Li-ion conductors for solid-state electrolytes from first-principles}} \\
\vspace{0.3cm} & \vspace{0.3cm} \\

 & \noindent\large{Tushar Singh Thakur$^{\ast}$\textit{$^{a}$}, Loris Ercole\textit{$^{a\dag}$}, and Nicola Marzari\textit{$^{a,b,c}$}} \\
\vspace{0.3cm} & \vspace{0.3cm} \\

 & \noindent\normalsize{We present a high-throughput computational screening for fast lithium-ion conductors to identify promising materials for application in all solid-state electrolytes.
Starting from more than 30,000 Li-containing experimental structures sourced from Crystallography Open Database, Inorganic Crystal Structure Database and Materials Platform for Data Science, we perform highly automated calculations to identify electronic insulators.
On these $\sim 1000$ structures, we use molecular dynamics simulations to estimate Li-ion diffusivities using the pinball model, which describes the potential energy landscape of diffusing lithium with accuracy similar to density functional theory while being 200-500 times faster.
Then we study the $\sim 60$ most promising and previously unknown fast conductors with full first-principles molecular dynamics simulations at several temperatures to estimate their activation barriers.
The results are discussed in detail for the 9 fastest conductors, including $Li_7NbO_6$ which shows a remarkable ionic conductivity of $\sim 5$ mS/cm at room temperature.
We further present the entire screening protocol, including the workflows where the accuracy of the pinball model is improved self-consistently, necessary to automatically running the required calculations and analysing their results.
} \\

\begin{center}
    \colorbox{cream}{
        \parbox[c][5.25cm][c]{0.95\textwidth}{
            \textbf{Broader context} \\
\footnotesize Solid-state electrolytes have emerged as a key component in the development of the next generation of energy storage devices. 
Their inherent safety and superior performance compared to conventional liquid electrolytes have attracted increased attention in the field of sustainable energy. 
Despite the tremendous attention, the design and discovery of a novel solid-state electrolyte with high Li-ion conductivity remains a significant challenge. 
While many structural families have been identified over the years, the progress has been slow and discovering new fast Li-ion conductors for solid-state electrolytes would have major impact.
Unlike all-experimental procedures that can be human intensive, computational methods for automated discovery are readily parallelisable and require much fewer resources. 
Nevertheless, a computational strategy relying on full first-principles methods can be exceptionally expensive, hence the need for methods that are sufficiently inexpensive to be able to run thousands of appropriate calculations while being accurate enough to yield meaningfully predictive results. 
This screening identifies fast Li-ion conductors by estimating Li-ion diffusivity with molecular dynamics simulations using the pinball model, which is typically two orders of magnitude faster than density functional theory while retaining a similar level of accuracy. 
We emphasise that we exclusively study experimentally known materials, ensuring that the fast ionic conductors we suggest are actually synthesisable and ready for in-depth experimental investigation.
        }
    }
\end{center}

\end{tabular}

 \end{@twocolumnfalse} \vspace{0.6cm}

  ]

\renewcommand*\rmdefault{bch}\normalfont\upshape
\rmfamily
\section*{}
\vspace{-1cm}


\footnotetext{$^{\ast}$~E-mail: \textit{tushar.thakur@epfl.ch} }
\footnotetext{\textit{$^{a}$~Theory and Simulation of Materials (THEOS), and National Centre for Computational Design and Discovery of Novel Materials (MARVEL), \'Ecole Polytechnique F\'ed\'erale de Lausanne, 1015 Lausanne, Switzerland }}
\footnotetext{$^{\dag}$~Current address: \textit{Centre Européen de Calcul Atomique et Moléculaire (CECAM), \'Ecole Polytechnique F\'ed\'erale de Lausanne, 1015 Lausanne, Switzerland }}
\footnotetext{\textit{$^{b}$~PSI Center for Scientific Computing, Theory and Data, Paul Scherrer Institute, 5232 Villigen PSI, Switzerland }}
\footnotetext{\textit{$^{c}$~Theory of Condensed Matter, Cavendish Laboratory, University of Cambridge, Cambridge CB3 0US, United Kingdom }}





\section{Introduction} \label{introduction}

All-solid-state Li-ion batteries (ASSLBs) have been intensively studied \cite{armand2008building, janek2016solid, Yao_2016} particularly for applications in electric vehicles \cite{schmuch2018performance, scrosati2010lithium} and mobile devices \cite{BATES200033}. 
This growing interest is primarily attributed to ASSLBs' higher energy densities and enhanced safety profiles compared to their conventional liquid counterparts \cite{janek2016solid, NISHI2001101, scrosati2010lithium}. 
Besides this, ASSLBs' lightweight nature facilitates improved battery miniaturisation and easier assembly process \cite{liu2020emerging}, and they exhibit superior mechanical, thermal and electrochemical stability \cite{bachman2016inorganic, lim2020review}. 
Despite the significant attention ASSLBs have received, no known solid-state material satisfies all of the desirable requirements needed for their application, including high ionic conductivity \cite{bachman2016inorganic, ohno2020materials}. 
While many structural families have been identified, progress remains slow, underscoring the importance of searching new materials for ASSLBs \cite{ZHENG2018198, meesala2017recent}.

In the past, materials discovery has relied on experimental approaches guided by chemical intuition \cite{zheng2018review}.
As a first example, phosphate based Li-containing materials were derived from NASICONs (Na Super Ionic CONductors) \cite{hagman1968crystal, guin2015survey} with structure formula $LiM_2(PO_4)_3$ (M = Ti, Zr) \cite{sudreau1989dimorphism, jian2017nasicon}. 
These so-called Li-NASICONs exhibit high Li-ion conductivity \cite{aono1990ionic} and continue to be subjects of ongoing research \cite{loutati2022nasicon, kaur2022solid}. 
Further examples include the gradual and systematic exploration of various inorganic families such as nitrides \cite{alpen1977ionic, adalati2022metal}, halides \cite{li2019water, nie2023halide}, hydrides \cite{matsuo2011lithium, mohtadi2016renaissance}, perovskites with the general formula $La_{3x}La_{2/3-x}TiO_3$ \cite{inaguma1993high, stramare2003lithium} and Li-argyrodites with the formula $Li_6PS_5X$ (X = Cl, Br, I) \cite{deiseroth2008li6ps5x}.
A final example is the development of Li-containing garnet structures, with chemical composition $Li_5La_3M_2O_{12}$ (M = Ta, Nb), which were identified to be promising conductors, albeit with limited ionic conductivity \cite{thangadurai2003novel}. 
However, the chemical substitution with aliovalent ions led to the discovery of $Li_7La_3Zr_2O_{12}$, commonly known as LLZO, that demonstrates significantly higher ionic conductivity \cite{murugan2007fast}.

The development of LLZO also serves as an example of chemical substitution in well-known ionic conductor families to explore the vast chemical space and identify new ionic conductors. 
Another example is the extensively studied family of LISICONs (Li-superionic conductors) with the formula $Li_{14}Zn(GeO_4)_4$ \cite{hong1978crystal}. 
Over time, numerous new LISICON-type materials were discovered \cite{hu1977ionic, khorassani1982new, rodger1985li, deng2015structural}, which can be represented with a more general formula of $Li_4XO_4$-$Li_3YO_4$ (X = Si, Ge, Ti; Y = P, As, V) \cite{zheng2018review}.
LISICONs also serve as the precursor to the thio-LISICON family \cite{kanno2001lithium}, which consists of a more polarisable sulphide anionic framework rather than an oxide sublattice, thereby enhancing their ionic conductivity \cite{tao2022thio}.
Further substitution of the cations led to the discovery of tetragonal-$Li_{10}GeP_2S_{12}$ (LGPS) \cite{kamaya2011lithium}, which is widely regarded as one of the best solid-state ionic conductors \cite{ohno2020materials} and has motivated the development of numerous promising derivative structures \cite{liang2021designing}.
To summarise, significant breakthroughs have primarily resulted from chemical intuition or by systematic substitution in known materials, motivated by the keen understanding of the underlying chemistry. 
Besides this, combinatorial methods \cite{koinuma2004combinatorial} and straightforward high-throughput experimental approaches \cite{xiang1995combinatorial, takeuchi2003identification} have also contributed to the discovery of new super ionic conductors, albeit with mixed success.

However, these experimental approaches do not scale as effectively as computational methods, which can be highly efficient in materials discovery by allowing for the exploration of a vast number of structural families within a short time frame \cite{castelli2012computational, pizzi2016aiida, mounet2018two}.
Furthermore, computer simulations have primarily been limited to understanding the underlying diffusion mechanism, which in turns contributes to developing deeper chemical intuition. 
As a result, many computational screenings are typically motivated by established chemical knowledge, focusing on specific ion-conduction mechanism or space-groups to propose new materials \cite{tufail2023}. 
For instance, Xiao \textit{et al.} \cite{Xiao2021} performed a computational screening motivated by the diffusion network in garnets and NASICONs type conductors; Muy \textit{et al.} \cite{muy2025optimizing} explored all the possible doping strategies within the argyrodite family. 
In contrast, a screening approach that is agnostic to the underlying chemistry of structures can probe a much more expansive chemical space and potentially identify novel materials that have no apparent connection to the existing materials.

\begin{figure}[ht]
\centering
  \includegraphics[width=\columnwidth]{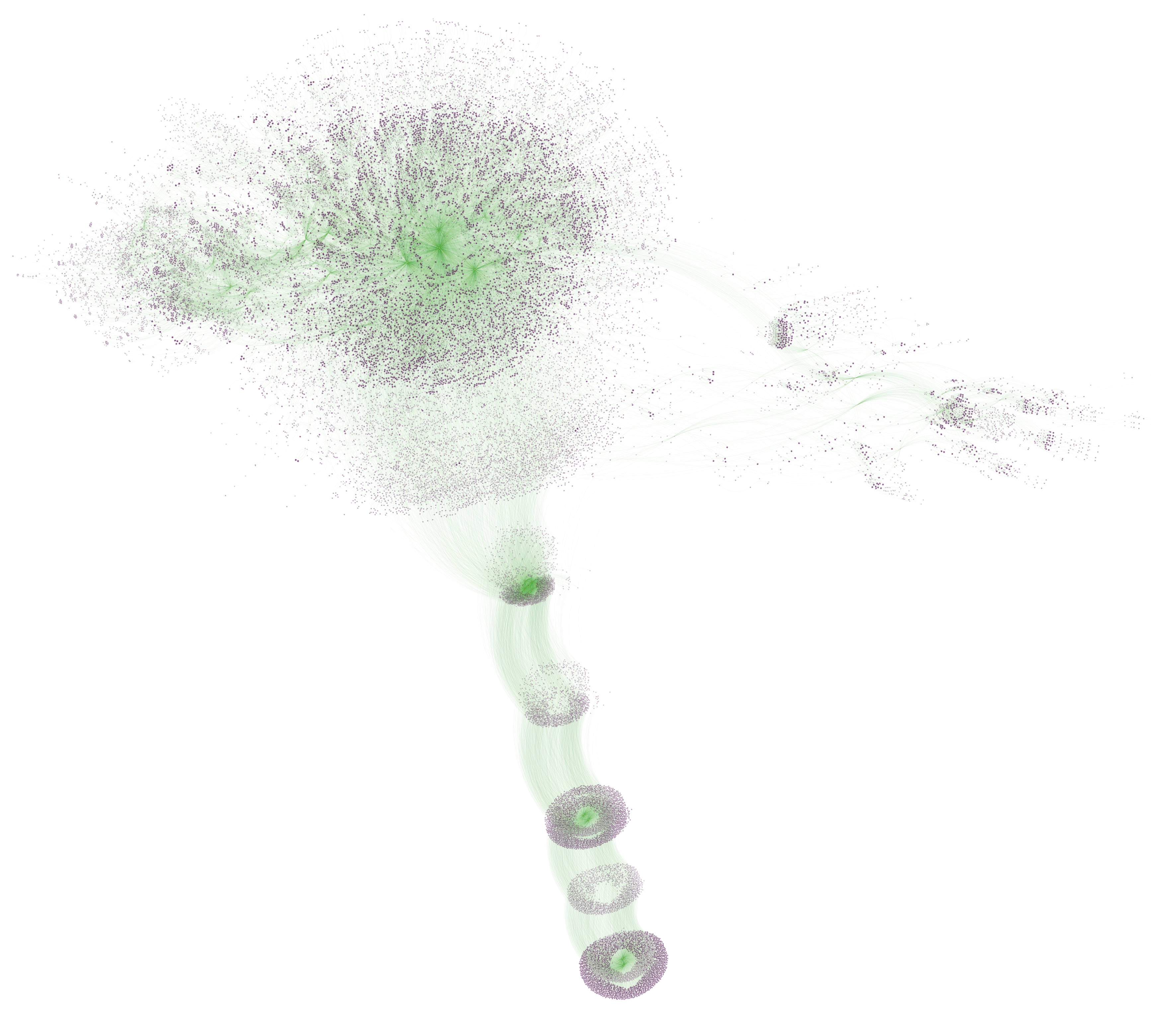}
  \caption{A segment of the AiiDA database spanning this screening is depicted, showcasing a small subset pertaining to single-point calculations performed on approximately 1500 structures at the level of DFT. Purple nodes represent either data instances (i.e., inputs and outputs of calculations) or the calculations themselves, while green links illustrate the logical provenance connecting these nodes.}
  \label{fgr:aiida_schematic}
\end{figure}

Consequently, it is essential to establish robust screening criteria motivated by physical properties to effectively identify the most suitable candidates for solid-state electrolytes.
To prevent self-discharge in a battery, an SSE ought to exhibit low electron mobility, which is determined by the material's electronic band gap.
The most accessible first-principles method for estimating band gaps is Kohn-Sham density functional theory (DFT) \cite{hohenberg1964inhomogeneous, kohn1965self}.
Although, more advanced approaches, such as GW \cite{shishkin2007accurate}, Koopmans-compliant functionals \cite{de2022bloch}, hybrid functionals \cite{miceli2018nonempirical}, Hubbard-corrected DFT \cite{himmetoglu2014hubbard} and many others \cite{marzari2021electronic}, can yield band gap values that are predictive, these methods are significantly more computationally demanding compared to single point DFT calculation.
Thus, for screening purposes, DFT offers a satisfactory balance between computational efficiency and accuracy for band gap estimates, despite its tendency to underestimate band gaps \cite{cohen2012challenges}. 
This was also utilised by the screenings studies of Muy \textit{et al.} \cite{muy2019high} and Sendek \textit{et al.} \cite{sendek2017holistic}, who calculated band gaps at the level of DFT-PBE \cite{perdew1996generalized}, and applied a filtering criterion treating any material with a band gap greater than 1 eV as an insulator.

Electrochemical stability can be estimated with first-principles calculations as well \cite{richards2016interface}, and it can be estimated in a high-throughput mode \cite{borodin2015towards}.
While a broad electrochemical stability window is desirable for SSEs, many currently utilised electrolytes exhibit narrow stability windows \cite{zheng2018review}.
A notable example is LGPS which is stabilised with interphases and protective coatings \cite{sun2017facile} \cite{jian2024interface}. 
In the same vein, although low interfacial resistance and high interfacial compatibility between electrolyte and electrode is important for optimum performance, higher resistance (and lower compatibility) can be mitigated by incorporating appropriate interfacial materials \cite{park2018design, jian2024interface}. 
Therefore, we emphasise that while electrochemical stability and interfacial compatibility are important considerations, they are not essential for a screening process, and thus, we have opted not to calculate these in this study.

Mechanical properties such as bulk and shear modulus can be readily obtained from simulations\cite{pokluda2015ab, materzanini2023solids}.
However, the relevance of this information remains somewhat ambiguous. 
For instance, preventing or retarding the unwanted growth of Li-dendrites is achieved not merely through the use of a high-modulus material, but rather through defect engineering \cite{ren2015direct, porz2017mechanism}.
Consequently, while bulk properties can be calculated easily, their utility as screening criteria is not well understood and as such we have chosen not to incorporate them in our study.

In summary, many challenges persist that limit the selection of materials for use as SSEs \cite{kerman2017practical, gao2018promises}; still, achieving high ionic conductivity remains the most critical criterion \cite{hull2004superionics, knauth2002solid}. 
Ionic diffusion can be estimated from atomistic simulations directly through MD \cite{alder1959studies, rahman1964correlations}, with the accuracy dependent on the underlying potential energy surface (PES), which can be computed using empirical or machine-learned force-fields or with first-principles methods\cite{frenkel2023understanding, tuckerman2010statistical, allen2017computer}. 
While empirical force-fields may be sufficiently accurate to model Li-diffusion \cite{deng2015structural, adams2012structural}, they require precise fitting of the parameters to the specific system under consideration, which limits their applicability in exploring a vast chemical space. 
DFT can provide highly accurate and general PES applicable to a wide variety of chemical compositions. 
However, first-principles molecular dynamics (FPMD) in the Born-Oppenheimer  \cite{born1985quantentheorie} approximation relies on performing single point DFT calculations at every MD step, rendering it prohibitively expensive \cite{tuckerman2002ab}. 
Another variant of FPMD, Car-Parrinello molecular dynamics \cite{car1985unified}, is computationally more efficient, but requires careful tuning to the system being studied. 
While this method can be highly useful for investigating diffusion mechanisms within a single system \cite{materzanini2021high}, it is non-trivial to calibrate its parameters across a multitude of systems. 
In addition to MD, ionic conductivity can be estimated in the simplest Arrhenius picture by calculating migration barriers for Li-diffusion, which can be obtained from inexpensive static calculations \cite{du2007li, lepley2013structures}. 
However, identifying barriers is a highly complex task that often requires human intervention and is thus challenging to automate \cite{marcolongo2017ionic, he2017origin, morgan2017lattice}. 
Other methods attempt to link diffusion to more easily accessible properties: for example, the bond-valence method \cite{brown1992chemical} has been used to inexpensively calculate Li-ion conductivity in several independent screenings \cite{anurova2008migration, avdeev2012screening, xiao2015candidate}, though with limited accuracy due to the limitations of the method \cite{swenson2000determining, muller2007comparison}.
Another approach involved deriving diffusion coefficients using specific phonon frequencies \cite{aniya1996phonons, wakamura1997roles}. 
In all cases, the aim to reduce computational costs goes directly against the requirement of reliable predictions across a broad range of materials. 

In the past few years, universal machine learning interatomic potentials (MLIP) have also emerged as one-stop solution for running cheap and accurate MD simulations, including MACE-MP0 \cite{batatia2022mace, batatia2023foundation}, M3Gnet \cite{chen2022universal}, CHGnet \cite{deng2023chgnet} and the proprietary GNOME \cite{merchant2023scaling}. 
These universal MLIPs are intended to be systems agnostic, can supposedly model most elements in the periodic table, and most importantly work out-of-the-box.
Before deployment, their suitability needs to be thoroughly tested. 
Besides the initial applications, few independent performance assessments of the universality have been performed \cite{riebesell2023matbench, yu2024systematic, focassio2024performance, wines2024chips}. 
Both Yu \textit{et al.} \cite{yu2024systematic} and Focassio \textit{et al.} \cite{focassio2024performance} concluded that universal MLIPs are not yet accurate enough to reproduce first-principles results and showed significant error in the estimation of properties under consideration. 
Both suggesting that the current best use case is as a foundation onto which a more appropriate model can be trained.
These shortcomings are also noted by the original authors \cite{batatia2023foundation}.
Nevertheless, these universal MLIPS promise a most promising way forward, and are starting to be employed in high-throughput screenings \cite{li2025hierarchical, lian2025high}.

Besides universal MLIPs, several other powerful predictive models exist \cite{Choudharymlreview}. 
The most common approach is to use descriptors to directly predict properties, like ionic conductivity, from the structures and\slash or chemical phase space \cite{Laskowski2023, C6EE02697D, MUY2019270}, by unsupervised or semi-supervised learning due to the lack of labelled data \cite{zhang2019unsupervised}, or atypically by training directly on experimental data \cite{Hargreaves2023}. 
Another approach that has garnered significant attention in the past year is inverse modelling, facilitated by artificial intelligence for materials discovery \cite{chen2024accelerating, merchant2023scaling}. 
These methods involve proposing hypothetical materials that may not necessarily be experimentally synthesisable \cite{chen2022universal, deng2023chgnet}. 
Nonetheless, predicting materials that are not merely synthesisable but also technologically relevant is highly non-trivial \cite{zunger2018inverse}, which suggests that the underlying premise may require further examination \cite{cheetham2024artificial}. 
This stands in direct contrast to the present work, where we screen experimentally known materials whose synthesis recipes are known.
It is important to note that several well-regarded screenings in the past few years \cite{muy2019high, kahle2020high, jun2022lithium, Laskowski2023} also utilised structures from the same repositories as ours. 
However, our workflow was able to identify promising conductors that were not highlighted in those earlier efforts, underscoring the effectiveness of our approach.

We conclude this brief review of computational methods for modelling ionic diffusion by noting that screening fast Li-ion conductors remains a challenging undertaking. 
This difficulty arises either from the limited transferability and/or accuracy of descriptors, force-fields and universal MLIPs or due to the cost of first-principles approaches.
Thus, accurately modelling the diffusion of Li-ions in a large-scale screening with MD simulations necessitates a computational approach that combines the low computational cost of force-fields with the precision and generality of DFT. 
In this study, this is achieved using the pinball model which describes the potential energy surface of lithium diffusing in an SSE, and is on average about 200-500 times faster than DFT, while offering often comparable accuracy \cite{kahle2018modeling}. 
It is based on two key assumptions: (1) all Li atoms are completely ionised, and are referred to as pinballs, and (2) the host lattice (all non-Li atoms along with the valence electrons of Li atoms) is fixed at the equilibrium positions, and the charge density is frozen. 
The pinball model forms the backbone of our screening, as detailed in Section \ref{method_pinball_md}, enabling the identification of promising Li-ion conductors for further investigation using full first-principles simulations. 

\begin{figure*}[ht]
 \centering
 \includegraphics[width=2\columnwidth]{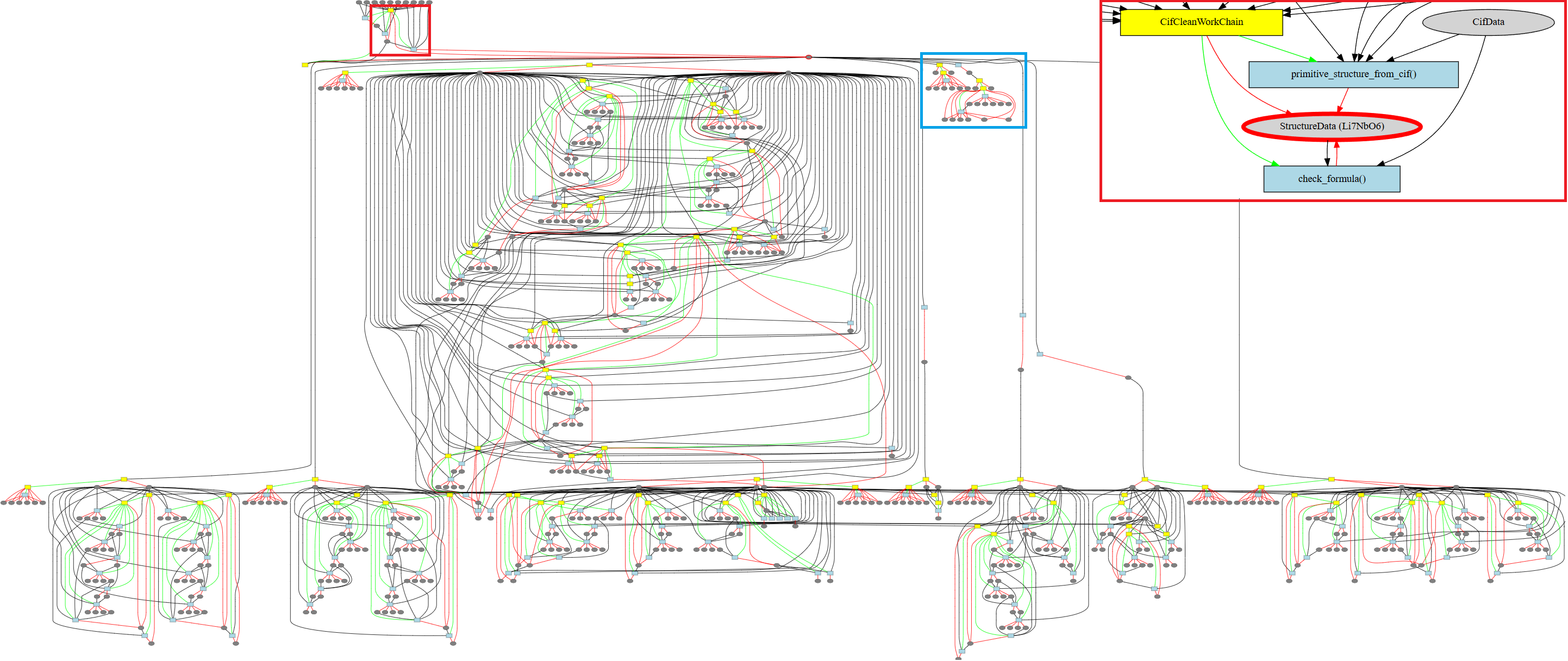}
 \caption{The provenance graph for one material, \(Li_7NbO_6\), illustrates AiiDA's meticulously tracking of each instance of input and output, along with all intermediate data and steps, as a directed acyclic graph. Nodes in the graph are colour-coded to denote different elements: workflows are highlighted in yellow, calculations in blue, and data instances in grey. Data instances, which can represent either inputs or outputs of calculations, are connected by black lines. Red lines signify logical provenance, i.e. a workflow outputting a data instance, while green lines denote operational provenance, illustrating the invocation of one workflow or calculation by another. The highlighted sub-graph provides a detailed view of the structure ingestion shown in a red box, the band gap calculation and variable-cell relaxation are given within the blue box, and the remaining graph corresponds to the self-consistent pinball MD simulations.}
 \label{fgr:provenance}
\end{figure*}

As a final note, we highlight a previous screening \cite{kahle2020high} conducted using a similar framework based on the pinball model.
The critical distinctions are as follows: (1) the present study utilises a more expansive database, encompassing over twice the number of structures, (2) we include non-local interactions within the pinball model, (3) we have implemented a self-consistent workflow that iteratively enhances the accuracy of the pinball model, and (4) we apply more stringent criteria across all filtering parameters, for instance by tightening the tolerances used to compare crystal structures, we classify nearly 30\% more structures as duplicates in this screening.
These differences and the advantages they offer are described in more detail along with methods in Section \ref{method}, followed by a discussion of results in Section \ref{result}. 
Last, we summarise this screening and present our conclusions, followed by an outlook on the development of a universal machine learning potential to model Li-ion diffusion in Section \ref{conclusion}.

\section{Methods} \label{method}

Any computational screening of this magnitude requires a robust framework to automatically launch and monitor calculations, handle errors on-the-fly, and link data generated during calculations \cite{curtarolo2013high, alberi20182019}. 
Furthermore, it is necessary that this infrastructure explicitly preserve the provenance for easy reproducibility, queryability, and shareability of the results \cite{buneman2001and, moreau2022provenance}. 
To achieve this twofold goal of automating and managing complex workflows and storing full provenance of all related data, we used the Automated Interactive Infrastructure and Database for Computational Science (AiiDA), which is a Python-based infrastructure and workflow manager \cite{huber2020aiida, uhrin2021workflows, pizzi2016aiida}. 
The key advantage of AiiDA over other workflow managers lies in its ability to preserve the provenance of a calculation in its entirety.
This includes storing the complete history of a calculation along with an exhaustive list of all inputs that led to the creation of that piece of data, as a directed acyclic graph within a relational database.
This feature allows one to query any data point as a graph node in an easy to navigate fashion and assess causal relationship between nodes.
Fig. \ref{fgr:provenance} illustrates this capability in an acyclic graph, taken from this work, that illustrates the entire screening path for one structure. 
This approach not only supports Open Science but goes beyond the well-known FAIR principle \cite{wilkinson2016fair}. 
Additionally, AiiDA facilitates a high degree of automation and parallelisation to easily run calculations on high-performance computing platforms, and every calculation in this screening was run using AiiDA.

\subsection{Preliminary filters}

\begin{figure*}
 \centering
 \includegraphics[width=2\columnwidth]{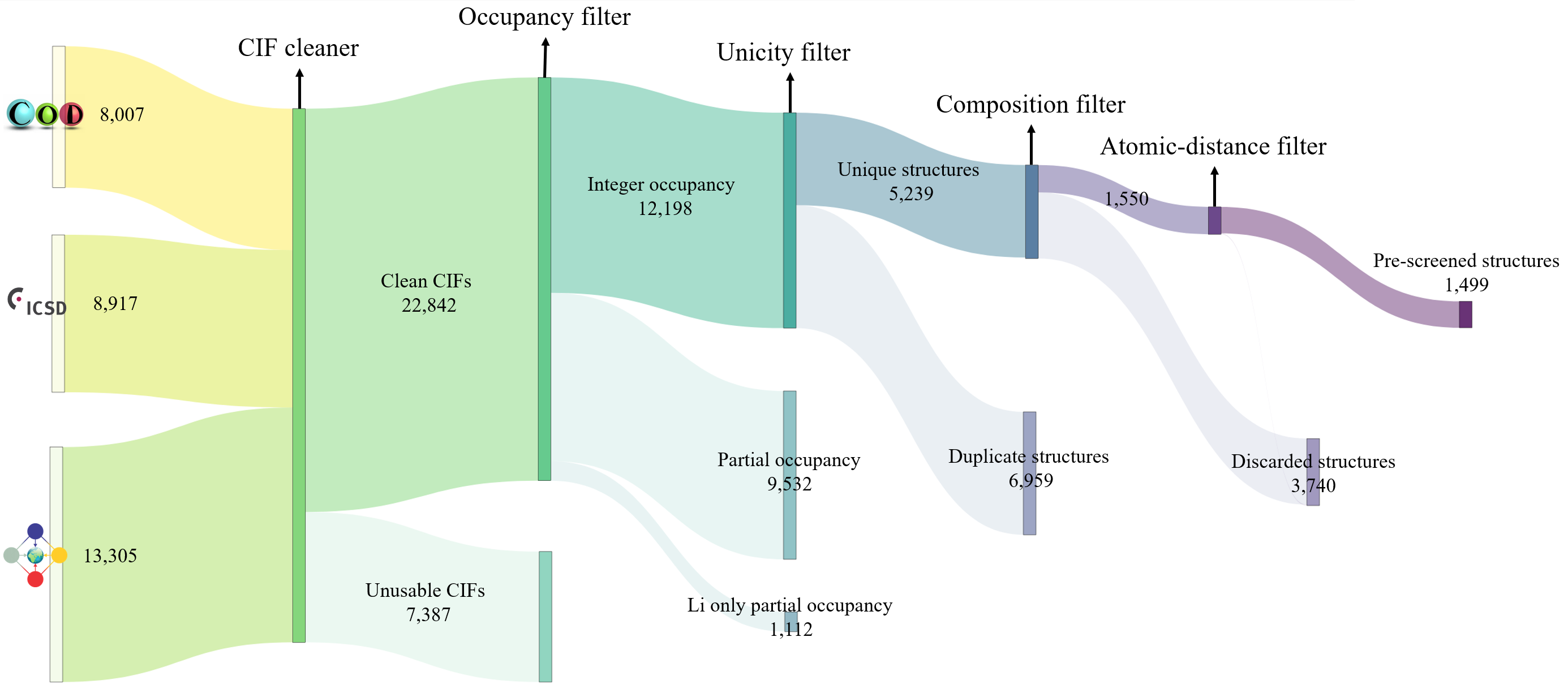}
 \caption{Flowchart illustrating the pre-screening workflow, beginning with all Li-containing structures sourced from COD, ICSD and MPDS, and culminating with \textit{ab initio} calculations. Each node represents a filter that eliminates undesirable structures (indicated by lighter shaded links), while potentially suitable structures advance to the next filter (indicated by darker shaded links). The link thickness corresponds to the number of structures passing through each filter. Beginning with over 30,000 experimental structures, the pre-screening narrows the selection down to 1,499 structures for subsequent electronic structure calculations.}
 \label{fgr:sankey1}
\end{figure*}

Starting from experimental structures sourced from the Crystallography Open Database (COD) \cite{codref}, Inorganic Crystal Structure Database (ICSD) \cite{Belskyicsd} and Materials Platform for Data Science (MPDS) \cite{Blokhin2018} repositories, we identify more than 30,000 lithium containing structures, which are imported as CIF files using AiiDA.
These files sometimes contain syntax errors or extraneous information that require correction before they can be used. 
The issues and their corresponding solutions are comprehensively described in the work by Mounet \textit{et al.} \cite{mounet2018two}.
We follow that protocol to clean, parse and standardise CIF files using COD-tools \cite{Vaitkus2021, Merkys2016, Grazulis2015}.
Finally, on the cleaned CIF files, we apply a sequence of filters to systematically narrow down the list of promising structures, as illustrated in Fig. \ref{fgr:sankey1}.

\paragraph*{Occupancy filter.~~} We remove structures with partial occupancies i.e. those whose stoichiometry doesn’t align with the reported atomic positions, as generating and modelling derivative configurations necessitates sampling strategies that can be highly non-trivial \cite{hart2008algorithm, grau2007symmetry, mustapha2013use}.

\paragraph*{Unicity filter.~~} Subsequently, we use the CMPZ algorithm \cite{hundt2006cmpz} implemented within the structure matcher function of pymatgen \cite{ong2013python} to compare crystal structures with the same stoichiometry, to eliminate equivalent structures and retain only unique ones.

\paragraph*{Composition filter.~~} Additionally, we exclude structures containing certain elements. Specifically, we filter out those with hydrogen, as elements lighter than lithium cannot be correctly modelled by the pinball approximation; those containing noble gas atoms; 3d-transition elements, due to their potential to changing oxidation states during simulations and become electronically conducting; and elements heavier than Polonium. Furthermore, we apply additional filtering criteria to ensure that each structure contains a specific selection of anions from the pnictogen, chalcogen and halogen families.

\paragraph*{Atomic-distance filter.~~} For each structure, we calculate the bond distances between every atom pair that is compatible with inorganic materials to filter out structures with bond lengths typically associated with organic molecules such as double bond with O or triple bond with N.

We note that thus far we have conducted data analysis. The subsequent sections describe the final two filters wherein we perform electronic-structure calculations.

\subsection{Electronic filter}

To classify the filtered structures as electronic insulators, we calculate the band gap at the level of DFT. 
As a rule of thumb, we categorise structures with a band gap greater than 1 eV as electronically insulating. 
Generally, DFT underestimates the band gap for most materials \cite{cohen2012challenges}.
All DFT calculations are performed using the pw.x code from the Quantum ESPRESSO distribution \cite{QE-2009, QE-2017}, using experimental geometry, and with the PBEsol \cite{perdew1996generalized, perdew2008restoring} exchange-correlation functional. 
Pseudopotentials and their corresponding cut-offs are sourced from the Standard Solid-State Pseudopotential (SSSP) Efficiency 1.2.1 library \cite{prandini2018precision}, which provides comprehensive validation of pseudopotentials across various libraries and methods \cite{willand2013norm, dal2014pseudopotentials, garrity2014pseudopotentials, topsakal2014accurate, schlipf2015optimization, van2018pseudodojo}.
For each SCF calculation, we use Marzari-Vanderbilt cold smearing \cite{marzari1999thermal} and increase the number of bands by 20\%, while the Brillouin zone is sampled with a Monkhorst-Pack grid of density 0.15 $\text{\normalfont\AA}^{-1}$.

Besides this, we perform variable-cell relaxation on about 20\% of the structures to investigate the effects of geometry optimisation on band gap estimation.

\begin{figure*}[ht]
 \centering
 \includegraphics[width=2\columnwidth]{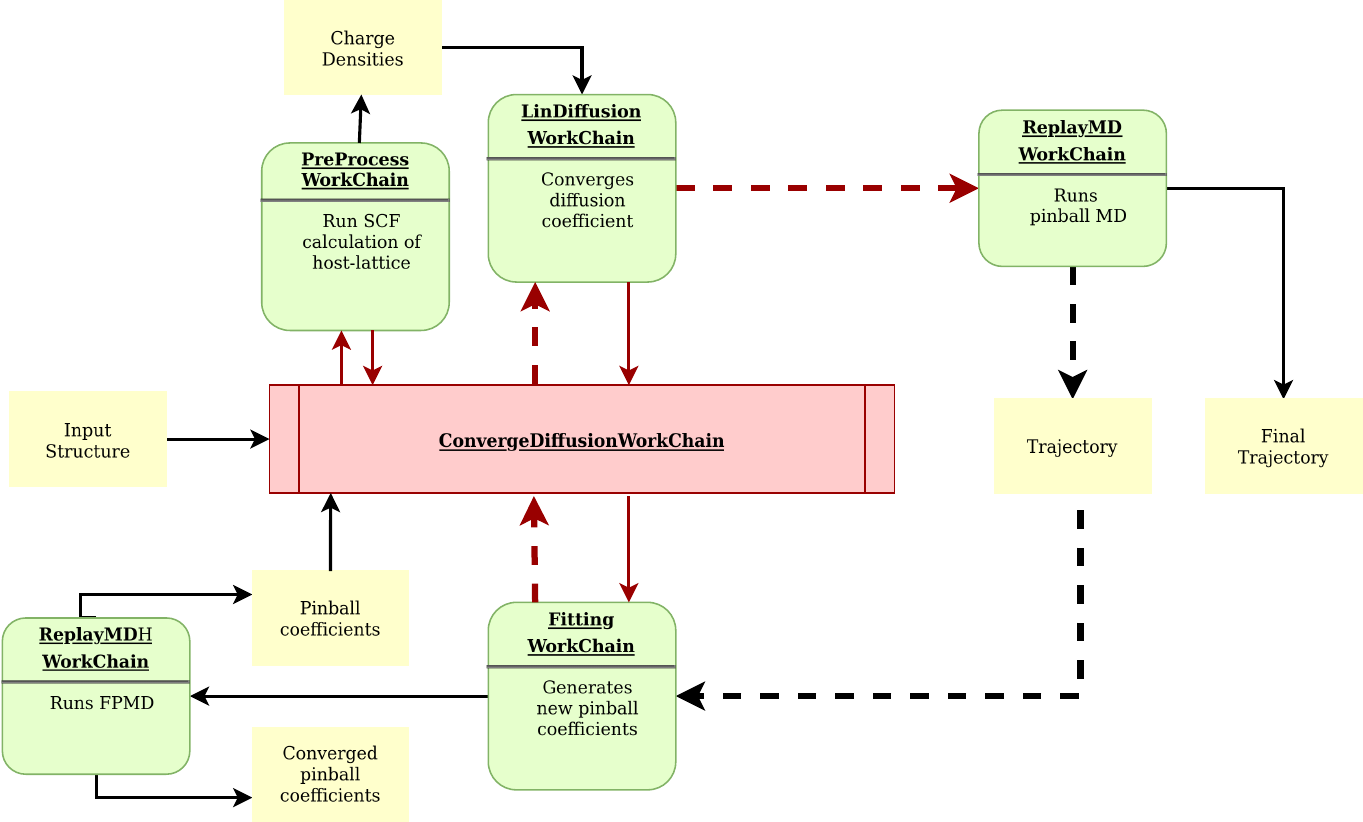}
 \caption{A schematic representation of the self-consistent workflow of aiida-flipper \cite{flipperwf}, the python package employed to run MD simulations using the pinball model \cite{kahle2018modeling} and compute ionic conductivity of lithium. The nomenclature depicted corresponds exactly to the Python classes within the plugin. The ConvergeDiffusion workchain initiates the process by launching the PreProcess workchain, which runs a single point calculation and stores the charge densities of the host lattice to be used in all subsequent pinball MD simulations. Next, the Fitting workchain is launched, generating sufficient snapshots with random displacement of Li-ions in the supercell to fit 10,000 force components through calculations at both the pinball and DFT levels. This initial estimate of pinball coefficients is then used to initiate the LinDiffusion workchain, which runs a long MD simulation at the pinball level to converge the diffusion coefficient to a predetermined threshold. From the trajectory of this MD run, uncorrelated configurations are extracted, and a new set of pinball coefficients is derived through linear regression of the DFT and pinball forces. This iterative cycle continues self-consistently until the pinball coefficients converge. Once convergence is achieved, a final MD simulation is performed using the converged pinball coefficients, and the final MD trajectory is used to compute the diffusion coefficient. This workflow ensures accurate and reliable computation of ionic conductivity, leveraging the self-consistent refinement of pinball coefficients through iterative MD simulations and force component fitting.}
 \label{fgr:sc-pinball}
\end{figure*}

\subsection{Diffusivity filter}

To run MD simulations, we generate supercells based on experimental geometries, ensuring a minimum separation of 8 $\text{\normalfont\AA}$ between opposite faces, using the supercellor package \cite{supercellor}. 
We run MD simulations with Born-Oppenheimer approximation \cite{born1985quantentheorie} in the canonical ensemble. 
Temperature is controlled with the stochastic velocity rescaling thermostat \cite{bussi2007canonical}. 

From the Einstein relation \cite{tuckerman2010statistical} we can write tracer diffusion coefficient $D_{tr}$ as: 
\begin{equation} \label{eqn1}
    \begin{split}
        D_{tr} &= \lim_{t \to \infty} \frac{1}{6t} \langle MSD(t) \rangle_{NVT} \\
        &= \lim_{t \to \infty} \frac{1}{6} \frac{d}{dt} \frac{1}{N} \sum_{i=1}^{N} \langle | \vec{r}_i(t+\tau) - \vec{r}_i(\tau)|^2 \rangle_{\tau}
    \end{split}
\end{equation}

\noindent which is a derivative of the average mean-square displacement of particles with respect to time. 
In this context, we are essentially substituting the ensemble average with a time average. 
By performing a linear regression of the mean square displacement MSD(\textit{t}) with time we can accurately estimate the diffusion coefficient from the slope of the MSD, ensuring sufficient statistical precision. 

The tracer diffusion coefficient is related to the charge diffusion coefficient with Haven's ratio as $H = D_{tr}/D_\sigma$, which is a measure of correlated motion of the particles \cite{haven1965diffusion}. 
In the dilute limit, we assume it to be 1, though in practice it is often less than 1, implying that correlated motion can enhance conductivity \cite{marcolongo2017ionic}. 
Consequently, we do not overestimate conductivities.
And from the Nernst-Einstein equation \cite{einstein1905molekularkinetischen}, we can calculate the ionic conductivity $\sigma$ as:

\begin{equation} \label{eqn2}
\sigma = \frac{N(Ze)^2}{\Omega k_B T} \frac{D_{tr}}{H}
\end{equation}

\noindent where $\Omega$ is the system volume, $T$ the temperature, and $Ze$ is an integer multiple of the elementary charge.

All analysis of trajectories including the calculation of MSD was done using the open-source tool Suite for Analysis of Molecular Simulations (SAMOS) \cite{samos}.

\subsubsection{Self-consistent pinball MD.~~} \label{method_pinball_md}

Based on the two assumptions of the pinball model \cite{kahle2018modeling}, the Hamiltonian reads as:

\begin{equation} \label{eqn3}
    \begin{split}
        \mathcal{H}_P &= \frac{1}{2} \sum_{p}^{P} M_p \dot{\vec{R}}^2_p + \alpha_1 E_N^{P-P} + \alpha_2 E_N^{H-P} \\
        &+ \beta_1 \int n_{R_{H_0}}(\vec{r}) V_{LOC}^P(\vec{r}) dr + \beta_2 \sum_i \langle \psi_{i, R_{H_0}} | \hat{V}_{NL}^P | \psi_{i, R_{H_0}} \rangle
    \end{split}
\end{equation}

\noindent where $\vec{R}$ and $\dot{\vec{R}}$ are respectively the positions and velocities of the pinballs i.e. the Li-ions, $E_N^{A-B}$ is the electrostatic interaction between the frozen core electrons of species $A$ and $B$, $V_{LOC/NL}^P$ are the local and non-local external pseudopotential component of pinballs which act on the charge density $n(\vec{r}) = \sum_i \psi_i^*(\vec{r})\psi_i$ which is frozen for the host lattice $H_0$. 
The final term is responsible for non-local interactions which further improves the accuracy of the model with additional computational cost. 
$\alpha_1$, $\alpha_2$, $\beta_1$ and $\beta_2$ are phenomenological coefficients (referred to as pinball coefficients) introduced to further improve the accuracy that can be computed by fitting the pinball forces with DFT forces.

For this screening, we designed and implemented a highly automated and powerful workflow in AiiDA as a plugin called aiida-flipper \cite{flipperwf} 
All supercells are passed to the diffusion workflow, which iteratively runs MD simulations with the pinball Hamiltonian and self-consistently refines the pinball coefficients, thereby progressively enhancing the accuracy in determining Li-ion conductivity. 
Fig. \ref{fgr:sc-pinball} illustrates the details of the workflow.

\begin{figure*}[ht]
 \centering
 \includegraphics[width=2\columnwidth]{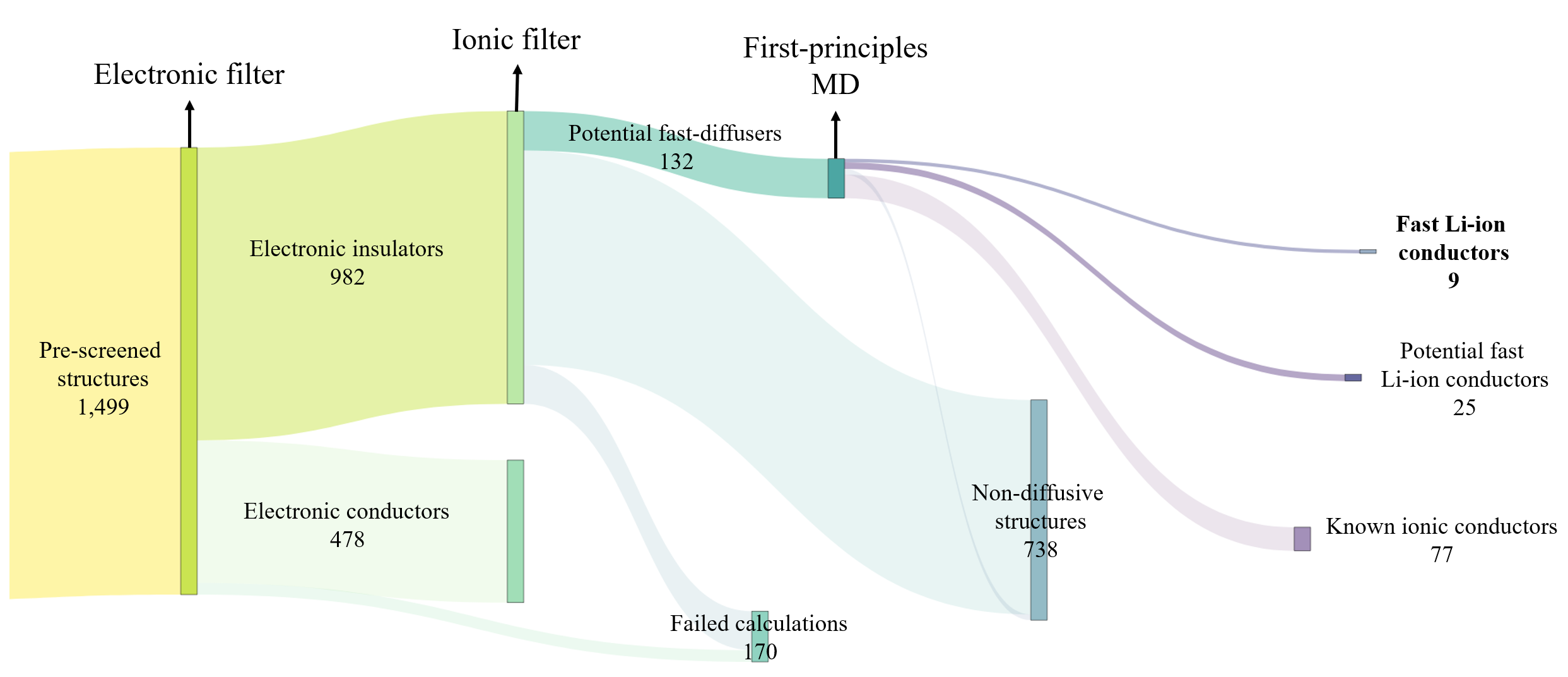}
 \caption{Flowchart of the remaining workflow that only shows electronic structure calculations. Beginning with 1,499 pre-screened structures, 9 most promising candidates are identified. Each node represents a filter based on \textit{ab initio} methods that eliminates undesirable structures (indicated by lighter shaded links), while potentially suitable structures advance to the next filter (indicated by darker shaded links). The link thickness corresponds to the number of structures passing through each filter.}
 \label{fgr:sankey2}
\end{figure*}

\subsubsection{First-principles MD.~~} 
As illustrated in Fig. \ref{fgr:sankey2}, the structures that exhibit high Li-ion diffusivity at 1000 K with the pinball model are subsequently studied with FPMD at the same temperature for 100 ps. 
However, structures already recognised in the literature as fast ionic conductors, detailed in Section \ref{known_conductors}, are excluded to prioritise the discovery of new Li-ion conductors. 
The structures validated by FPMD as fast ionic conductors are then studied at three lower temperatures: 750 K , 600 K and 500 K for 125 ps, 150 ps and 180 ps respectively.
Longer simulation times are chosen to account for comparatively slower equilibration at lower temperatures.
These temperatures are selected to be equidistant on the inverse temperature scale. 
Based on equation \ref{eqn1}, we determine the diffusion coefficient and quantify the statistical variance in diffusivity \cite{he2018statistical}. 
The activation barrier for these structures is estimated from a linear fit of the Arrhenius behaviour \cite{allen2017computer} and the error is obtained with Bayesian propagation \cite{sivia2006data}.
For the most promising structures, we plot Li-ion probability density  to better illustrate the Li-ion diffusion channels. 

\section{Results and discussion} \label{result}

The pre-screening phase, which does not involve any electronic structure calculations, is illustrated in Fig. \ref{fgr:sankey1}. 
Starting with approximately 8,000, 9,000, and 13,000 experimental structures sourced from COD \cite{codref}, ICSD \cite{Belskyicsd}, and MPDS \cite{Blokhin2018} respectively, we extract nearly 23,000 clean CIF files, discarding the unsalvageable ones. 
All subsequent filters are applied to structures derived from these clean CIF files. 
We eliminate approximately 10,600 structures with partial occupancies, and from the remaining 12,000 structures with integer atomic occupancies, 5,200 are identified as unique using the structure matcher algorithm of pymatgen \cite{ong2013python}. 
Further filtering removes structures containing unwanted elements and those with unwanted bond lengths, resulting in 1,499 structures that advance to the next phase of the screening.


\begin{figure}[ht]
\centering
  \includegraphics[width=\columnwidth]{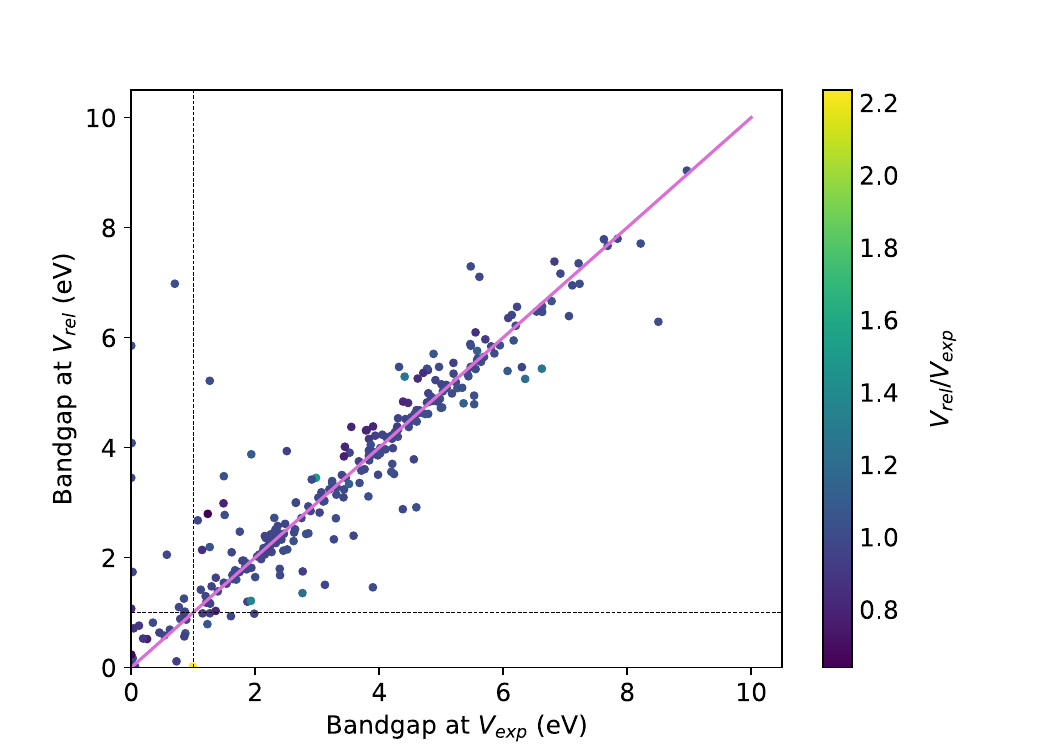}
  \caption{Comparison of band gaps at optimised geometry ($V_{rel}$) and experimental geometry ($V_{exp}$). For the majority of the structures, the classification as insulators remains unchanged upon relaxation.}
  \label{fgr:bandgap}
\end{figure}

We perform single-point calculations on these structures at the level of DFT-PBEsol \cite{perdew2008restoring}. 
Out of these, 251 calculations fail to converge due to issues in the self-consistent electronic cycle. 
These are subsequently rerun using the non-linear conjugate gradient method within SIRIUS \cite{sirius} enabled Quantum ESPRESSO. 
Following this, we calculate the band gap for all structures and classify a structure as an electronic insulator if its band gap exceeds 1 eV.
Out of the 1,499 unique structures, 982 are identified as electronic insulators, and 39 calculations failed, representing the first filter illustrated in Fig. \ref{fgr:sankey2}.
To assess the impact of geometry optimisation on our filtering criterion, we performed additional variable-cell relaxation on 25\% of these 1,499 structures, of which 316 finished successfully. 
Fig. \ref{fgr:bandgap} compares the band gaps between relaxed and experimental geometries. 
Our findings indicate that only 6 out of the 316 structures, or less than 2\%, are identified as insulators when calculated using experimental geometry instead of performing full variable-cell relaxation, representing false positive results. 
The reverse scenario, where metallic structures turn into insulators upon relaxation (false negatives), is slightly more common.
Given that all MD simulations are performed at experimental geometries, we opt not to relax any other structures, considering a less than 2\% false positive rate acceptable given the significantly higher computational cost of variable-cell relaxation and the additional failure due to issues in ionic convergence cycle.

\begin{figure}[ht]
\centering
  \includegraphics[width=\columnwidth]{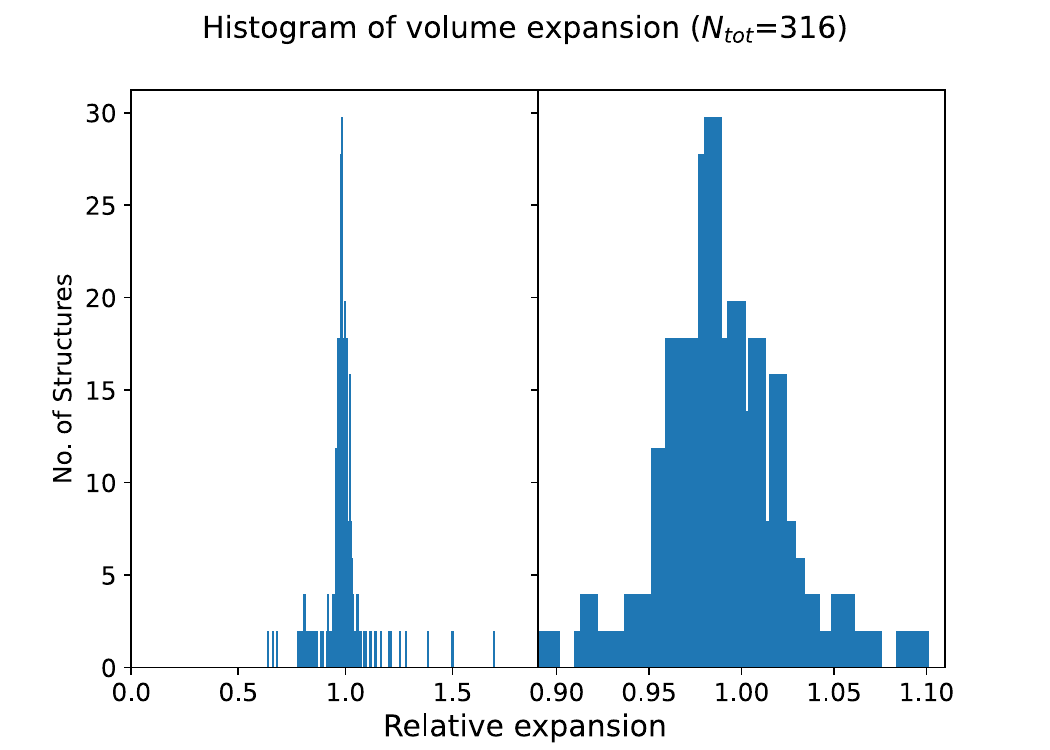}
  \caption{Histogram of relative volume expansion between optimised and experimental geometry at the level of DFT-PBEsol. The left panel displays the complete histogram while right panel provides a zoomed in view of the range from 0.9 to 1.1.}
  \label{fgr:volhist}
\end{figure}

Fig. \ref{fgr:volhist} presents a histogram of the relative volume change upon geometry optimisation, defined as the optimised volume divided by the experimental volume. 
Utilising the PBEsol functional, we achieve a narrow and uniform distribution of volume changes, maintaining lattice parameters that more closely match experimental values. 
This contrasts with the standard PBE functional \cite{perdew1996generalized}, where structures are more likely to exhibit expansion \cite{de2011performance}, as observed by Kahle \textit{et al.} \cite{kahle2020high}.

\subsection{Pinball MD}

\begin{figure}[ht]
\centering
  \includegraphics[width=\columnwidth]{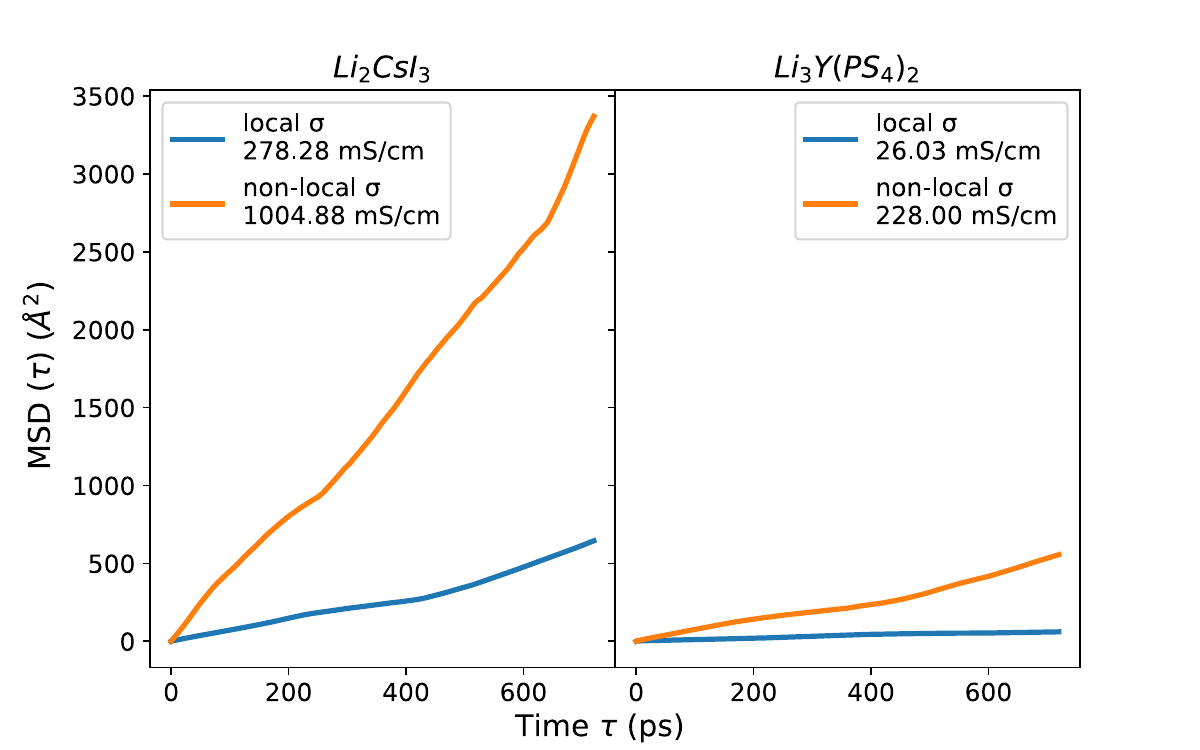}
  \caption{MSD plot of two materials comparing Li-diffusion with and without non-local interactions within the pinball model at 1000 K. Using only local projectors typically leads to an underestimation of Li-ion diffusion. Based on first-principles simulations $Li_3Y(PS_4)_2$ and $Li_2CsI_3$ show ionic conductivity of 2.16 mS/cm at 300 K \cite{zhu2017li3y} and 0.22 mS/cm at 500 K \cite{kahle2020high} respectively. }
  \label{fgr:non-loc-msd}
\end{figure}

All the MD simulations are performed on the supercells generated from the 982 insulators identified in the previous step. 
To evaluate the significance of including non-local interactions within the pinball model, we conducted tests on a few systems both with and without non-local interactions. 
The MSD plots of this comparison, shown in Fig. \ref{fgr:non-loc-msd}, reveal that using only local projectors typically leads to an underestimation of Li-ion diffusion. 
Consequently, we opt to include it in our screening, despite the higher computational cost. 

\begin{figure*}[htbp]
 \centering
 \includegraphics[width=2\columnwidth]{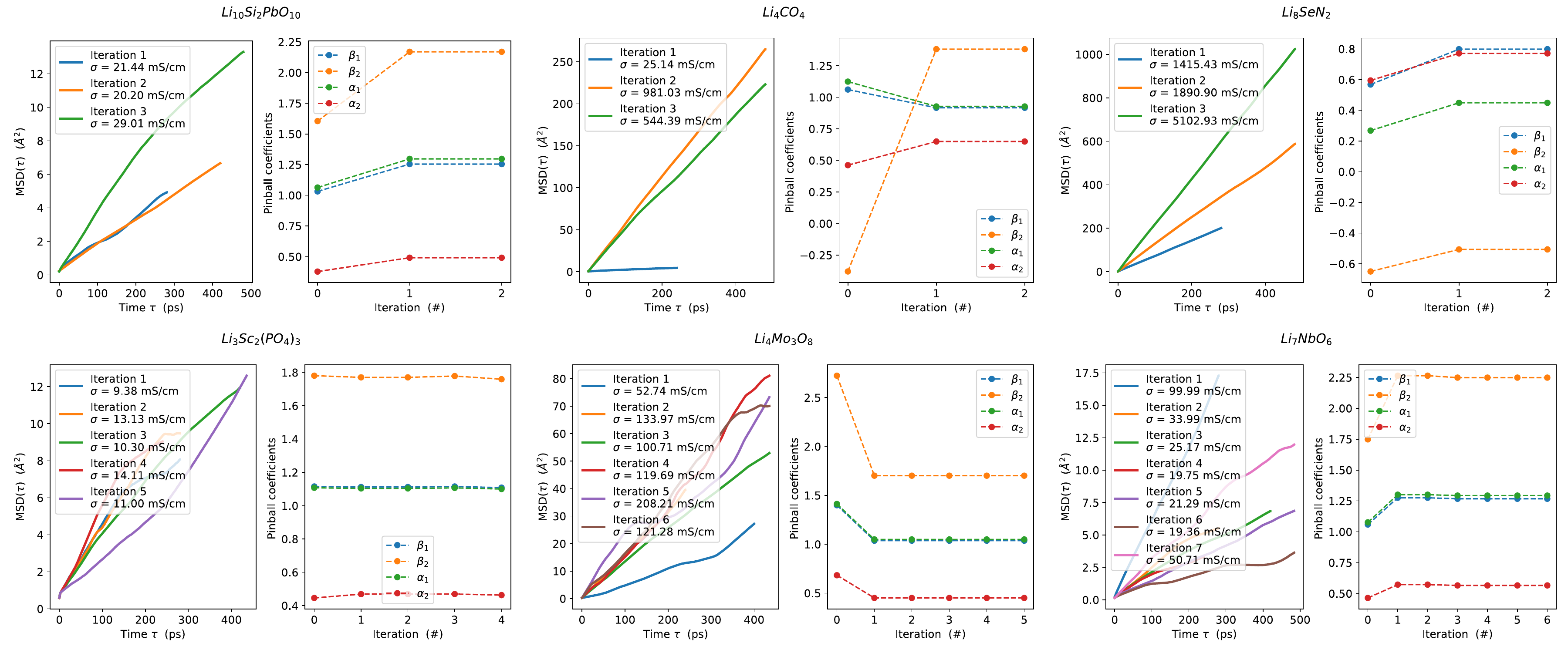}
 \caption{Self-consistent iterations for the MSD plots of Li, along with the convergence of pinball coefficients for several fast Li-ion conductors. The zeroth pinball coefficients are derived by fitting DFT and pinball forces on randomly rattled structures, which are then used to perform the first MD iteration. Force fitting is subsequently performed on configurations obtained from the first MD iteration to obtain the first pinball coefficients, which are then used in the second MD iteration. For most structures pinball coefficients converge after two of these self-consistent iterations, with the estimate of Li-diffusion remaining largely unchanged. For a select few structures, additional iterations are performed after the convergence of the pinball parameters to verify that no divergence occurred in subsequent steps, ensuring the robustness of the workflow. We attribute the slight change in dynamics in some of the MD simulations to the inherent stochasticity of the thermostat that is used \cite{bussi2007canonical}.}
 \label{fgr:msd-converge}
\end{figure*}

Next, we derive an initial estimate of the pinball coefficients through the linear regression of forces calculated at both DFT and pinball levels for all supercells. 
The quality of these coefficients is evaluated using the \( r^2 \) correlation between DFT and pinball forces. 
We ensure that the \( r^2 \) correlation for the converged pinball coefficients exceeds 0.95, with the majority of cases exceeding 0.99. 
If this criterion is not met, additional self-consistent pinball MD iterations are performed, allowing for the extraction of further uncorrelated configurations from these extended MD simulations. 
These serve as additional data points for improving the fit until full convergence is achieved, as indicated by stable pinball coefficients and an \( r^2 \) value approaching 1.
Out of 982 structures, we achieve convergence for 914, with failures occurring due to issues in the self-consistent electronic cycle when calculating DFT forces. 
An additional 63 structures failed the pinball MD simulations due to drift in the constant of motion, leading to 851 structures with a final iteration of the pinball MD run with converged coefficients and a total simulation time of 22.1 $\mu$s. 
As illustrated in Fig. \ref{fgr:msd-converge}, the pinball coefficients readily converge for most structures.
Based on the slope of the MSD plot from the final MD iteration and equation \ref{eqn2}, we estimate Li-ion conductivity. 
Ionic conductivity of 1 mS/cm at 1000 K is chosen as the threshold to categorise potential fast ionic conductors at the pinball level. 
At the end of this process, 132 structures are identified for further study using first-principles calculations.

\subsection{First-principles MD}

We classify the 132 structures obtained from the self-consistent pinball workflow into four categories: 1) structures already identified in the literature as Li-ion conductors, 2) structures that do not exhibit diffusion within FPMD, 3) structures that show negligible diffusion at lower temperatures but may still be of interest, and 4) fast Li-ion conductors.

\subsubsection{Known Li-ion conductors.~~} \label{known_conductors}

For all the structures that are identified as fast Li-ion conductors using the pinball model, we conduct an extensive literature review to assess those that have already been studied and reported as fast ion conductors. 
Out of these 132 structures, we rediscover 77 known Li-ion conductors and as such we exclude them from FPMD investigations. 

In the following short review, we report these 77 structures and their current use case if applicable. 
$Li_2Ti_6O_{13}$ is a known ionic conductor \cite{kataoka_ion-exchange_2011} and was recently proposed as cathode material \cite{fernandez-gamboa_theoretical_2022}, while sodium substituted $Li_2Ti_6O_{13}$ is used as intercalation anode \cite{kuhn_comprehensive_2018}.
$Li_7P_3S_{11}$ is a well-known superionic conductor \cite{chu2016insights, yamane2007crystal}.
$Li_2TeO_4$ is a known superionic conductor \cite{xiao_lithium_2021} and was proposed as an electrode material \cite{kirklin_high-throughput_2014}.
$Li_6NBr_3$ was experimentally shown to be a fast ionic conductor \cite{howard2015hydrogen}, but worse than the well known $Li_3N$ \cite{boukamp_lithium_1976, boukamp_fast_1978}, which we also identified. 
$LiI$ is well known ionic conductor \cite{shannon1977new}, while $LiBr$ and $LiCl$ show negligible Li-ion conduction without doping \cite{adelstein_role_2016}.
$Li_2Se$ is used as cathode material \cite{wu_nanostructured_2016} and also as an interface material \cite{yang_li2se_2022}.
$Li_3BN_2$ is a well known fast ionic conductor \cite{yamane1987high}.
$Li_3BS_3$ was reported in the computational screening by Laskowski \textit{et al.} \cite{laskowski_identification_2023}, despite an earlier computational study \cite{BIANCHINI2018186} that proposed it as a potential ionic conductor.
$Li_4SnSe_4$ is a known ionic conductor \cite{kaib2013lithium}.
$Li_2SiN_2$ is used as anode material \cite{ulvestad_silicon_2018} and Li anode coatings to increase electrochemical stability \cite{snydacker_electrochemically_2017}.
$Li_2SiP_2$ is a known ionic conductor \cite{toffoletti_lithium_2016} and was proposed as a potential solid-state electrolyte material \cite{yeandel_enhanced_2019}.
$Li_2SiS_3$ is a know ionic conductor \cite{hang_lithium_2010, huang_anomalously_2022}.
$LiBF_4$ has been used as non-aqueous electrolyte for two decades long time \cite{xu2004nonaqueous}.
$Li_3BrO$ is a known superionic conductor \cite{emly_phase_2013, zhao_superionic_2012}.
$Li_3Y(PS_4)_2$ was proposed in a computational study with very high ionic conductivity \cite{zhu2017li3y}.
$Li_3PS_4$ is known ionic conductor \cite{tachez_ionic_1984} and has been engineered with much better properties in the past decade \cite{liu_anomalous_2013}.
$Li_4PN_3$ was recently discovered with first principles simulations \cite{al-qawasmeh_computational_2020, obeidat_structural_2022}.
$Li_5AlS_4$  was experimentally reported to have low ionic conductivity at room temperature \cite{lim_structure_2018}, but is otherwise known in the argyrodite family \cite{huang_superionic_2019}.
$Li_5NCl_2$ has been known as ionic conductor for a long time \cite{weppner_consideration_1981}, but was recently studied in greater detail by Landgraf \textit{et al.} \cite{landgraf_li5ncl2_2023}.
$Li_7BiO_6$ has been known for a long time as an ionic conductor \cite{nomura_ionic_1984}.
$LiGa(SeO_3)_2$ was proposed recently by Jun \textit{et al.} \cite{jun2022lithium}.
$LiHf_2(PO_4)_3$ is a known ionic conductor \cite{aono1993electrical, martinez-juarez_ionic_1996}, but Al substituted $Li_{1+x}Al_xHf_{2-x}(PO4)_3$ showed more promise \cite{chang_spark_2005}.
$LiInS_2$ is a known ionic conductor \cite{tell_ionic_1977} and was recently studied within $LiXS_2$ family as cathode material \cite{rao_first_2016}.
$LiS$ \cite{rosenman_review_2015} is a part of Li-S battery system, while $Li_2S$ is used as cathode material \cite{su2018toward}.
$Li_3InO_3$ is a known ionic conductor \cite{eichinger1981decomposition}.
$LiZnPS_4$ is a poor ionic conductor, but with defect engineering shows more promise \cite{richards2016design, kaup2018correlation}.
$LiTi_2(PO_4)_3$ is used as cathode material in aqueous batteries especially when doped as $LiMn_xTi_{2-x}(PO_4)_3$ \cite{JIANG201694, wang2018synthesis}, further doping has yielded promising results as an SSE \cite{huang2019influence}.
$LiNbO_3$ is used as a coating on cathode materials \cite{walther2021working}, and also as anode material in Li-ion capacitors \cite{JIANG2020125207}.
$LiAlCl_4$ is not well studied, despite being a known ionic conductor for a long time \cite{spiesser1983physical}.
$Li_2O$ is a well known ionic conductor \cite{chadwick1993diffusion, hull2004superionics}.
$Li_2Mo_4O_{13}$ was recently proposed as anode material \cite{VERMA20171445}.
$LiSn_2(PO_4)_3$ is well known anode material with various different synthesis methods \cite{Burba_2005, cui2012synthesis, tian2017improved}.
$Li_4SnS_4$ is known ionic conductor \cite{kaib2012new}.
$Li_9S_3N$ is a known ionic conductor \cite{marx2006li9ns3} and was proposed as barrier coating between electrolyte and $Li$ metal anode \cite{miara2015li}.
$Li_4GeS_4$ is a well known ionic conductor \cite{murayama2002structure}.
$LiCF_3SO_3$ is a known ionic conductor along with sodium, caesium and rubidium substitutes \cite{pompetzki2004system, van2005cation}.
$Li_5NBr_2$ and $Li_{10}N_3Br$ were investigated recently in the halogen-nitride system $Li_{3a+b}N_aX_b$, with $Li_{10}N_3Br$ found to be an excellent ionic conductor \cite{sang2020theoretical}.
$Li_3In_2(PO_4)_3$ is a known superionic conductor \cite{kravchenko1994lithium}.
$Li_2B_6O_9F_2$ is a known ionic conductor \cite{pilz2011li2b6o9f2}.
$Li_2SrTa_2O_7$ is a known ionic conductor but other substitution compounds are more promising \cite{FANAH2019106014}.
$Li_7SbO_6$ is a known ionic conductor \cite{muehle2004new}.
Besides this, Kahle \textit{et al.} \cite{kahle2020high} proposed following as fast Li-ion conductors: $Li_5Cl_3O$, $Li_7TaO_6$, $LiGaI_4$, $LiGaBr_3$, and $Li_3CsCl_4$ and $Li_2CsI_3$ which are theoretical structures \cite{PENTIN2008804}, $Li_2WO_4$ which is used to improve conductivity of other materials either as solid mixtures \cite{ahmad2002structural} or in solid solutions \cite{dissanayake1991ionic}, and $LiAlSiO_4$, whose suitability was systematically studied with Al doping by Ryu \textit{et al.} \cite{ryu2023synthesis}.
Last, FPMD simulations performed by Kahle \textit{et al.} \cite{kahle2020high} showed insignificant diffusion in the following structures at lower temperatures:  $LiAlSe_2$, $Li_4Re_6S_{11}$, $LiPO_3$, $Li_3Sc_2(PO_4)_3$, $Li_4P_2O_7$, $(LiI)_2Li_3SbS_3$, $Li_6PS_5I$, $Li_5P(S_2Cl)_2$, $Li_3P_7$, $Li_3SbS_3$, $Li_2B_3O_4F_3$, $Li_2Mg_2(SO_4)_3$, $Li_3AsS_3$, $Li_2Si_2O_5$, $Li_2NaB(PO_4)_2$, $Li_6Y(BO_3)_3$, and $LiAuF_4$.

\subsubsection{Non diffusive structures.~~} \label{non-diff}

We find 18 materials that do not exhibit Li-ion diffusion in our FPMD simulations at 1000 K. This absence of diffusion suggests that they are unlikely to demonstrate Li-ion conductivity in experiments unless significantly doped. 
The materials in question include oxides, halides, sulphides and selenides, all of which are detailed in Table \ref{tbl:slow1000K} along with their respective experimental references.

\begin{table}[ht]
\small
  \caption{\ The structures that were found to be conducting at the level of pinball, but show insignificant diffusion with FPMD at 1000 K. Consequently, these were not studied at lower temperatures. We report their stoichiometry, the repository and identifier from where they originated along with the corresponding experimental reference}
  \label{tbl:slow1000K}
  \begin{tabular*}{0.48\textwidth}{@{\extracolsep{\fill}}lll}
    \hline
    Structure & Database & Database-id \\
    \hline
    $Li_2Te_2O_5$ & ICSD & 26451, 26452 \cite{CACHAUHERREILLAT1981352} \\
    $Li_2CsCl_3$ & MPDS & S1022277 \cite{pentin2012} \\
    $LiKSe$ & ICSD & 67277 \cite{orst1989interalkalimetallselenide} \\
    $LiYS_2$ & MPDS & S537670 \cite{ohtani1987synthesis} \\
    $LiInSe_2$ & MPDS & S1214509 \cite{greuling1987pressure} \\
    $LiAlS_2$ & ICSD & 608360 \cite{HELLSTROM1979881} \\
    $LiLuS_2$ & MPDS & S307222 \cite{tromme1971synthese} \\
    $Li_7Te_3O_9F$ & MPDS & S1533619 \cite{feng2017li7} \\
    $Li_5SiP_3$ & MPDS & S1145472 \cite{eickhoff2017synthesis} \\
    $Li_6RbBiO_6$ & MPDS & S1408313 \cite{CARLSON1992332} \\
    $LiAuF_6$ & MPDS & S1904723 \cite{graudejus2000m} \\
    $Li_3Na_3Ga_2F_{12}$ & MPDS & S1836948 \cite{chassaing1968contribution} \\
    $LiZrS_2$ & MPDS & S301115 \cite{whittingham1975lithium} \\
    $Li_2CdSnSe_4$ & MPDS & S1952801 \cite{zhang2017li} \\
    $LiBa_4Ga_5Se_{12}$ & MPDS & S1021504 \cite{yin2012ba} \\
    $Li_3Na_3Rh_2F_{12}$ & MPDS & S307582 \cite{de1969quelques} \\
    $Li_2HgO_2$ & MPDS & S1702887 \cite{hoppe1964oxomercurate} \\
    $Li_2Ca_2Ta_3O_{10}$ & ICSD & 88497 \cite{toda1999synthesis} \\
    \hline
  \end{tabular*}
\end{table}

\subsubsection{Potential fast Li-ion conductors.~~} \label{pot-diff}

We have identified 25 structures that exhibit significant diffusion at 1000 K in our FPMD simulations, but do not display the same behaviour at lower temperatures. 
These structures are listed in Table \ref{tbl:fast1000K}, ranked according to their likelihood of exhibiting diffusion at lower temperatures. 
It is important to emphasise that these structures may indeed show significant diffusion at lower temperatures in experiment conditions. 
The inability of our simulations to detect diffusion at these temperatures is likely due to the prohibitively long simulation times required to observe Li-ion hoping at lower temperatures. 
For instance, Materzanini \textit{et al.} report ionic conductivities of 28 mS/cm and 6 mS/cm for tetragonal-LGPO at 500 K and orthorhombic-LGPO at 600 K \cite{materzanini2021high}, respectively, corresponding to MSDs of approximately 0.04 $\text{\normalfont\AA}^2$/ps and 0.005 $\text{\normalfont\AA}^2$/ps. 
This indicates that in the tetragonal phase, a Li-ion travels an average distance of 1 $\text{\normalfont\AA}$ within 25 ps, whereas in the orthorhombic phase, it would require 200 ps to cover the same distance.
Similarly, cubic-LLZO exhibits an ionic conductivity of 20 mS/cm, which despite being 10,000 times greater than that of the tetragonal phase \cite{wolfenstine2012high}, corresponds to an MSD of 0.01 $\text{\normalfont\AA}^2$/ps.  
Therefore, simulations with durations of 100-200 ps are insufficient to accurately resolve diffusion in such systems. 
Consequently, the activation barriers, that we report, may not be entirely accurate due to lack of sufficient statistics at lower temperatures. 
Similar to previous section, we observe materials that include oxides, halides, phosphides and additional nitrides. 

\begin{figure}
\centering
  \includegraphics[width=\columnwidth]{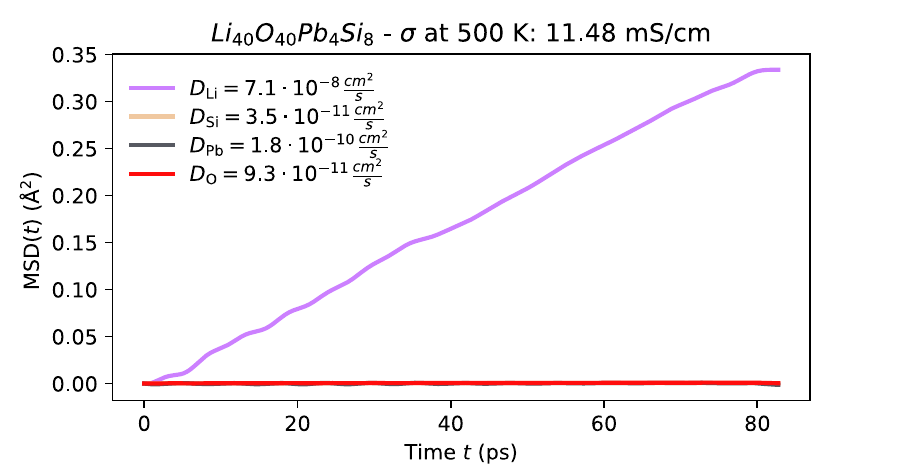}
  \includegraphics[width=\columnwidth]{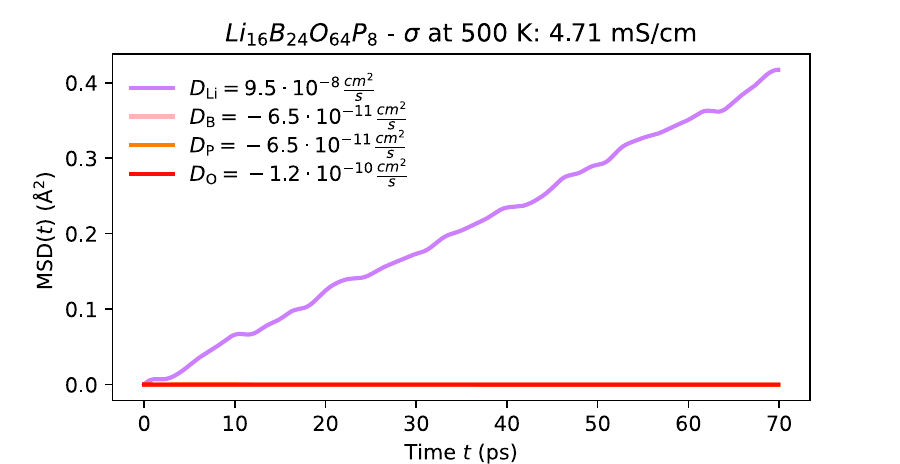}
  \caption{MSD plot of Li along with host-lattice species of $Li_{10}Si_2PbO_{10}$ and $Li_2B_3PO_8$ at 500 K from FPMD.}
  \label{fgr:msd_potential_fast}
\end{figure}

\paragraph*{$Li_{10}Si_2PbO_{10}$.~~} Originally synthesised in 1994 by Brandes \textit{et al.} \cite{brandes1994li10si2pbiio10}, this material has received limited attention, particularly in the context of fast Li-ion conduction.  
Lead-silicate glasses, including this compound, are known for diverse optical properties, such as transparency, refractive index, colouration, electrical conductivity, and chemical durability \cite{caurant2024structure}; and have found application in areas such as the monitoring of radioactive materials \cite{kacem2017structure}.  
However, their potential as solid-state electrolytes remains largely unexplored. 
In our simulations, $Li_{10}Si_2PbO_{10}$ displays Li-ion diffusion at lower temperatures, as shown in Fig. \ref{fgr:msd_potential_fast}; yet, its relatively high activation energy of 0.35 eV results in an estimated room-temperature ionic conductivity of less than 0.1 mS/cm, limiting its viability as a potential electrolyte material.

\paragraph*{$Li_2B_3PO_8$.~~} Synthesised relatively recently in 2014 by Hasegawa \textit{et al.} \cite{hasegawa2014crystal}, this material has yet to receive significant attention, particularly in the context of Li-ion batteries.  
Borophosphates of this kind have primarily been investigated in the semiconductor industry for their magnetic coupling mechanisms, optical characteristics, and catalytic behaviour \cite{tomashyk2014quaternary, li2016new}.  
Similar to $Li_{10}Si_2PbO_{10}$, our simulations are able to resolve Li-ion diffusion at lower temperatures for $Li_2B_3PO_8$, as shown in Fig. \ref{fgr:msd_potential_fast}, the relatively high activation energy of 0.28 eV renders it less suitable for room-temperature applications, where the estimated ionic conductivity falls well below 0.1 mS/cm.

\paragraph*{$Li_2BeF_4$.~~} This material was first synthesised in 1952 by Novoselova \textit{et al.} \cite{novoselova1952thermal} and exhibits one of the highest Li-diffusion at elevated temperatures.
However, due to the inability to resolve diffusion at lower temperatures, accurately quantifying its activation barrier remains challenging.
We anticipate that with sufficiently long simulations, on the order of several nanoseconds, it would be possible to quantify diffusion at lower temperatures as well, making this materials an excellent candidate for further investigation with machine learning techniques.
Furthermore, it has notably been used as a coolant in nuclear reactors \cite{kelleher2016observed}, highlighting the established interest in its synthesisability within experimental settings. 
Given these factors and the toxic nature of beryllium, it remains an interesting case study. 

\paragraph*{\normalfont $Li_8SeN_2$ and $Li_8TeN_2$.~~} Both selenium and tellurium nitrides, which demonstrate excellent Li-ion diffusion at higher temperatures were first synthesised in 2010 by Br{\"a}uling \textit{et al.} \cite{braunling2010synthesis}.
They are three-dimensional diffusers, but at lower temperatures they do not show high diffusion.
Furthermore, nitrides are generally among the most challenging materials to process due to relatively high-temperature synthesis routes \cite{xie2016transition}, we refrain from classifying them as the most promising candidates within this screening.

\begin{table*}[ht]
\small
  \caption{\ \ The structures that show significant diffusion at 1000 K but not at lower temperatures with FPMD. We report their stoichiometry, the repository and identifier from where they originated along with the experimental reference, band gap at the level of DFT-PBEsol, and ionic conductivity at 1000 K with pinball MD and FPMD. }
  \label{tbl:fast1000K}
  \begin{tabular*}{\textwidth}{@{\extracolsep{\fill}} l l l S[table-format=1.2] S[table-format=4.0] S[table-format=4.0]}
    \hline
    Structure & Database & Database-id & \multicolumn{1}{c}{Bandgap (eV)} & \multicolumn{1}{c}{Ionic conductivity} & \multicolumn{1}{c}{Ionic conductivity} \\
    & & & \multicolumn{1}{c}{DFT-PBEsol} & \multicolumn{1}{c}{pinball (mS/cm)} & \multicolumn{1}{c}{FPMD (mS/cm)} \\
    \hline
    $Li_2BeF_4$ & MPDS & S1935520 \cite{novoselova1952thermal, roy1954fluoride} & 7.49 & 10 & 1822 \\
    $Li_2Ti_4O_9$ & MPDS & S559372 \cite{dion1978tetratitanates} & 3.2 & 1042 & 1348 \\
    $LiY_2Ti_2S_2O_5$ & COD & 4124533 \cite{hyett2004electronically} & 1.25 & 74 & 1251 \\
    $Li_{10}BrN_3$ & MPDS & S1614518 \cite{marx1995reindarstellung} & 1.79 & 4247 & 886 \\
    $Li_2Cs_3Br_5$ & ICSD & 245978 \cite{PENTIN2008804} & 3.79 & 106 & 594 \\
    $Li_8SeN_2$ & MPDS & S1931016 \cite{braunling2010synthesis} & 1.88 & 467 & 588 \\
    $Li_8TeN_2$ & MPDS & S1931019 \cite{braunling2010synthesis} & 2.28 & 214 & 446 \\
    $LiCF_3SO_3$ & ICSD & 110018 \cite{Boltecv6022} & 6.78 & 40 & 384 \\
    $Li_2ZnBr_4$ & COD & 1517836 \cite{pfitznerneue1993} & 3.75 & 8 & 357 \\
    $LiBeP$ & ICSD & 670551 \cite{YADAV2015388}, 42037 \cite{ELMASLOUT1975213} & 2.75 & 18 & 356 \\
    $Li_5Br_2N$ & ICSD & 78836 \cite{marx1995reindarstellung} & 2.29 & 1351 & 344 \\
    $Li_{10}Si_2PbO_{10}$ & ICSD & 78326 \cite{brandes1994li10si2pbiio10} & 2.86 & 29 & 342 \\
    $Li_2ZnGeSe_4$ & COD & 7031897 \cite{zhang2015infrared} & 1.89 & 63 & 291 \\
    $LiCs_2I_3$ & ICSD & 245984 \cite{PENTIN2008804} & 3.41 & 1410 & 280 \\
    $LiSr_2Br_5$ & MPDS & S1941469 \cite{mahendran2003studies} & 3.53 & 176 & 263 \\
    $LiGaSe_2$ & COD & 1531591 \cite{isaenko2003growth} & 2.23 & 5 & 173 \\
    $LiP_7$ & ICSD & 23621 \cite{v1972lithiumphosphide} & 1.56 & 11 & 133 \\
    $LiMoPO_6$ & COD & 7701361 \cite{C9DT03451J} & 2.52 & 8 & 132 \\
    $LiY(MoO_4)_2$ & COD & 1008103 \cite{le1980structure, peng2019liy} & 3.22 & 660 & 68 \\
    $Li_{10}B_{14}Cl_2O_{25}$ & MPDS & S1803375 \cite{huang2001syntheses} & 6.30 & 13 & 65 \\
    $Li_2P_2PdO_7$ & COD & 1000333 \cite{laligant1992crystal} & 1.39 & 513 & 48 \textsuperscript{**} \\
    $Li_2B_3PO_8$ & MPDS & S1614518 \cite{hasegawa2014crystal} & 5.49 & 45 & 41 \\
    $Li_2B_2Se_5$ & COD & 1510746 \cite{lindemann2001first} & 1.84 & 84 & 20 \\
    $Li_8Bi_2(MoO_4)_7$ & ICSD & 54021 \cite{klevtsova1997synthesis} & 2.95 & 7 & 12 \\
    $Li_3AuS_2$ & COD & 4319430 \cite{huang2001syntheses} & 1.86 & 895 & 8 \\
    \hline
  \end{tabular*}
\end{table*}

\footnotetext{\textsuperscript{**}at 600 K, since we only performed FPMD simulations at one temperature due to high computational costs for this structure.}

\begin{table*} [ht]
\small
  \caption{\ \ The most promising structures that were found to be conducting with FPMD at lower temperatures. We report their stoichiometry, the repository and identifier from where they originated along with the experimental reference, band gap at the level of DFT-PBEsol, ionic conductivity at 500K, 750K and 1000 K, estimated activation energy using Arrhenius plot. As a comparison LGPS and LLZO have an ionic conductivity of 1101 mS/cm \cite{kahle2020high} and 295 mS/cm \cite{jalem2013concerted} respectively at 1000 K.}
  \label{tbl:fast500K}
  \begin{tabular*}{\textwidth}{@{\extracolsep{\fill}} l l l S[table-format=1.2] S[table-format=3.0] S[table-format=4.0] S[table-format=4.0] S[table-format=1.2]}
    \hline
    Structure & Database & Database-id & \multicolumn{1}{c}{Bandgap (eV)} & \multicolumn{1}{c}{Ionic conductivity} & \multicolumn{1}{c}{Ionic conductivity} & \multicolumn{1}{c}{Ionic conductivity} & \multicolumn{1}{c}{Activation} \\
    & & & \multicolumn{1}{c}{DFT-PBEsol} & \multicolumn{1}{c}{at 500 K (mS/cm)} & \multicolumn{1}{c}{at 750 K (mS/cm)} & \multicolumn{1}{c}{at 1000 K (mS/cm)} & \multicolumn{1}{c}{energy (eV)} \\
    \hline
    $Li_4CO_4$ & ICSD & 245389 \cite{vcanvcarevic2007possible} & 5.26 & 235 & 551 & 726 & 0.15 \\ 
    $LiCsI_2$ & ICSD & 245986 \cite{dvoryanova2010lif, manyakova2018stable} & 3.32 & 203 & 340 & 698 & 0.16 \\
    $Li_3CsBr_4$ & ICSD & 245982 \cite{PENTIN2008804} & 3.73 & 456 & 1035 & 1616 & 0.17 \\
    $Li_3Cs_2Br_5$ & ICSD & 245980 \cite{PENTIN2008804} & 3.90 & 181 & 358 & 844 & 0.18 \\
    $Li_7NbO_6$ & MPDS & S1818764 \cite{hauck1969lithiumhexaoxometallate} & 3.58 & 77 & 288 & 418 & 0.21 \\
    $Li_3Cs_2I_5$ & ICSD & 245987 \cite{MEYER19831353, NAGARKAR2015161} & 3.43 & 161 & 554 & 1112 & 0.23 \\
    $LiCs_3Cl_4$ & ICSD & 245969 \cite{PENTIN2008804, ortner2019old} & 4.38 & 35 & 112 & 283 & 0.23 \\
    $Li_4Mo_3O_8$ & MPDS & S1614518 \cite{hibble1997structure} & 1.17 & 6 & 32 & 54 & 0.25 \\
    $Li_5NaN_2$ & ICSD & 92313 \cite{SCHON2000449} & 1.49 & 279 & 1268 & 3609 & 0.28 \\
    \hline
  \end{tabular*}
\end{table*}

\subsubsection{Fast Li-ion conductors.~~} \label{diff}

\begin{figure}[ht]
\centering
  \includegraphics[width=\columnwidth]{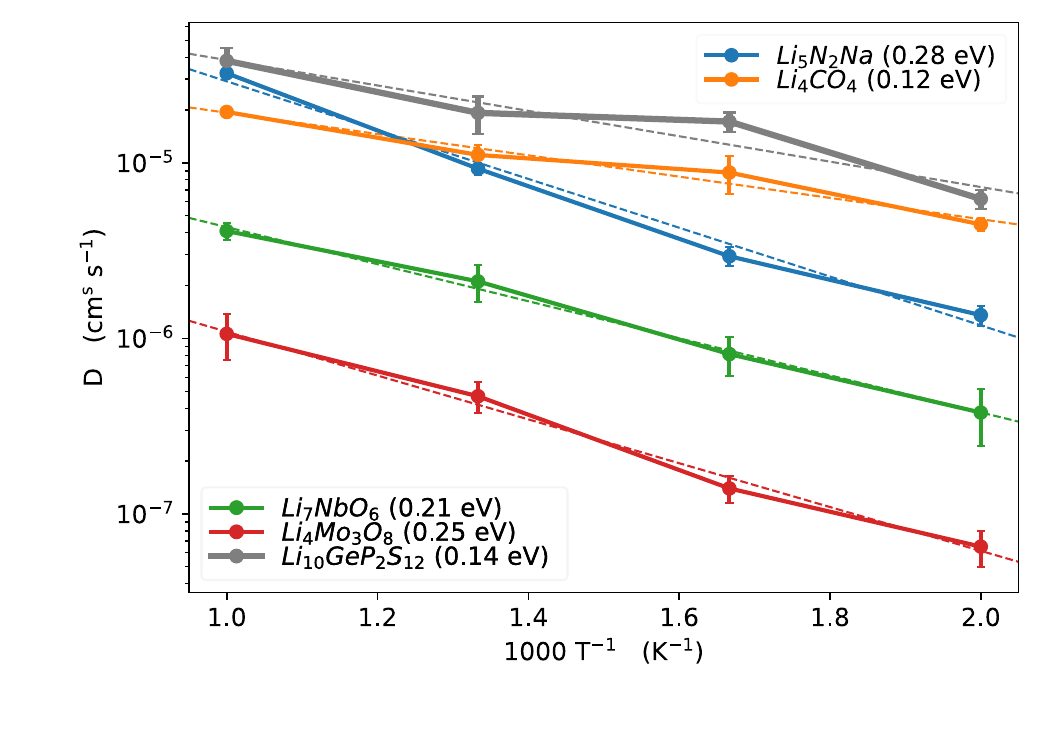}
  \caption{Diffusion coefficients derived from our FPMD simulations for the most promising oxides and nitrides. The dashed line represents the best-fit line, with the slope corresponding to the activation barriers, indicated in brackets (eV). We additionally show LGPS for comparison, with data taken from the work of Kahle \textit{et al.} \cite{kahle2020high}.}
  \label{fgr:activation1}
\end{figure}

\begin{figure}[htb]
\centering
  \includegraphics[width=\columnwidth]{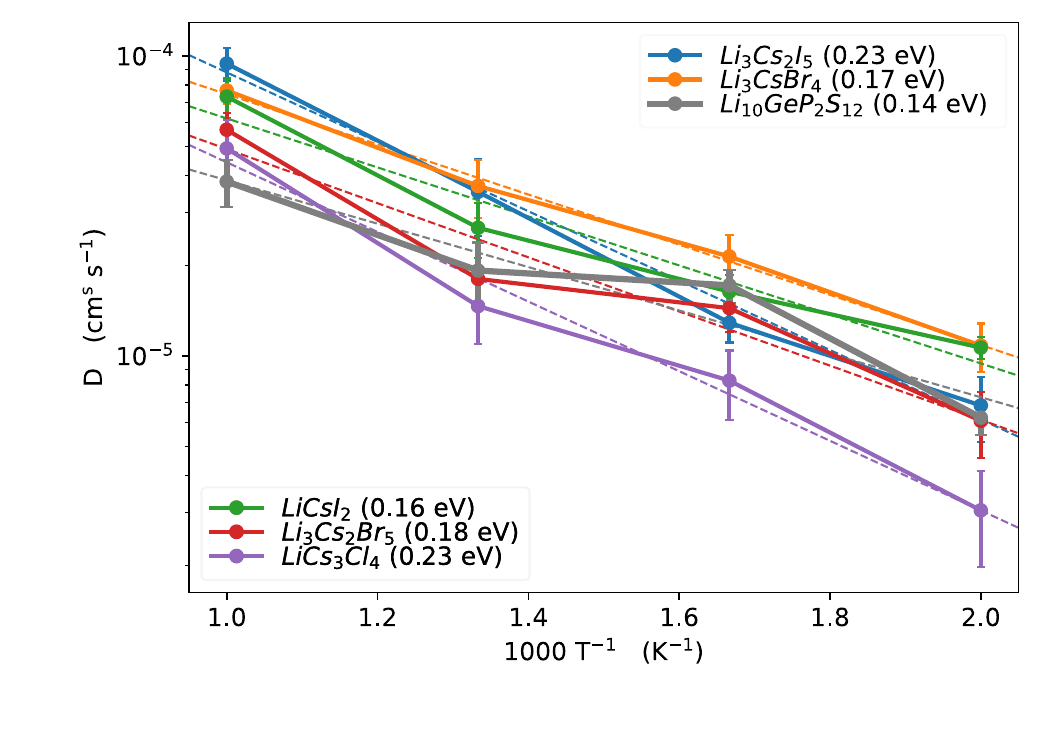}
  \caption{Diffusion coefficients derived from our FPMD simulations for the most promising halides. The dashed line represents the best-fit line, with the slope corresponding to the activation barriers, indicated in brackets (eV). We additionally show LGPS for comparison, with data taken from the work of Kahle \textit{et al.} \cite{kahle2020high}.}
  \label{fgr:activation2}
\end{figure}

In this section we discuss the most promising materials identified as candidates for solid-state electrolytes. 
These materials are of particular interest due to their potential applications, characterised by their fast ionic conduction, which allows us to resolve Li-ion diffusion even at low temperatures and estimate activation barriers, as illustrated in Fig. \ref{fgr:activation1} and \ref{fgr:activation2}. 
For comparison, we also include tetragonal-LGPS, with data taken from the work of Kahle \textit{et al.} \cite{kahle2020high}.
Although the materials identified in our screening, with the exception of the Cs-doped Li-halides, do not exhibit ionic conductivities as high as LGPS, their excellent activation energies suggest that they could perform very well as conductors at room temperature.
We list these 9 materials in Table \ref{tbl:fast500K} along with details on their provenance, band gap, ionic conductivity at 1000 K and projected conductivity at room temperature, and activation barrier.

\begin{figure} [H]
\centering
  \includegraphics[width=\columnwidth]{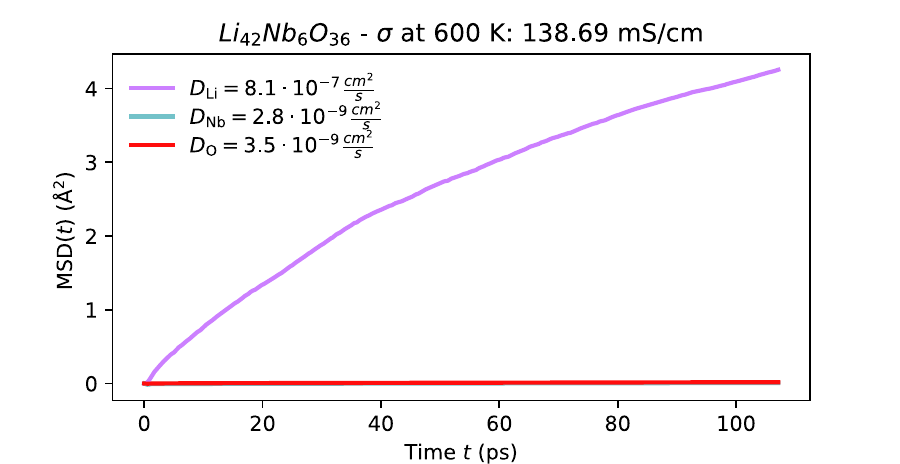}
  \includegraphics[width=\columnwidth]{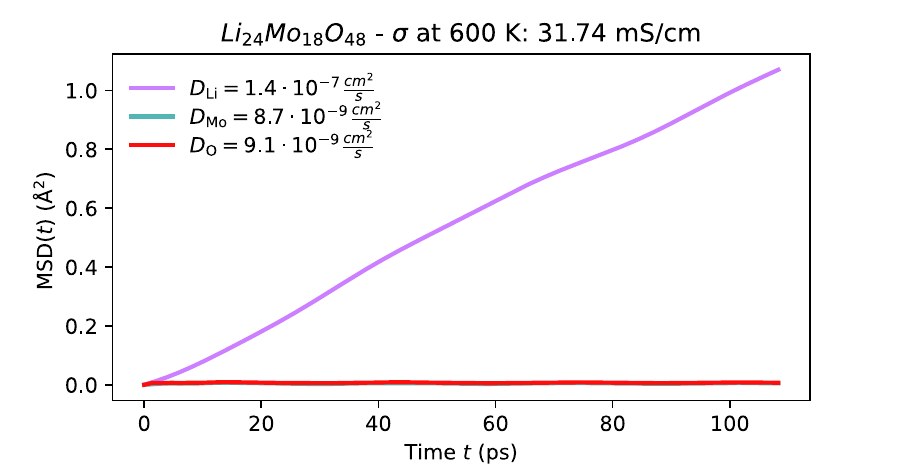}
  \includegraphics[width=\columnwidth]{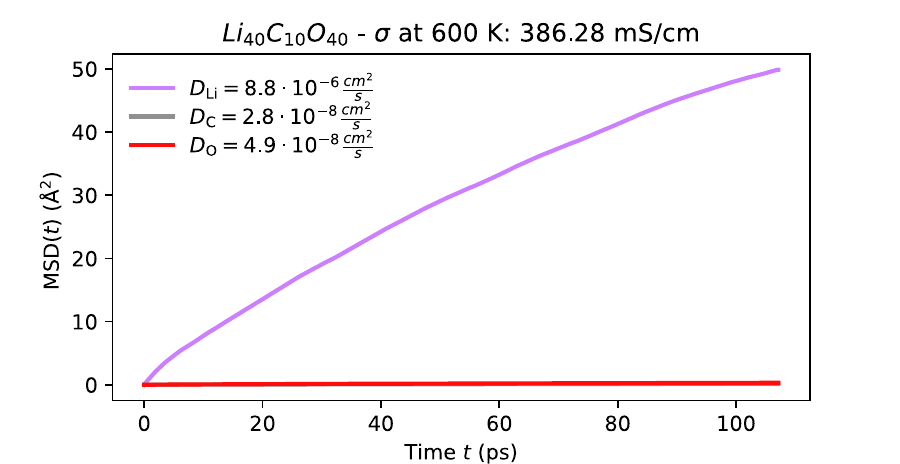}
  \includegraphics[width=\columnwidth]{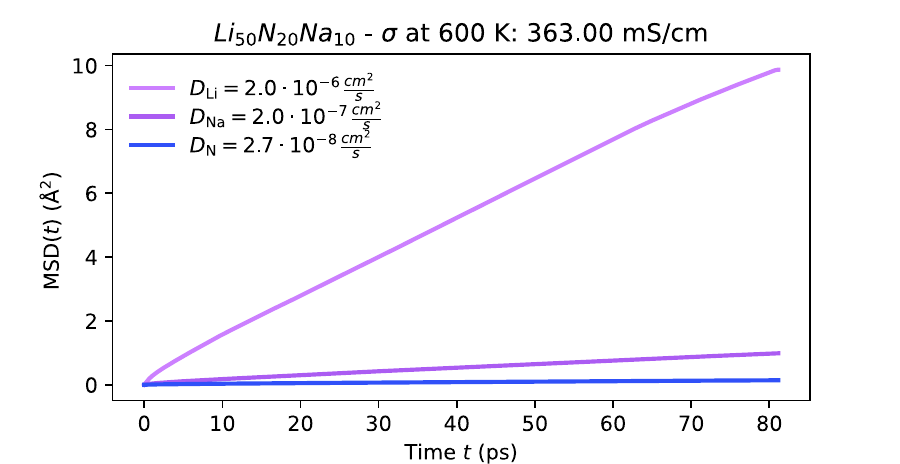}
  \caption{MSD plot of Li along with host-lattice species of the oxides $Li_7NbO_6$, $Li_4Mo_3O_8$ and $Li_4CO_4$, and the nitride $Li_5NaN_2$ at 600 K from FPMD.}
  \label{fgr:msd_lno}
\end{figure}

\paragraph*{$Li_7NbO_6$.~~} First synthesised in 1969 \cite{hauck1969lithiumhexaoxometallate}, this material has been the subject of multiple experimental studies \cite{sorescu1995excimer, muehle2004new}; however it has never been investigated as a Li-ion conductor until He \textit{et al.} proposed $Li_7NbO_6$ as a potential ionic conductor \cite{he2019crystal}. 
Subsequent investigations by Feng \textit{et al.}, reported a low ionic conductivity of 0.008 mS/cm along with a significantly higher activation barrier and lower diffusion than our predictions \cite{feng2024theoretical}, as illustrated in Fig. \ref{fgr:msd_lno}.
These discrepancies arise from structural differences, as the two structures possess different space groups and lattice parameters despite sharing the same stoichiometry.
Feng \textit{et al.} further investigated doped with tungsten to enhance the conductivity at room temperature to 0.28 mS/cm, which remains an order of magnitude lower than our estimated value of 5 mS/cm.
Nevertheless, these findings support doping as an effective strategy to further improve the ionic conductivity. 
Given these promising properties and the substantial experimental background already established, we propose that $Li_7NbO_6$ holds significant potential as an excellent electrolyte for future applications.

\paragraph*{\normalfont $Li_4Mo_3O_8$.~~} This molybdenum oxide exhibit high Li-ion conductivity, as illustrated in Fig. \ref{fgr:msd_lno}. 
The yttrium-doped variant was first synthesised in 1980, while the undoped form was synthesised in 1999. 
Our FPMD simulations indicate that both materials possess low activation barriers, with the yttrium-doped version performing slightly better. 
Based on the activation energies, we estimate the ionic conductivities at room temperature to be 0.2 mS/cm.
We strongly recommend further experimental studies to validate these findings and confirm the potential as solid-state electrolytes. 

\paragraph*{$Li_5NaN_2$.~~} The well known $Li_3N$ was first proposed in 1935 \cite{zintl1935konstitution} and has since spawned an broad class of Li-ion conductors that continue to attract attention today \cite{li2025nitride}. 
While studying $Li_3N$ in 2000, Sch\"on \textit{et al.} proposed the metastable $Li_5NaN_2$, as a derivative of $Li_3N$, with relatively low formation energy\cite{SCHON2000449}.
Our calculations are able to resolve Li-ion diffusion at lower temperatures as shown in Fig. \ref{fgr:msd_lno} and indicate a relatively higher activation barrier of 0.28 eV, which corresponds to an estimated room-temperature ionic conductivity of 4 mS/cm. 
Given the success of doping in enhancing the conductivity of $Li_3N$ \cite{lapp1983ionic, tapia2013chemistry}, we are optimistic that similar strategies could improve the performance of $Li_5NaN_2$, making it a promising candidate. 
However, a notable challenge is the concurrent diffusion of Na-ions, highlighting the need for targeted compositional or structural engineering to restrict Na-ion mobility.

\paragraph*{$Li_4CO_4$.~~} We examined this material in four distinct crystal structures with the same stoichiometry, all of which exhibited excellent Li-ion diffusion. 
However, this material remains a theoretical structure that exists at high pressure and appears to simply be a variant of Li-doped carbonates which may decompose at ambient temperature and pressure \cite{vcanvcarevic2007possible}. 
Given these uncertainties, we are cautious about its potential as an electrolyte. 
Despite its low activation barrier, we have opted not to list it as a promising candidate until further validation can be conducted, and it is established that these materials can exist at normal temperature and pressure without decomposing into simple carbonates.
Based on the activation energy, we estimate the ionic conductivity at room temperature to be 37 mS/cm.

\begin{figure}[H]
\centering
  \includegraphics[width=\columnwidth]{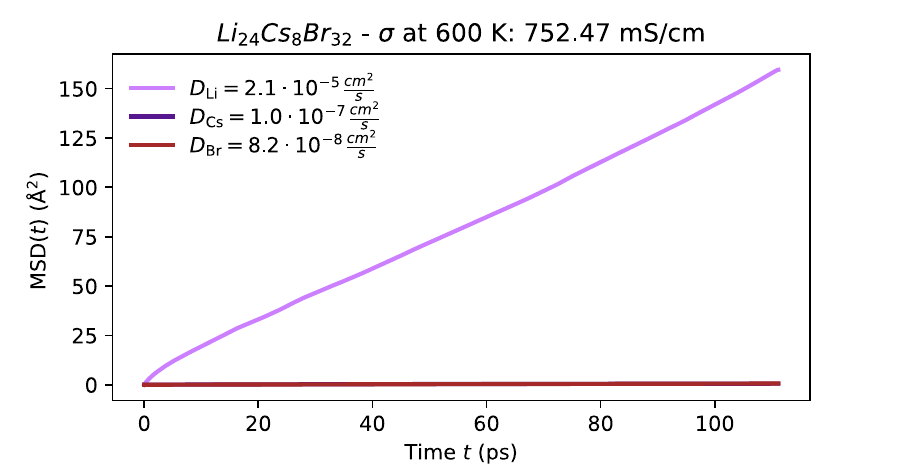}
  \includegraphics[width=\columnwidth]{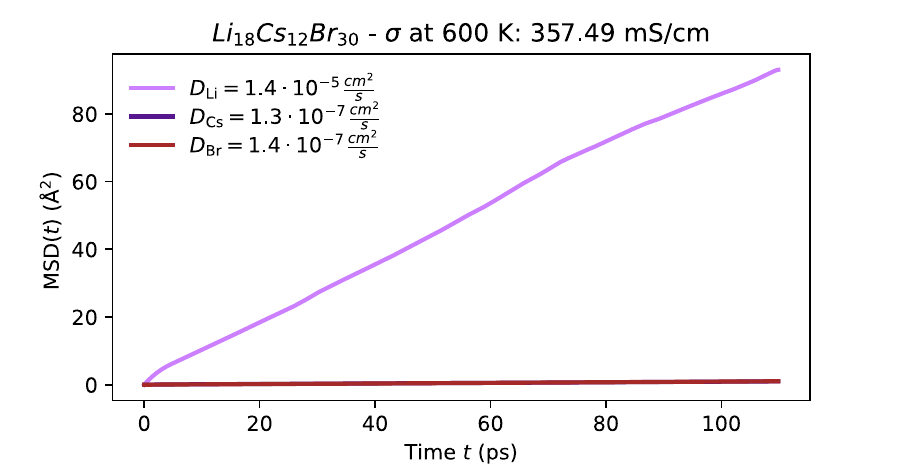}
  \includegraphics[width=\columnwidth]{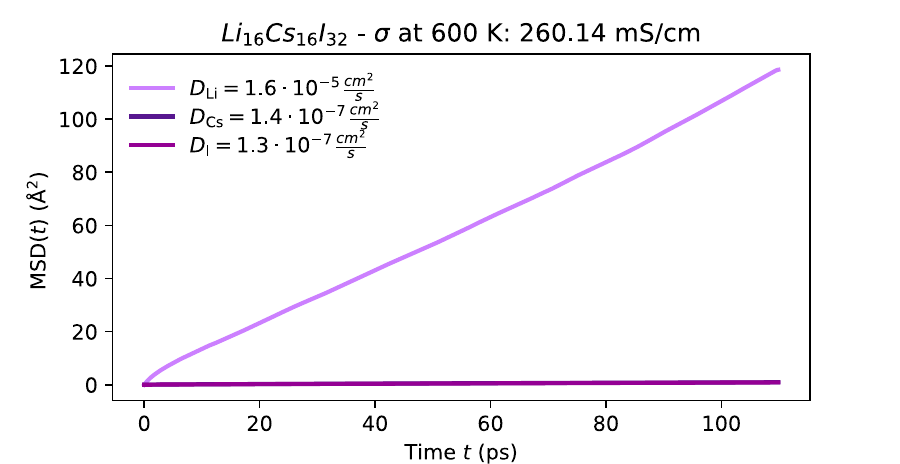}
  \includegraphics[width=\columnwidth]{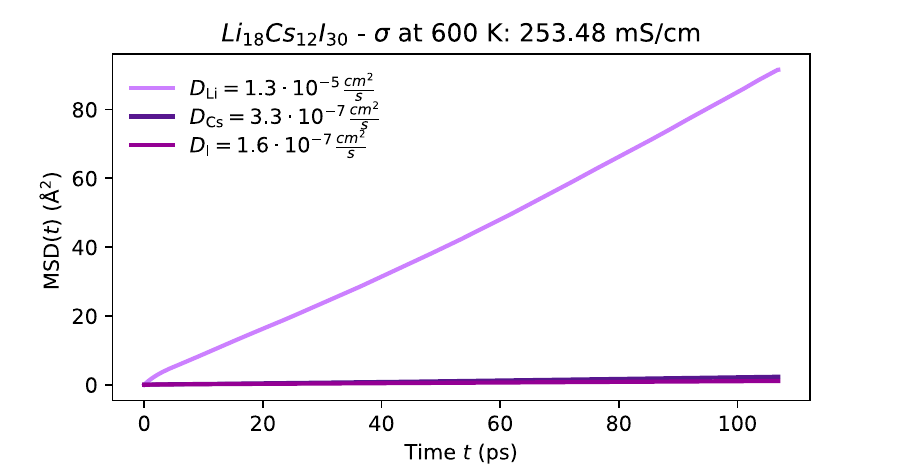}
  \includegraphics[width=\columnwidth]{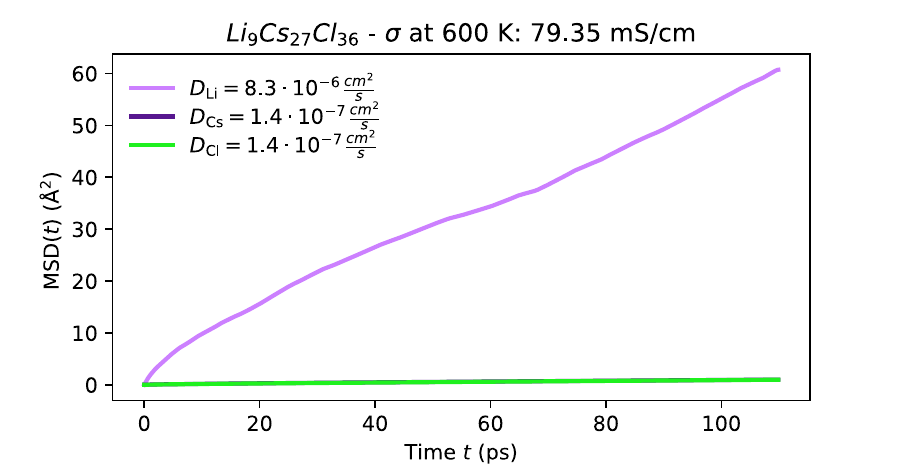}
  \caption{MSD plot of Li along with host-lattice species in Cs-doped bromides at 600 K from FPMD.}
  \label{fgr:msd_lcs}
\end{figure}

\paragraph*{\normalfont Cs-doped Li-halides.~~} Amongst the most promising materials we identify are $Li_3CsBr_4$, $Li_3Cs_2Br_5$, $LiCs_3Cl_4$, $LiCsI_2$ and $Li_3Cs_2I_5$, some of which were first proposed by Pentin \textit{et al.} using \textit{ab initio} methods \cite{PENTIN2008804}. 
Each of these materials demonstrates high ionic conductivity as shown in Fig. \ref{fgr:msd_lcs}; and low activation barrier ranging from 0.15 to 0.25 eV as illustrated in \ref{fgr:activation2}. 
Although experimental validation for these materials is still pending, their synthesis appears feasible.
Most notably, $LiCsI_2$ \cite{manyakova2018stable}, $Li_3Cs_2I_5$ \cite{NAGARKAR2015161} and $Li_2CsI_3$ \cite{MEYER19831353} (which was also proposed by Kahle \textit{et al.} \cite{kahle2020high}) have all been successfully synthesised, and $Li_3Cs_2Br_5$  may also be synthesised following a similar approach to $Li_3Cs_2I_5$.
While these compounds have not yet been explored as ionic conductors, our screening suggests significant potential for future experimental validation.
Though the synthesisability of the other Cs-Li-halides remains uncertain, they may depend on methodologies similar to those used for this ternary system \cite{MEYER19831353}. 
Given these considerations, we hesitate from designating these materials as the top candidates within this screening, pending further experimental investigations.
However, it is important to emphasise that despite these uncertainties, this system represents a promising avenue for further exploration and warrants both experimental and theoretical pursuits.

\begin{figure}[ht]
\centering
  \includegraphics[width=\columnwidth]{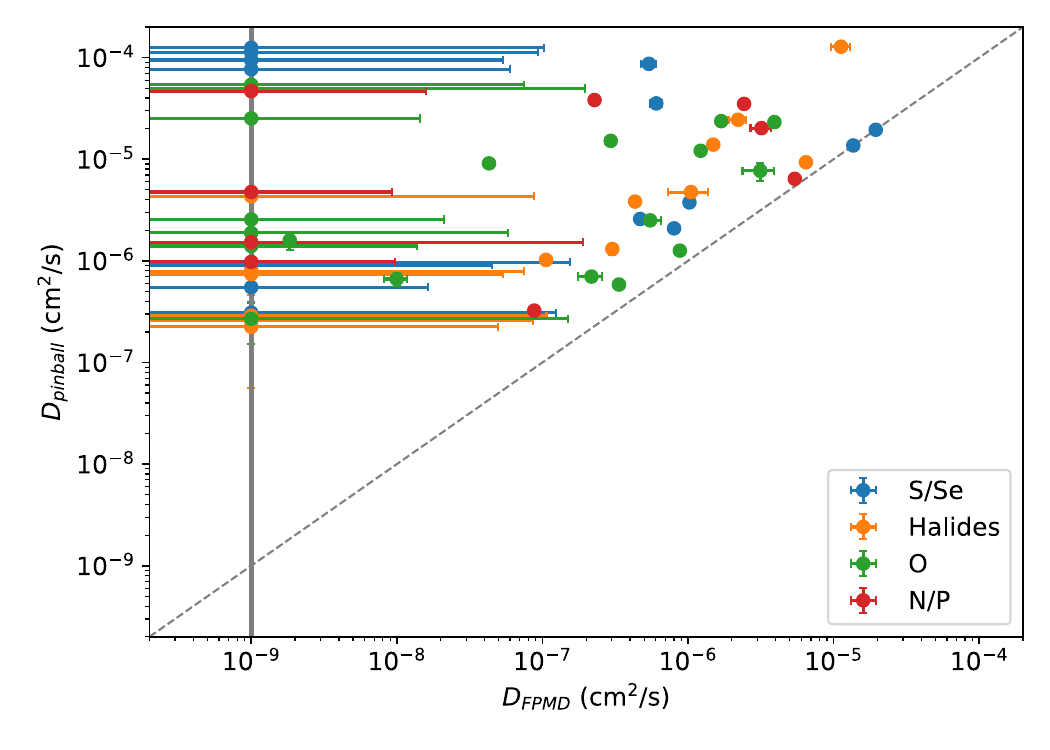}
  \caption{Comparison of diffusion coefficients at 1000 K, obtained with the pinball model and FPMD, categorised by the predominant anion. The bold-grey line represents the threshold below which MSD convergence cannot be achieved with FPMD, serving as the lower bound for diffusion. The dashed-grey line denotes the identity line, with all of the structures lie on or above it, suggesting that the pinball model typically overestimates the actual ionic conductivity. }
  \label{fgr:diff_pbfp}
\end{figure}

As a final validity of the pinball model, we compare the diffusion of structures in sections \ref{non-diff}, \ref{pot-diff} and \ref{diff} computed with FPMD and the pinball model as illustrated in Fig. \ref{fgr:diff_pbfp}. 
On the surface, first-principles diffusion is not well reproduced by the pinball model, which generally tends to overestimate the diffusion coefficient. 
Interestingly, it does not underestimate the ionic conductivity, i.e. the number of false negatives is low.
However, the true number of false negatives would be much higher than illustrated in Fig. \ref{fgr:diff_pbfp}, as there are many materials that were classified as non-conducting at the level of pinball MD that may be conducting \cite{xie2024high}, which underscores the difficulty in quantifying the true predictive power of the pinball model.  
Considering, that the pinball model rediscovers 77 ionic conductors, the true number of false positives is low, which sets an upper bound of \textasciitilde85\% as the overall predictive rate of our workflow.

\section{Conclusions and outlook} \label{conclusion}

We conducted a high-throughput computational screening of over 30,000 lithium containing experimental structures sourced from the MPDS, ICSD, and COD repositories.
Through the application of several structural filters, we identified approximately 1,500 unique crystal structures suitable for electronic structure calculations.
We determined the band gaps for these structures at the level of DFT with the PBEsol functional, and identified nearly 1,000 as electronic insulators.
To investigate Li-ion diffusion, we implemented a self-consistent MD workflow in AiiDA, utilising the computationally efficient and highly accurate pinball model.
From these simulations, we identified 132 fast Li-ion diffusers, 77 of which were previously recognised in the literature as Li-ion conductors.
The remaining 55 materials were further examined using full first-principles MD simulations, leading to the discovery of seven promising materials, including the oxides $LiY(MoO_4)_2$, $Li_4Mo_3O_8$ and $Li_7NbO_6$, the nitrides $Li_8SeN_2$ and $Li_8TeN_2$, and Cs-doped iodides $LiCsI_2$ and $Li_3Cs_2I_5$.
These materials demonstrated excellent activation barriers and Li-ion diffusion near room temperature comparable to or exceeding that of LGPS, a well-known Li-ion superconductor. 
However, it is important to note that this estimation is based on the extrapolation of the Arrhenius plot, where a change of slope is possible.
Additionally, we identified five other materials with similar levels of ionic conductivity, although their synthesisability remains uncertain.
Furthermore, we identified 25 potential fast Li-ion conductors, including $Li_2BeF_4$ and $Li_8SeN_2$ that exhibit high Li-ion diffusion at elevated temperatures.
However, due to the limited timescales accessible to first-principles MD simulations, we were unable to resolve their diffusion behaviour at lower temperatures.
These materials may be promising candidates for further study using machine learning techniques, which could enable more extended simulations at lower temperatures.

Finally, we expect that the extensive first-principles data generated through this study will play a crucial role in training the next generation of machine learning interatomic potentials (MLIP).
To facilitate this, we have made all our first-principles data, along with comprehensive provenance, publicly available on the open-source Materials Cloud archive platform \cite{mca_data2024}.
This dataset could be particularly instrumental in developing a "universal-Li" MLIP, which has the potential to unlock new and intriguing systems in the future, and serve as a foundational tool for the study of next-generation solid-state Li-ion batteries.

\section*{Author Contributions} \label{contri}

We use in the following the CRediT (Contributor Roles Taxonomy) author statement.
T.S.T.: conceptualisation, data curation, formal analysis, methodology, software, validation, visualisation,  writing – original draft; 
L.E.: methodology, software; 
N.M.: funding acquisition, project administration, supervision. 
All authors: writing – review \& editing.

\section*{Conflicts of interest} \label{confli}

There are no conflicts to declare.

\section*{Acknowledgements} \label{acknwldg}

This project has received funding from the European Union’s Horizon 2020 research and innovation programme under grant agreement No 957189. 
The project is part of BATTERY 2030+, the large-scale European research initiative for inventing the sustainable batteries of the future.
We acknowledge support from the NCCR MARVEL, a National Centre of Competence in Research, funded by the Swiss National Science Foundation (grant number 205602).
This work was supported by a grant from the Swiss National Supercomputing Centre (CSCS) under project ID 465000416 (LUMI-G) and project ID mr33 (EIGER).



\balance


\phantomsection
\addcontentsline{toc}{section}{\refname}
\bibliography{biblio} 
\bibliographystyle{rsc} 



\newpage
\FloatBarrier

\twocolumn[
  \begin{@twocolumnfalse}
\vspace{1em}
\sffamily
\begin{tabular}{m{0cm} p{17cm}}

 & \noindent\LARGE{\textbf{Supplementary information$^\dag$}} \\
\vspace{0.3cm} & \vspace{0.3cm} \\

 & \noindent\LARGE{\textbf{Novel fast Li-ion conductors for solid-state electrolytes from first-principles}} \\
\vspace{0.3cm} & \vspace{0.3cm} \\

 & \noindent\large{Tushar Singh Thakur,$^{\ast}$\textit{$^{a}$} Loris Ercole,\textit{$^{a}$} and Nicola Marzari\textit{$^{a,b,c}$}} \\
\vspace{0.3cm} & \vspace{0.3cm} \\

\end{tabular}

 \end{@twocolumnfalse} \vspace{0.6cm}

  ]

\noindent\normalsize{
The following sections exhaustively illustrate the MSD plots derived from FPMD simulations of all the structures discussed in the main text.
}

\bigskip
\FloatBarrier 

\renewcommand{\thesection}{S\arabic{section}}
\renewcommand{\theHsection}{S\arabic{section}}%
\setcounter{section}{0}

\renewcommand{\thefigure}{S\arabic{figure}}
\setcounter{figure}{0}

\renewcommand{\thetable}{S\arabic{table}}
\setcounter{table}{0}

\phantomsection
\section{Fast Li-ion conductors} \label{S1_fast}

We discover 9 novel fast Li-ion conductors that exhibit significant diffusion at low temperatures along with desirable activation energy.
We show the MSD plots at 1000 K, 750 K, 600 K and 500 K.

\begin{figure}[H]
\centering
  \includegraphics[width=\columnwidth]{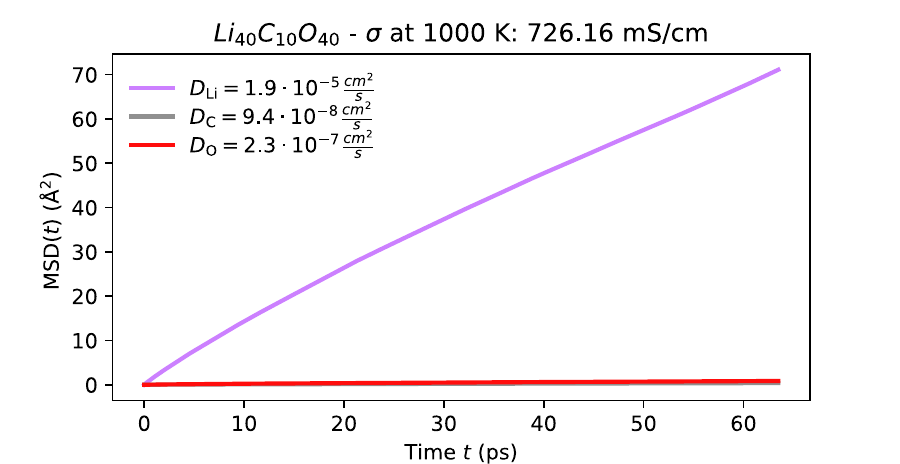}
  \includegraphics[width=\columnwidth]{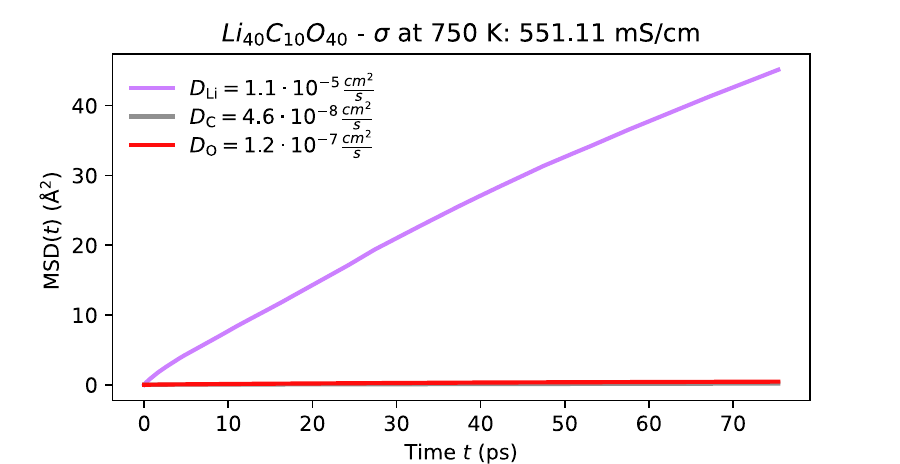}
  \caption{MSD(t) plot of Li along with host-lattice species of $Li_4CO_4$ at all temperatures studied with FPMD.}
\end{figure}

\begin{figure}[H]
\centering
  \includegraphics[width=\columnwidth]{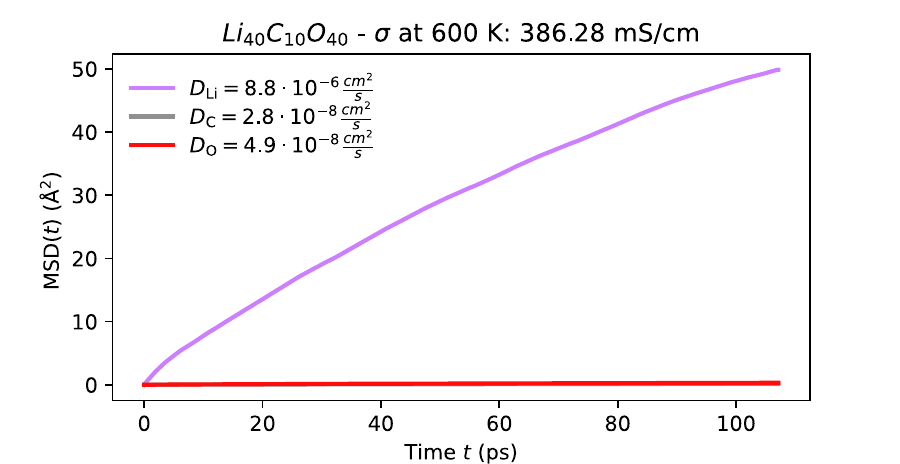}
  \includegraphics[width=\columnwidth]{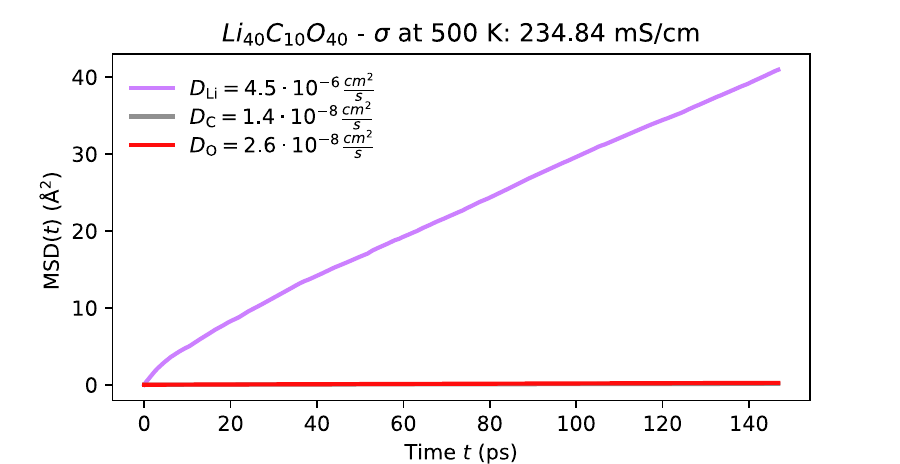}
  \caption{MSD(t) plot of Li along with host-lattice species of $Li_4CO_4$ at all temperatures studied with FPMD.}
\end{figure}

\footnotetext{\textit{$^{\ast}$~E-mail: tushar.thakur@epfl.ch }}
\footnotetext{\textit{$^{a}$~Theory and Simulation of Materials (THEOS), and National Centre for Computational Design and Discovery of Novel Materials (MARVEL), \'Ecole Polytechnique F\'ed\'erale de Lausanne, CH-1015 Lausanne, Switzerland }}
\footnotetext{\textit{$^{b}$~PSI Center for Scientific Computing, Theory and Data, Paul Scherrer Institute, 5232 Villigen PSI, Switzerland }}
\footnotetext{\textit{$^{c}$~Theory of Condensed Matter, Cavendish Laboratory, University of Cambridge, Cambridge CB3 0US, United Kingdom }}

\begin{figure}[H]
  \centering
  \includegraphics[width=\columnwidth]{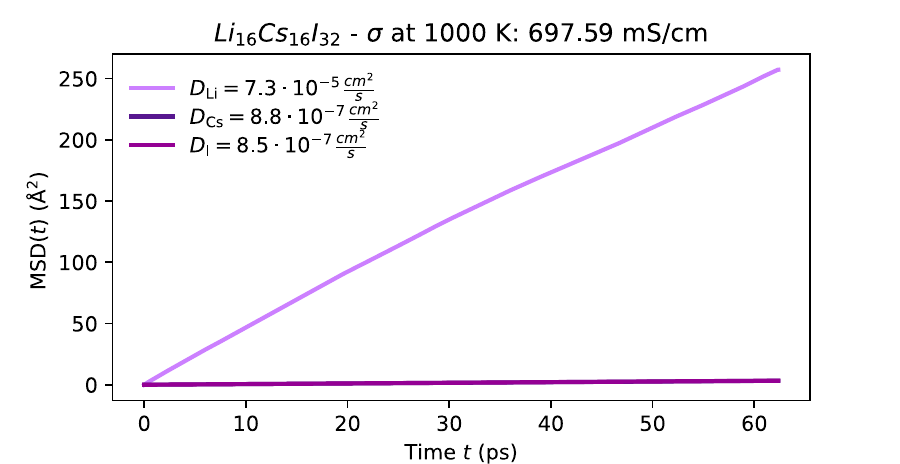}
  \includegraphics[width=\columnwidth]{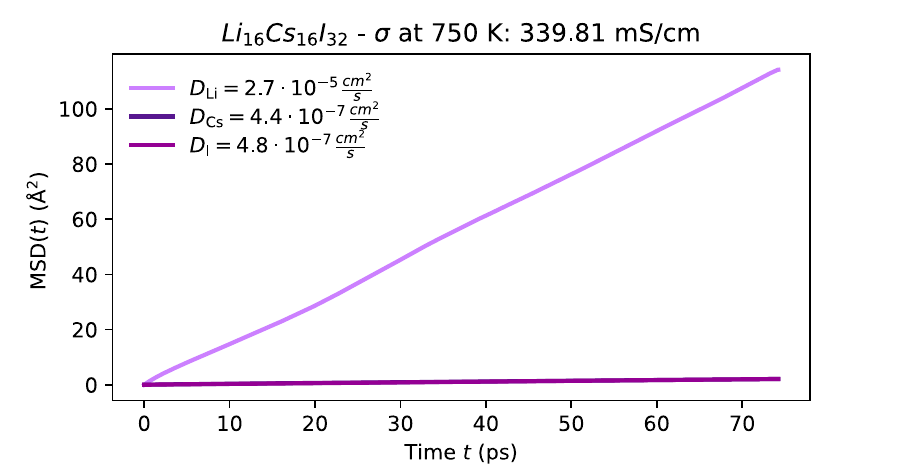}
  \includegraphics[width=\columnwidth]{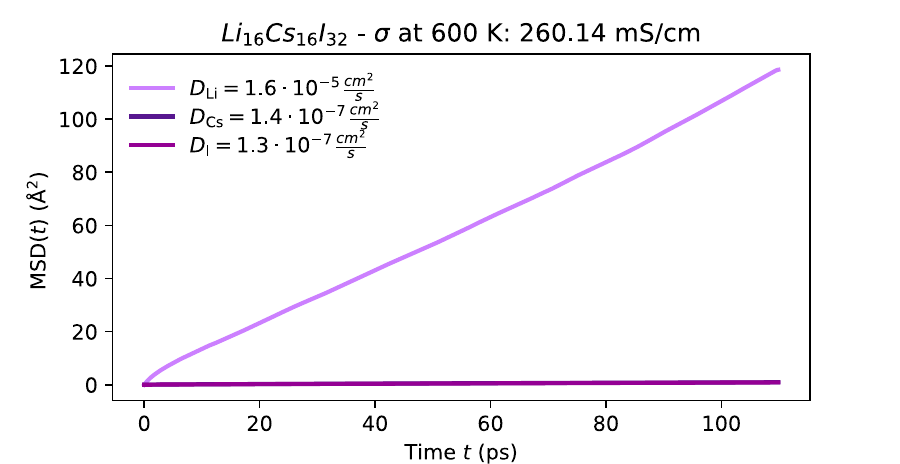}
  \includegraphics[width=\columnwidth]{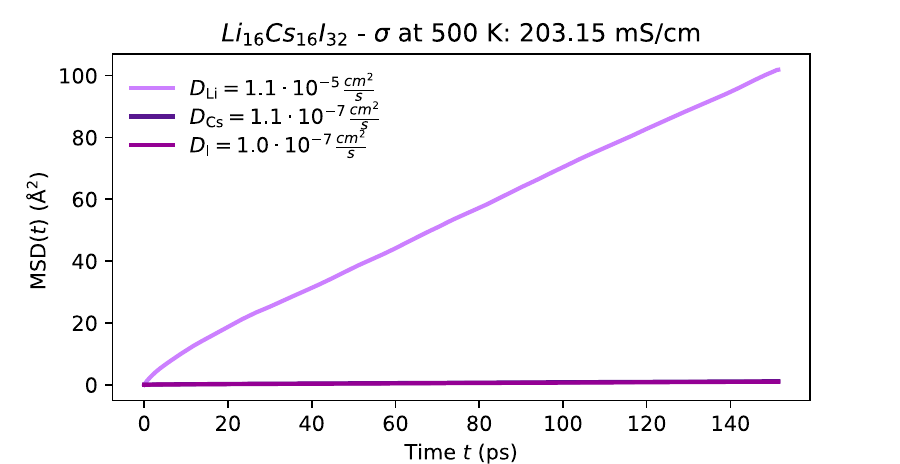}
  \caption{MSD(t) plot of Li along with host-lattice species of $LiCsI_2$ at all temperatures studied with FPMD.}
\end{figure}

\begin{figure}[H]
  \centering
  \includegraphics[width=\columnwidth]{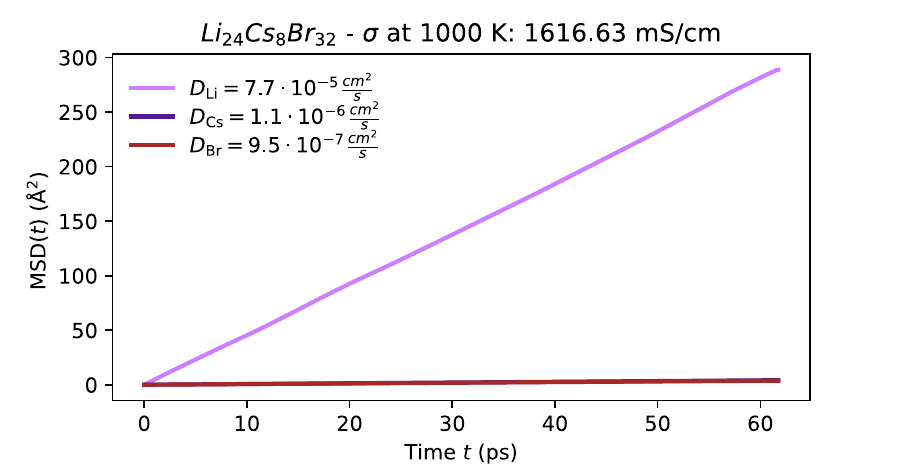}
  \includegraphics[width=\columnwidth]{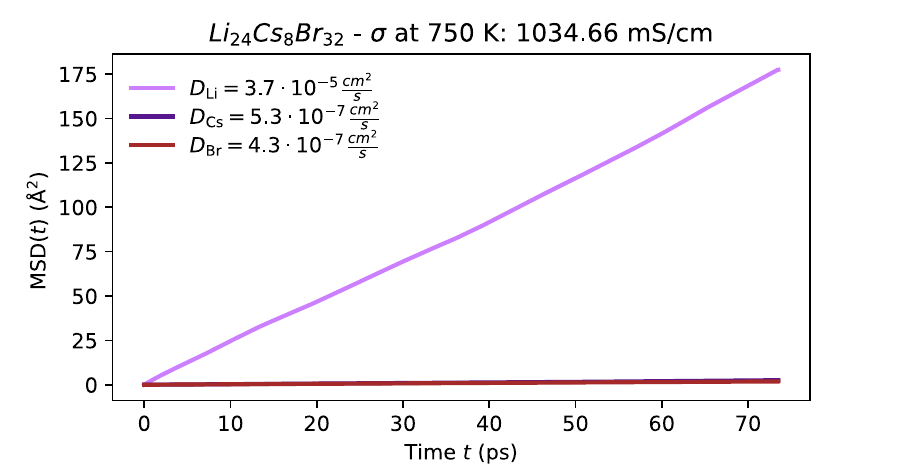}
  \includegraphics[width=\columnwidth]{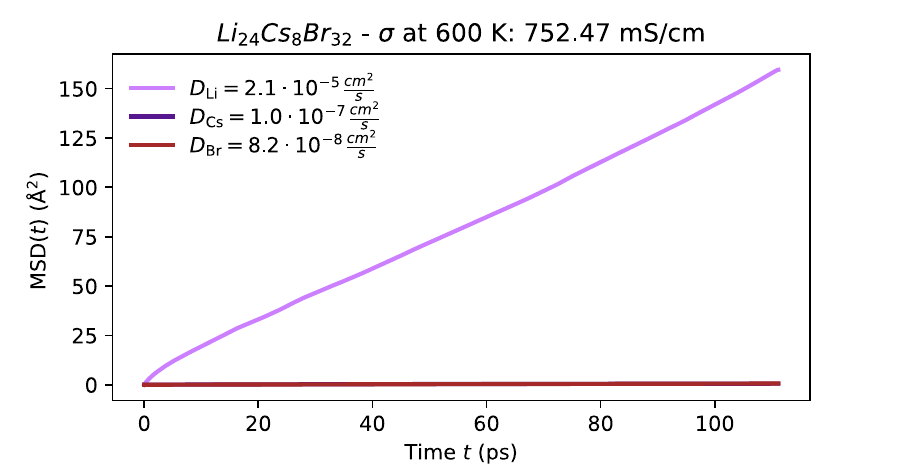}
  \includegraphics[width=\columnwidth]{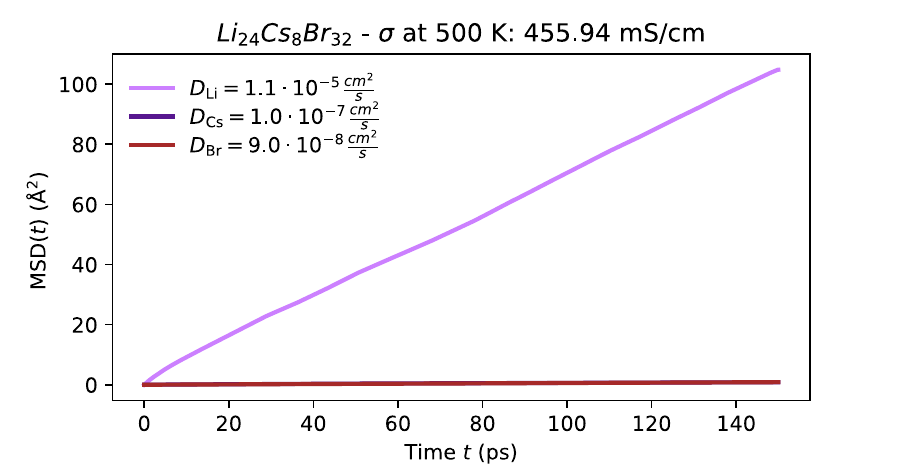}
  \caption{MSD(t) plot of Li along with host-lattice species of $Li_3CsBr_4$ at all temperatures studied with FPMD.}
\end{figure}

\begin{figure}[H]
  \centering
  \includegraphics[width=\columnwidth]{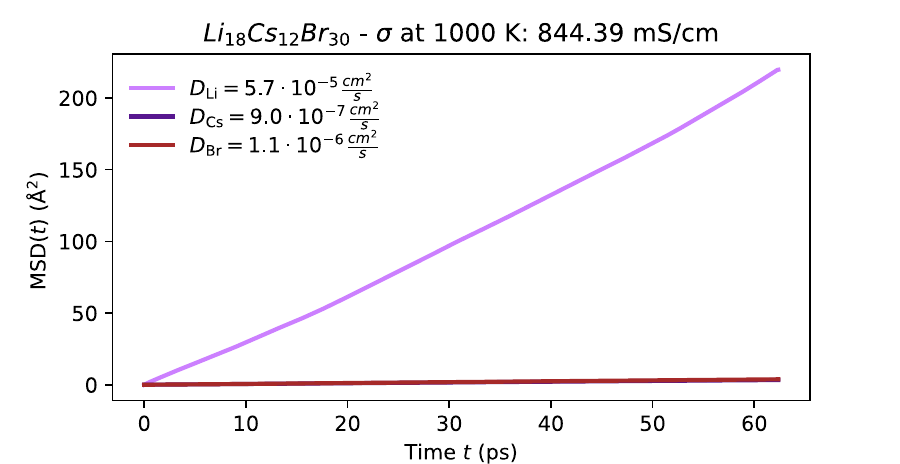}
  \includegraphics[width=\columnwidth]{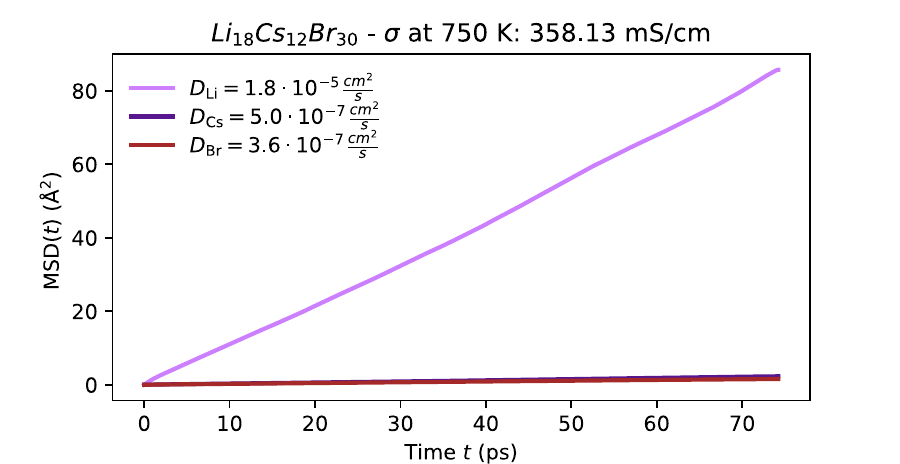}
  \includegraphics[width=\columnwidth]{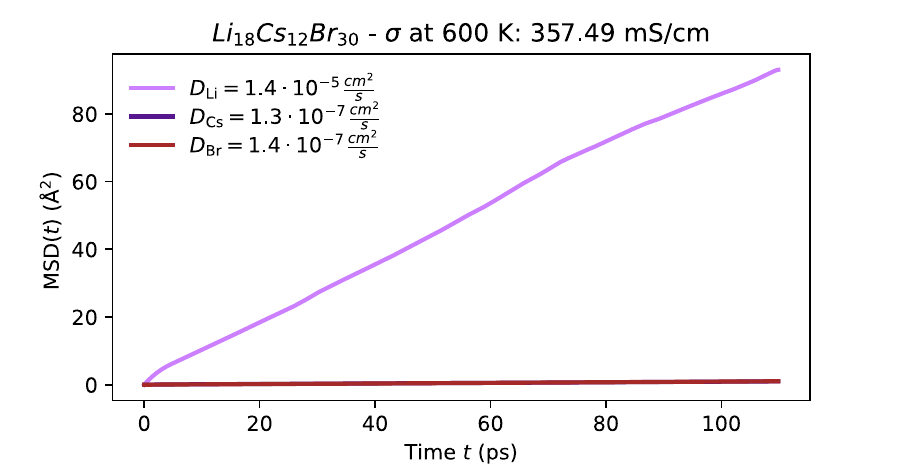}
  \includegraphics[width=\columnwidth]{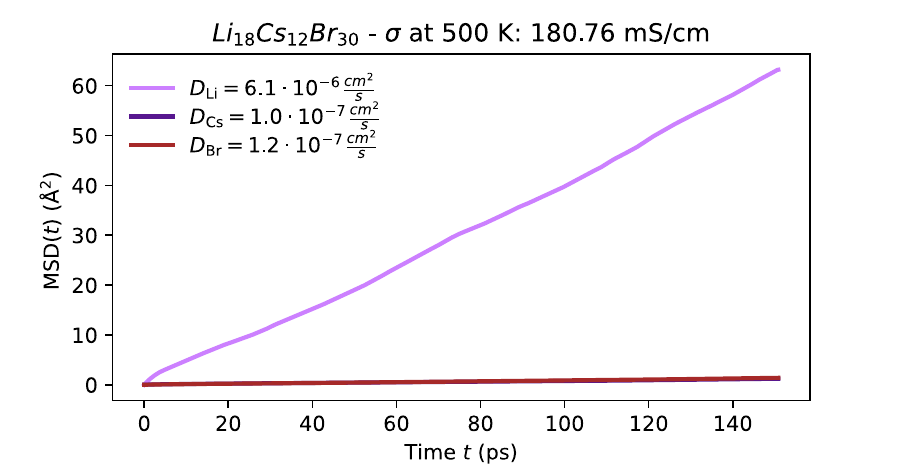}
  \caption{MSD(t) plot of Li along with host-lattice species of $Li_3Cs_2Br_5$ at all temperatures studied with FPMD.}
\end{figure}

\begin{figure}[H]
\centering
  \includegraphics[width=\columnwidth]{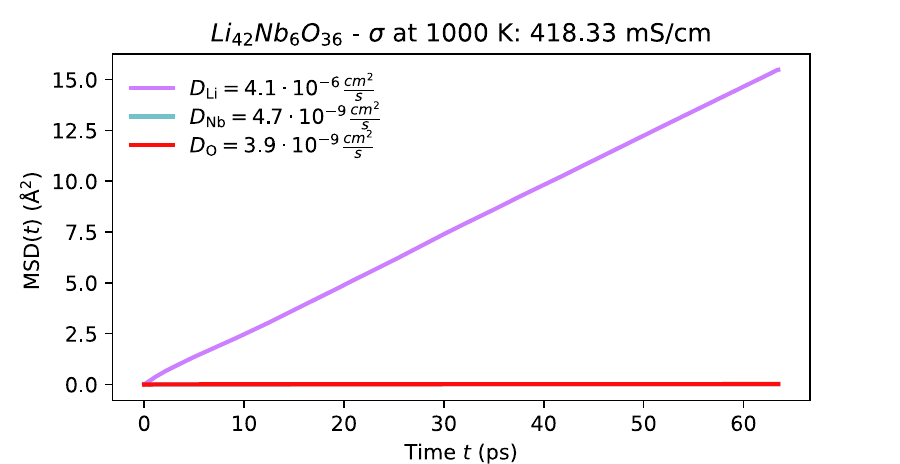}
  \includegraphics[width=\columnwidth]{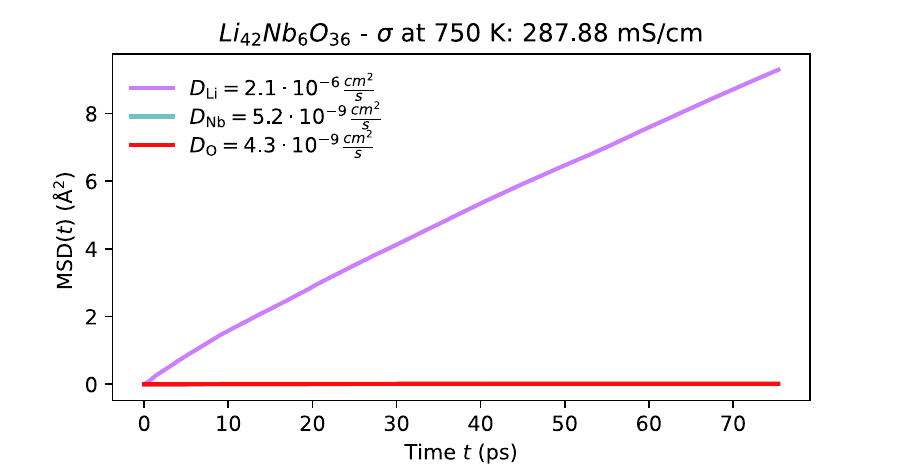}
  \includegraphics[width=\columnwidth]{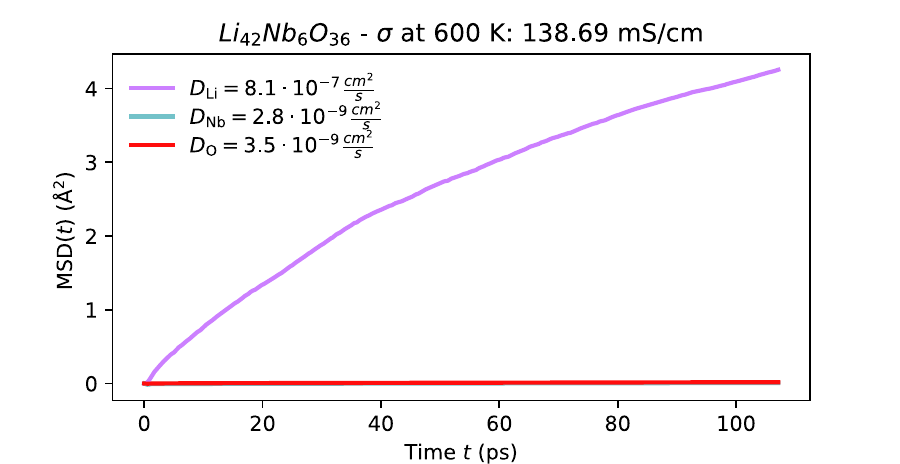}
  \includegraphics[width=\columnwidth]{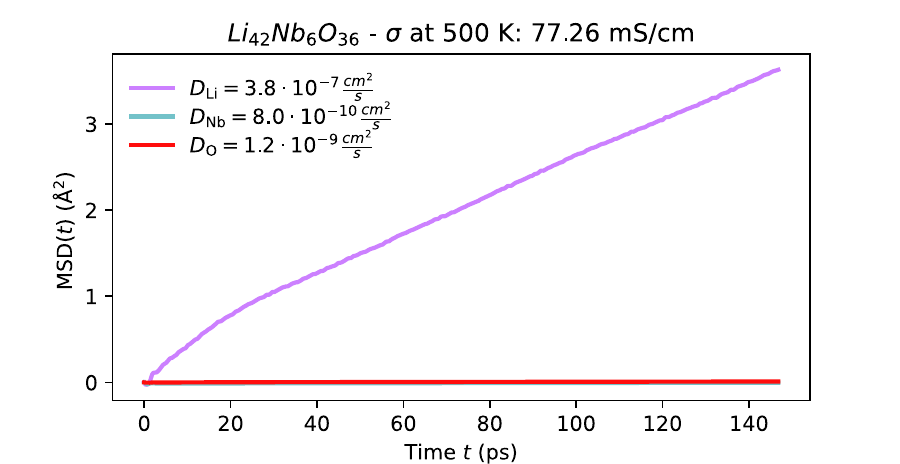}
  \caption{MSD plot of Li along with host-lattice species of $Li_7NbO_6$ at all temperatures studied with FPMD}
\end{figure}

\begin{figure}[H]
  \centering
  \includegraphics[width=\columnwidth]{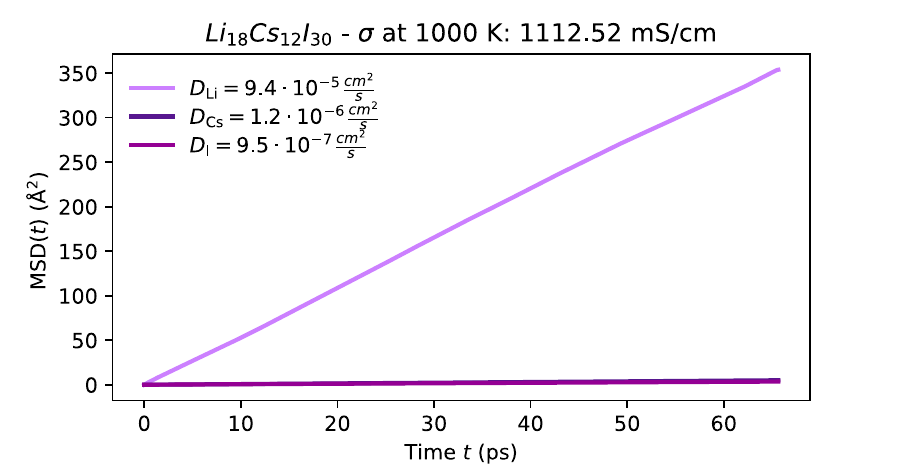}
  \includegraphics[width=\columnwidth]{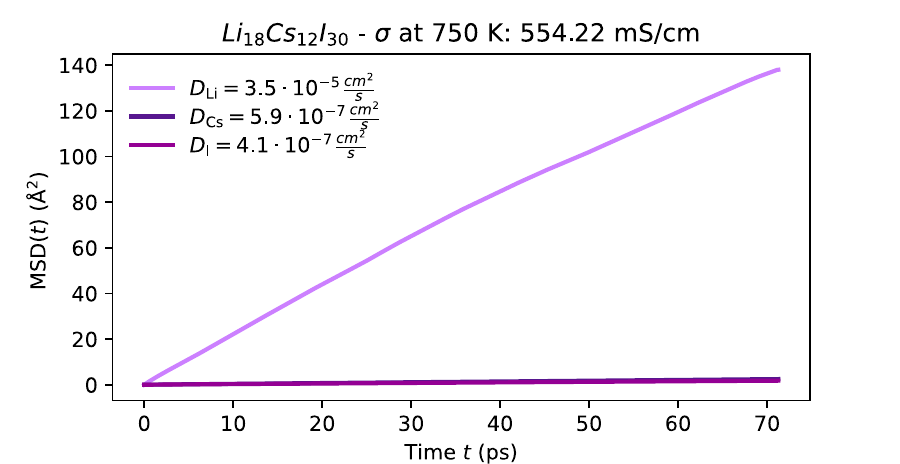}
  \includegraphics[width=\columnwidth]{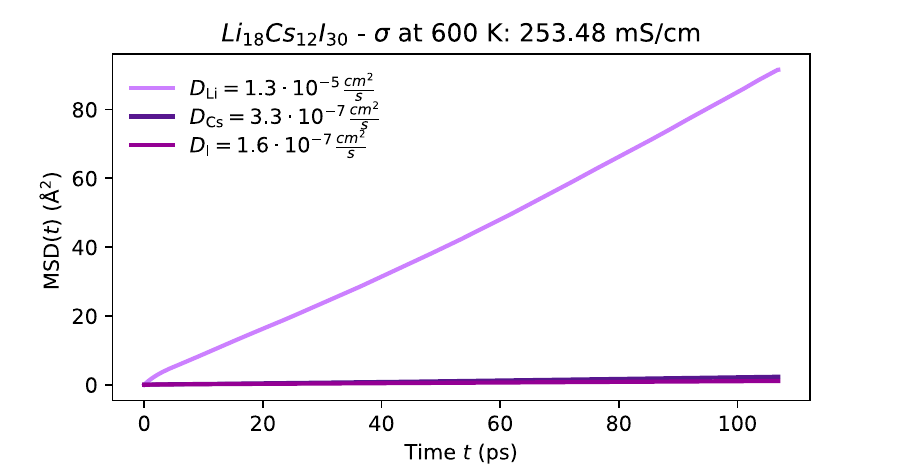}
  \includegraphics[width=\columnwidth]{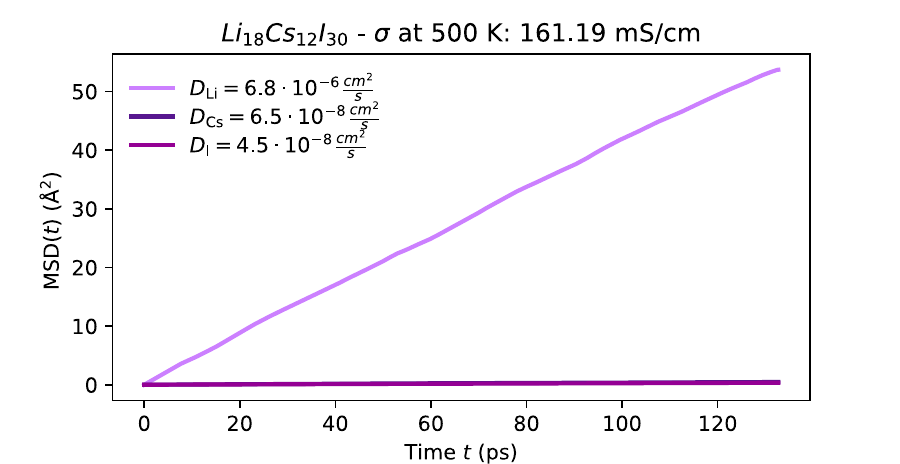}
  \caption{MSD plot of Li along with host-lattice species of $Li_3Cs_2I_5$ at all temperatures studied with FPMD}
\end{figure}

\begin{figure}[H]
  \centering
  \includegraphics[width=\columnwidth]{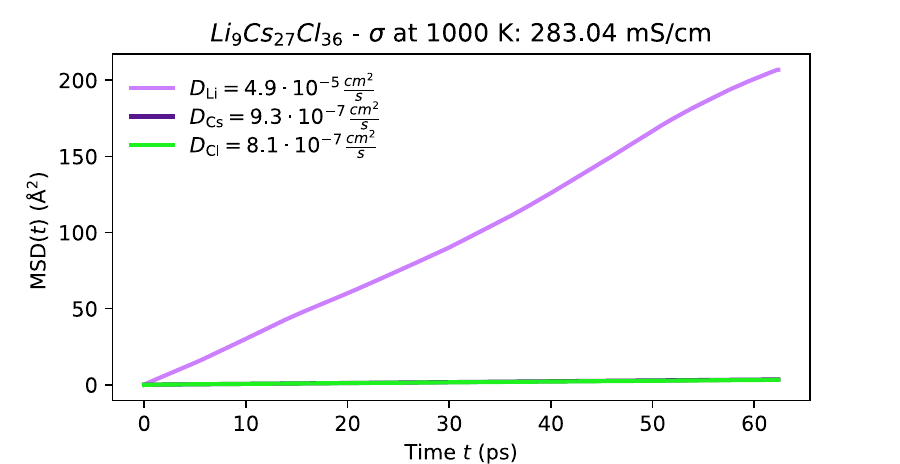}
  \includegraphics[width=\columnwidth]{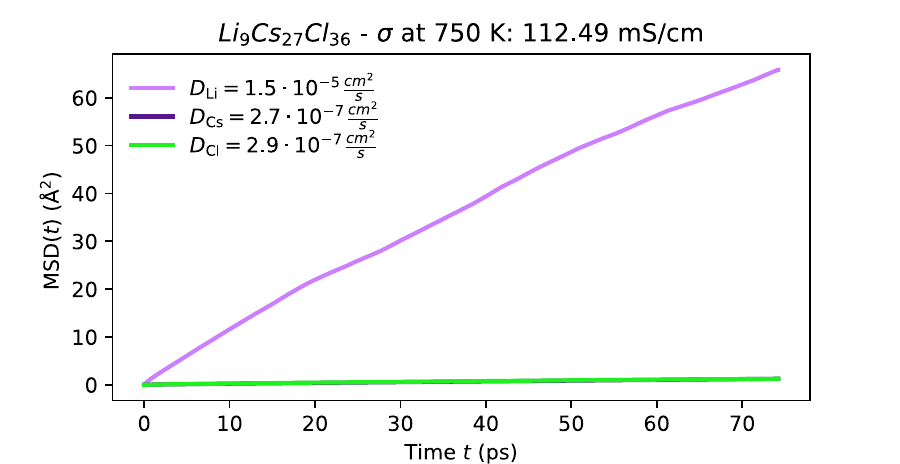}
  \includegraphics[width=\columnwidth]{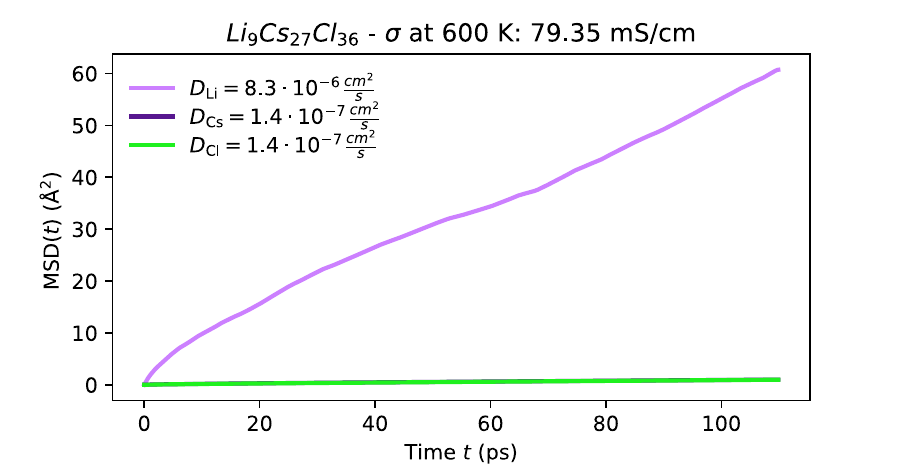}
  \includegraphics[width=\columnwidth]{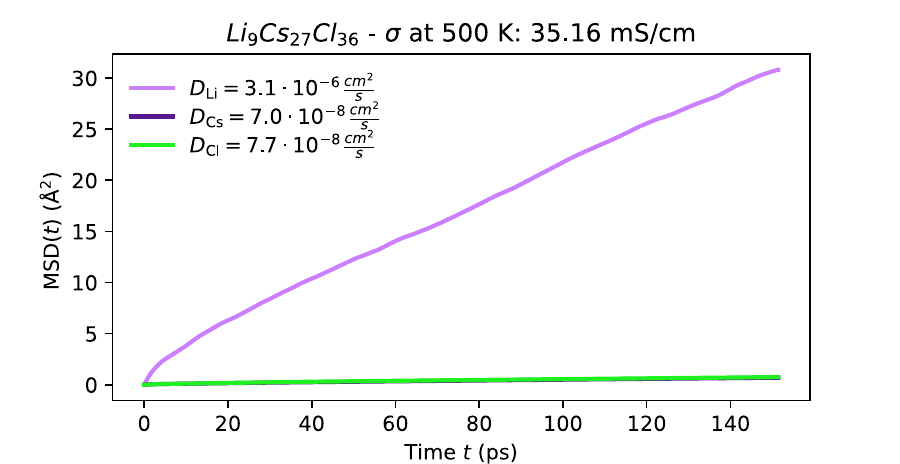}
  \caption{MSD plot of Li along with host-lattice species of $LiCs_3Cl_4$ at all temperatures studied with FPMD}
\end{figure}

\begin{figure}[H]
\centering
  \includegraphics[width=\columnwidth]{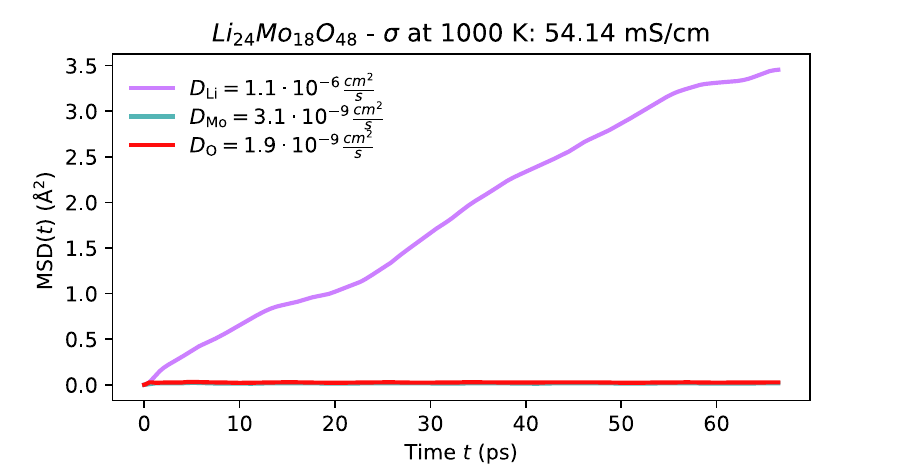}
  \includegraphics[width=\columnwidth]{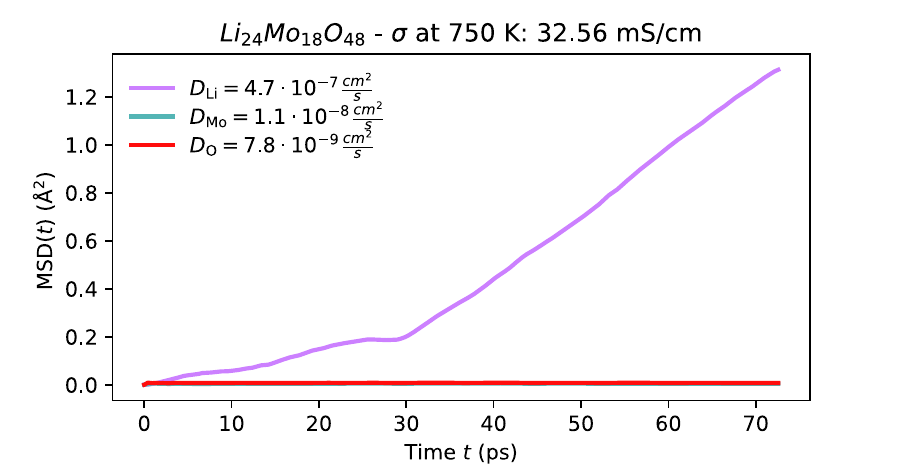}
  \includegraphics[width=\columnwidth]{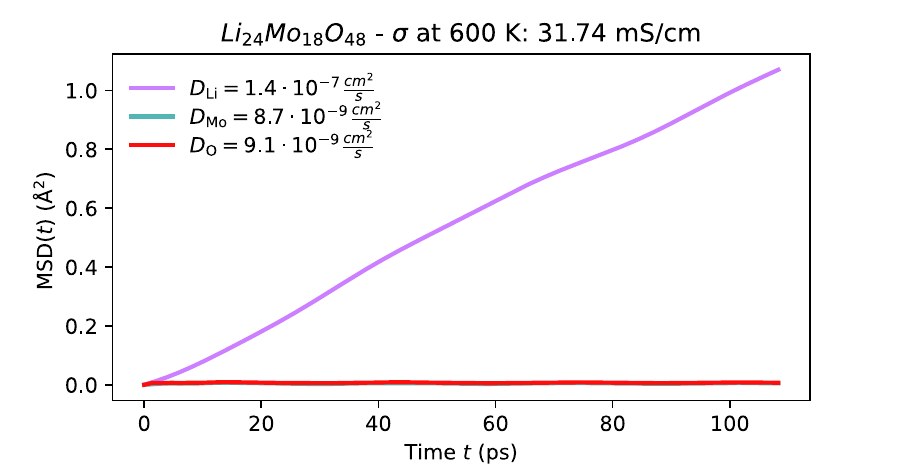}
  \includegraphics[width=\columnwidth]{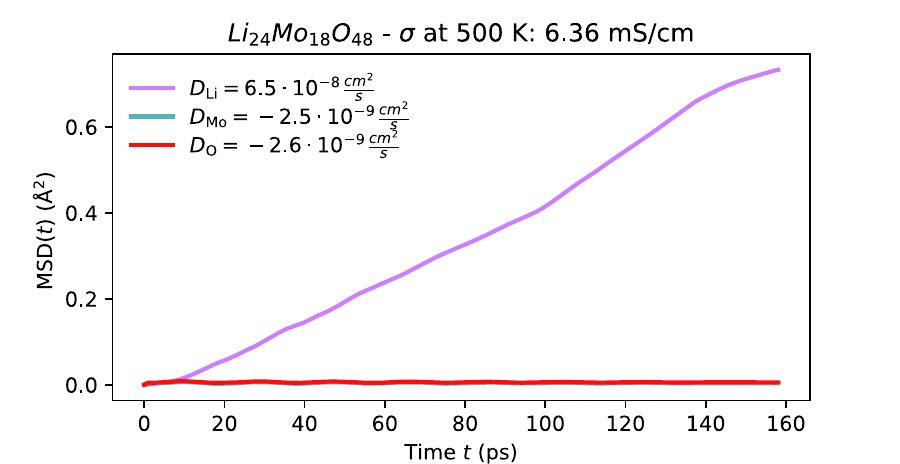}
  \caption{MSD plot of Li along with host-lattice species of $Li_4Mo_3O_8$ at all temperatures studied with FPMD}
\end{figure}

\begin{figure}[H]
\centering
  \includegraphics[width=\columnwidth]{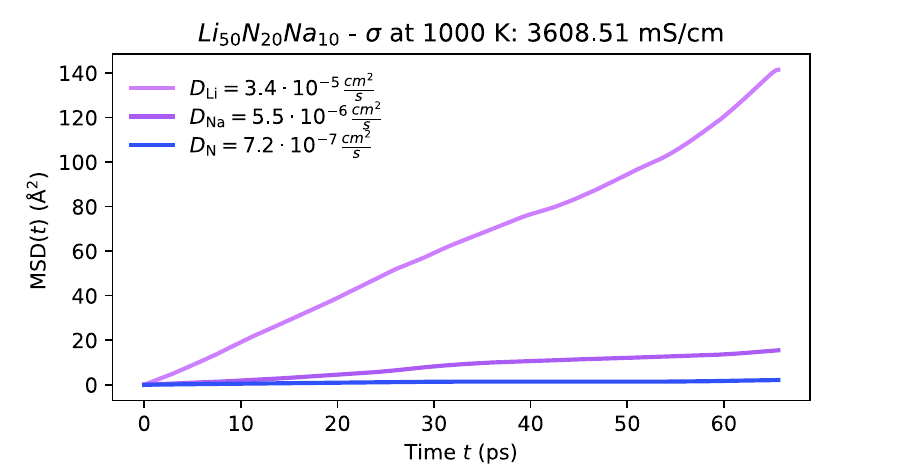}
  \includegraphics[width=\columnwidth]{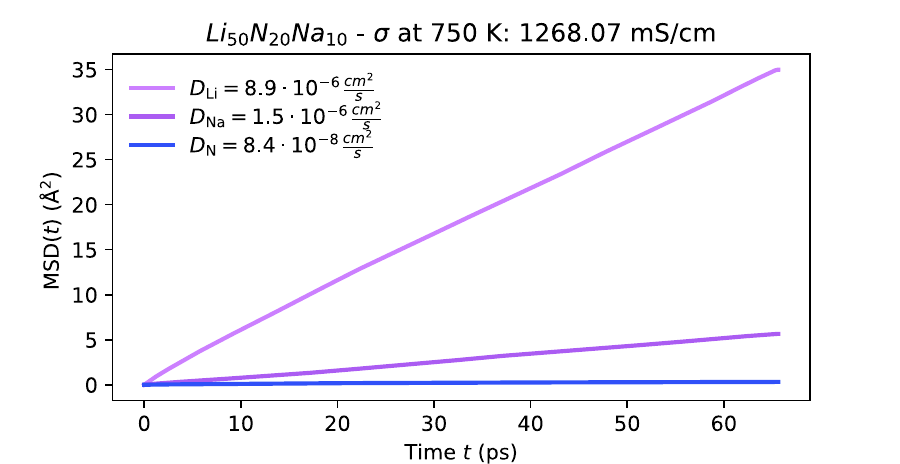}
  \includegraphics[width=\columnwidth]{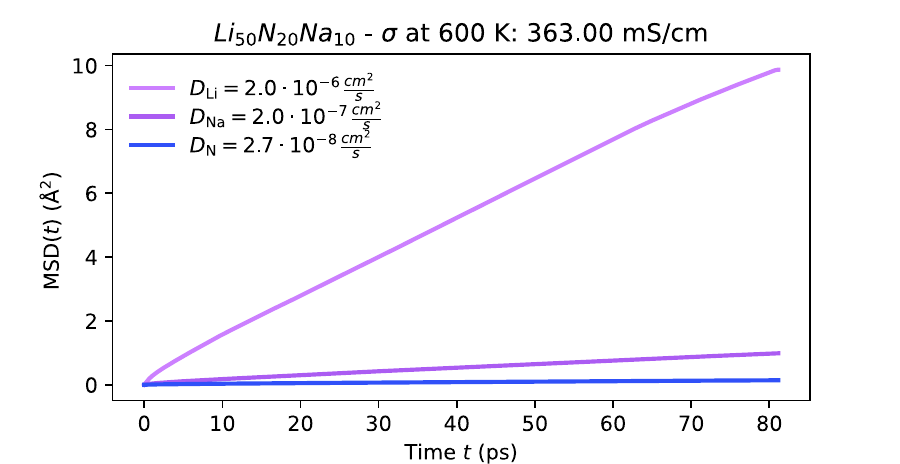}
  \includegraphics[width=\columnwidth]{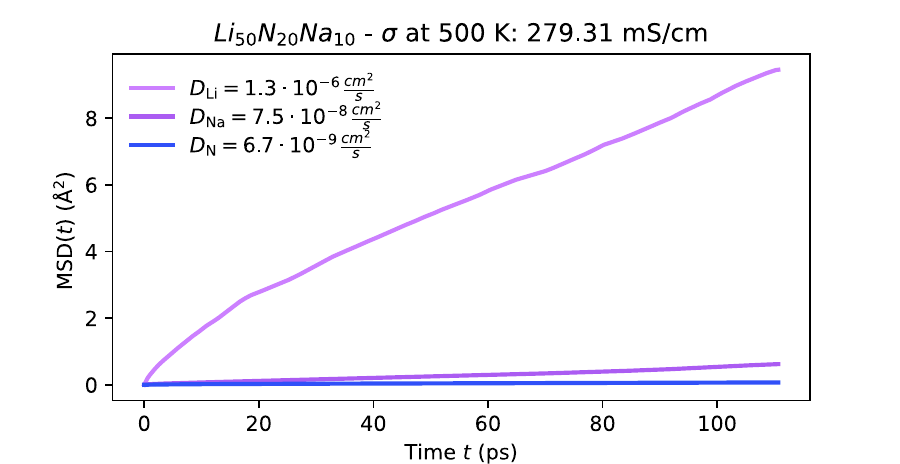}
  \caption{MSD plot of Li along with host-lattice species of $Li_5NaN_2$ at all temperatures studied with FPMD}
\end{figure}

We show the iso-surface plots at 600 K for the oxides and nitrides.

\begin{figure}
\centering
  \includegraphics[width=0.75\columnwidth]{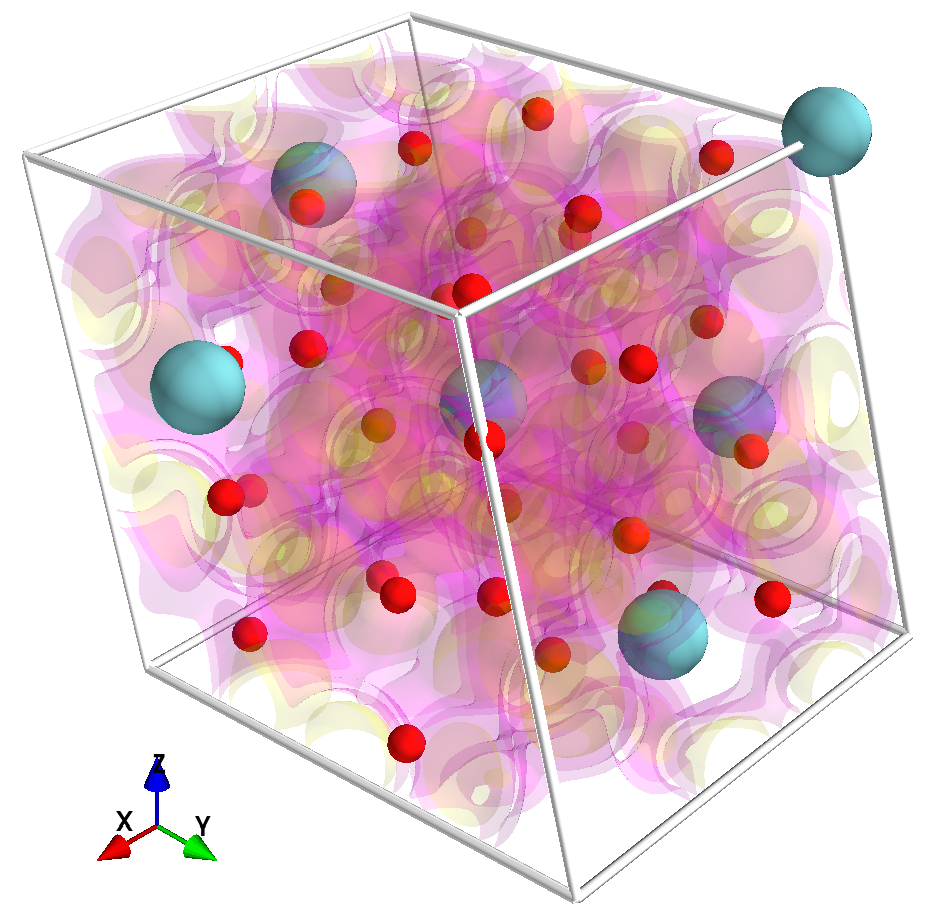}
  \caption{Li-ion density of $Li_7NbO_6$ at 600 K from FPMD. The pink and yellow clouds depict Li-ion diffusion in a dispersed manner, establishing this materials as a robust 3-dimensional ionic conductor at lower temperatures.}
  \label{fgr:iso_lnbo}
\end{figure}

\begin{figure}
\centering
  \includegraphics[width=0.75\columnwidth]{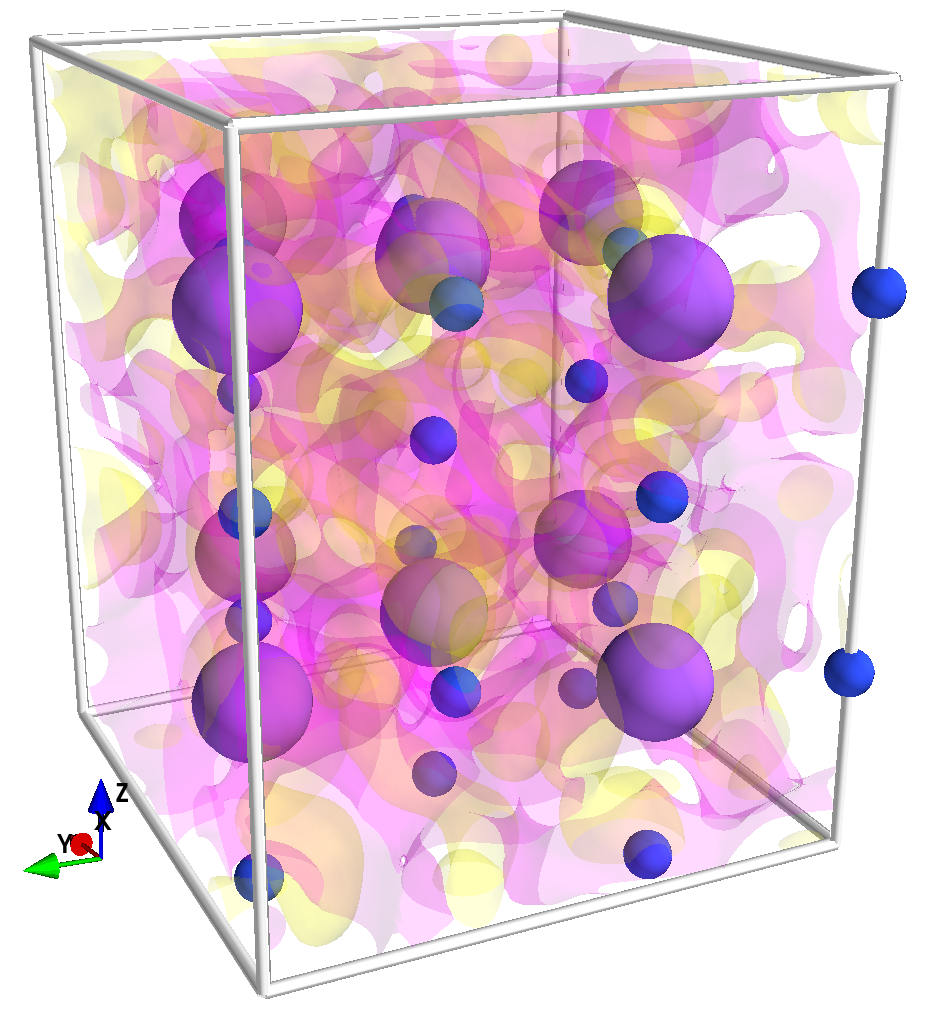}
  \caption{Li-ion density of $Li_5NaN_2$ at 600 K from FPMD. The pink and yellow clouds illustrate Li-ion diffusion in a dispersed fashion, establishing this materials as a promising 3-dimensional ionic conductor at lower temperatures.}
  \label{fgr:iso_lnn}
\end{figure}

\begin{figure}
\centering
  \includegraphics[width=\columnwidth]{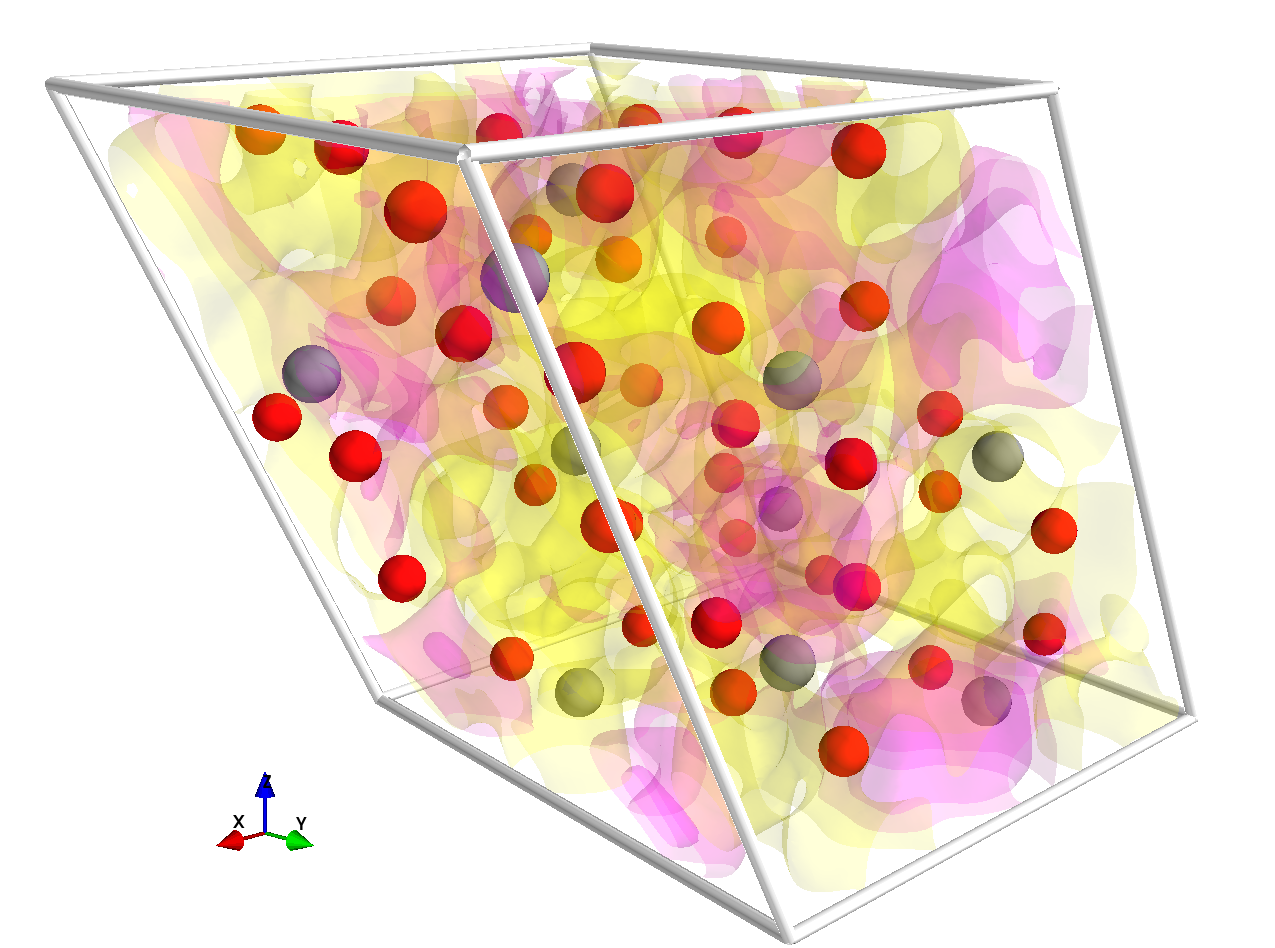}
  \caption{Li-ion density of $Li_4CO_4$ at 600 K from FPMD. The pink and yellow channels clearly illustrate Li-ion diffusion in this material, establishing this materials as a 3-dimensional Li-ion conductor even at lower temperatures.}
  \label{fgr:iso_lco}
\end{figure}

\begin{figure}
\centering
  \includegraphics[width=\columnwidth]{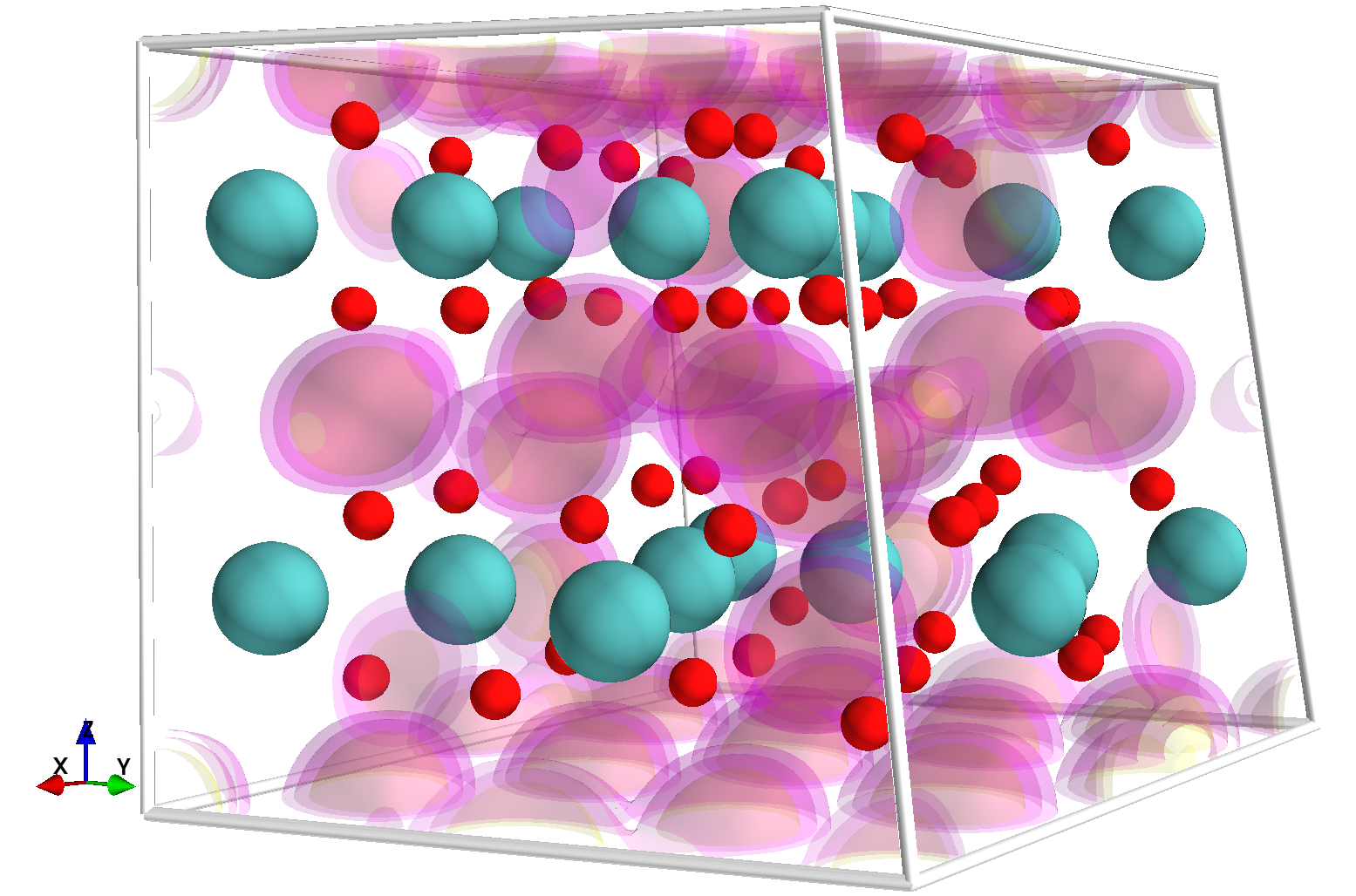}
  \caption{Li-ion density of $Li_4Mo_3O_8$ at 600 K from FPMD. The pink clouds illustrate Li-ion diffusion in this material in a layered fashion, establishing this materials as a 3-dimensional Li-ion conductor even at lower temperatures.}
  \label{fgr:iso_lmo}
\end{figure}

\newpage
\FloatBarrier

\phantomsection
\section{Potential fast Li-ion conductors} \label{S2_potential}

We identify 25 structures that exhibit significant diffusion at 1000 K in our FPMD simulations, but do not display the same behaviour at lower temperatures. 
We show the MSD plots at 1000 K, 750 K, 600 K and 500 K.

\begin{figure}[H]
\centering
  \includegraphics[width=\columnwidth]{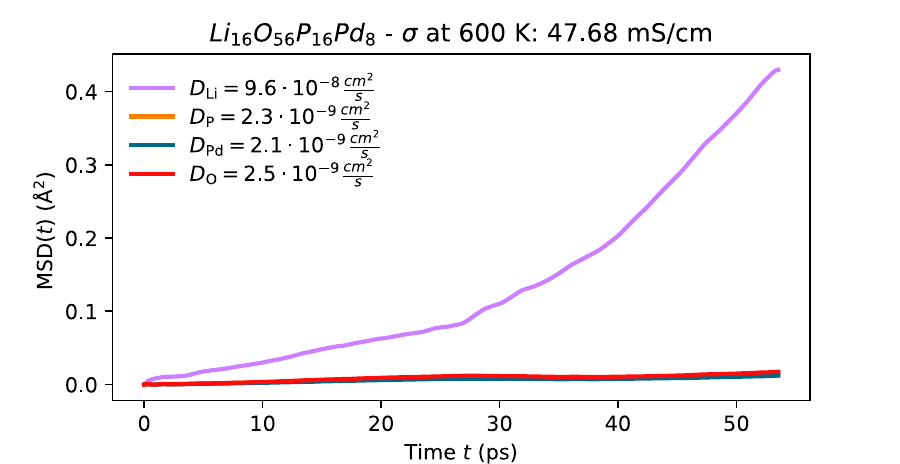}
  \caption{MSD plot of Li along with host-lattice species of $Li_2P_2PdO_7$ at 600 K. This structure was too expensive to simulate with full-first principles so we studied it at only 600 K with FPMD.}
\end{figure}

\begin{figure}[H]
\centering
  \includegraphics[width=\columnwidth]{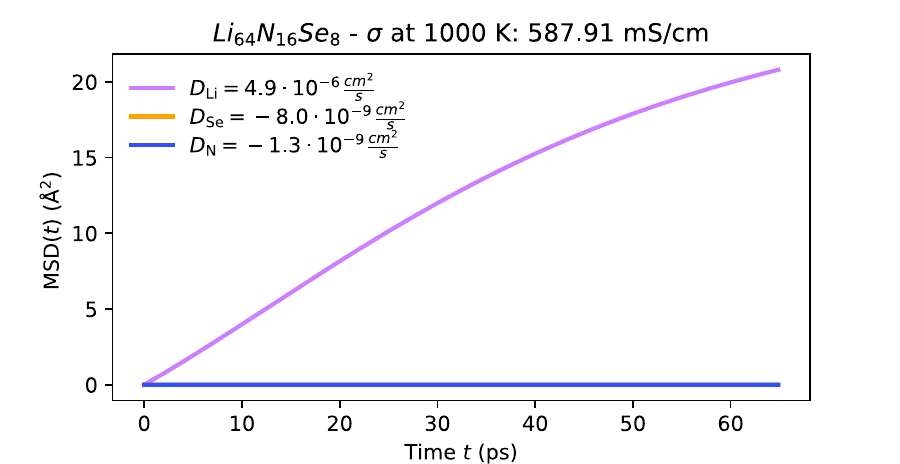}
  \caption{MSD(t) plot of Li along with host-lattice species of $Li_8SeN_2$ at all temperatures studied with FPMD.}
\end{figure}

\begin{figure}[H]
\centering
  \includegraphics[width=\columnwidth]{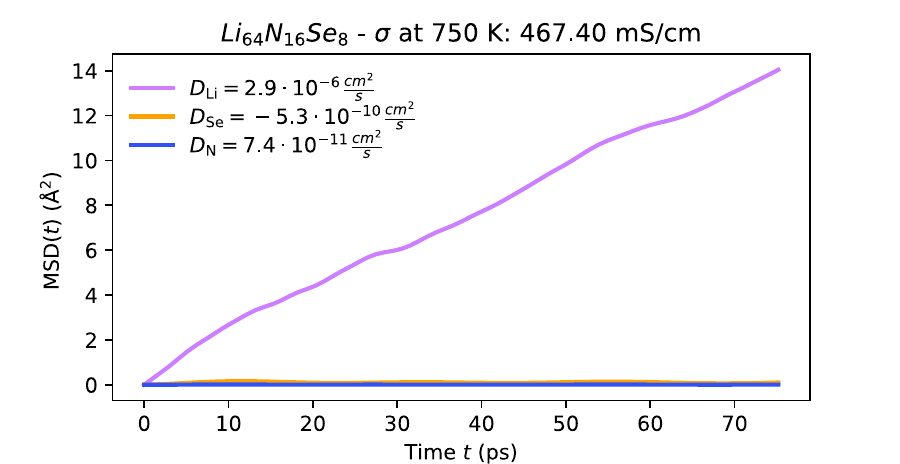}
  \includegraphics[width=\columnwidth]{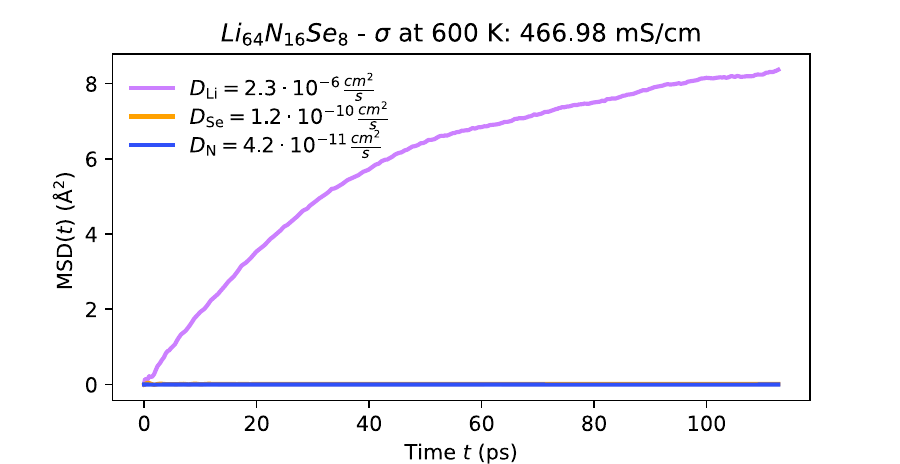}
  \includegraphics[width=\columnwidth]{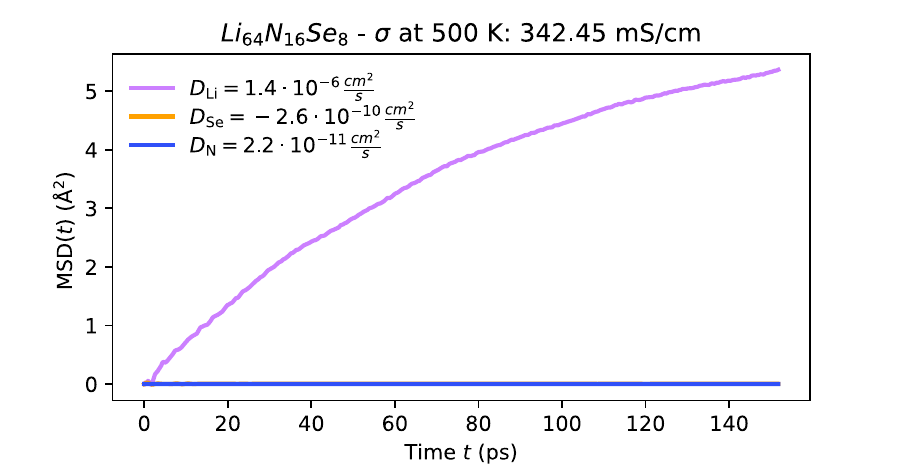}
  \caption{MSD(t) plot of Li along with host-lattice species of $Li_8SeN_2$ at all temperatures studied with FPMD.}
\end{figure}

\begin{figure}[H]
\centering
  \includegraphics[width=\columnwidth]{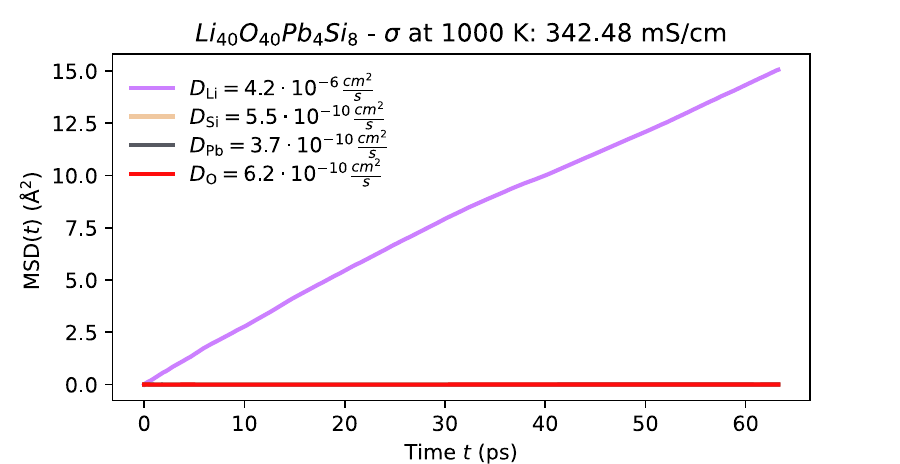}
  \includegraphics[width=\columnwidth]{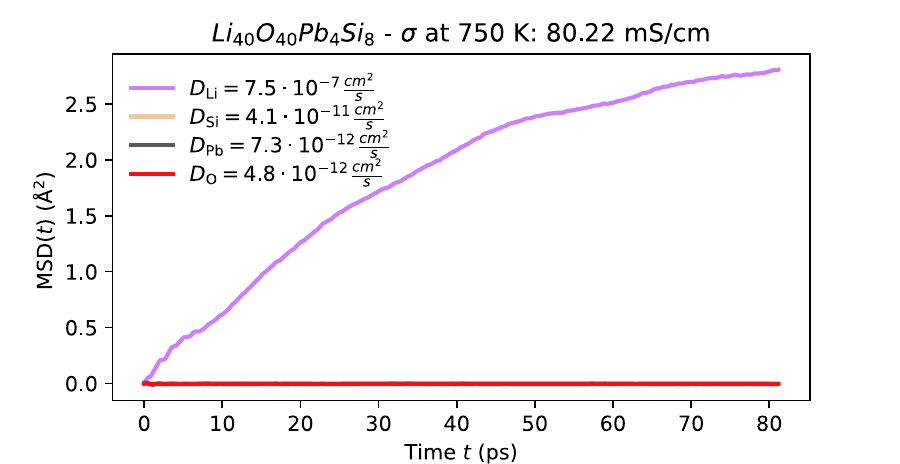}
  \includegraphics[width=\columnwidth]{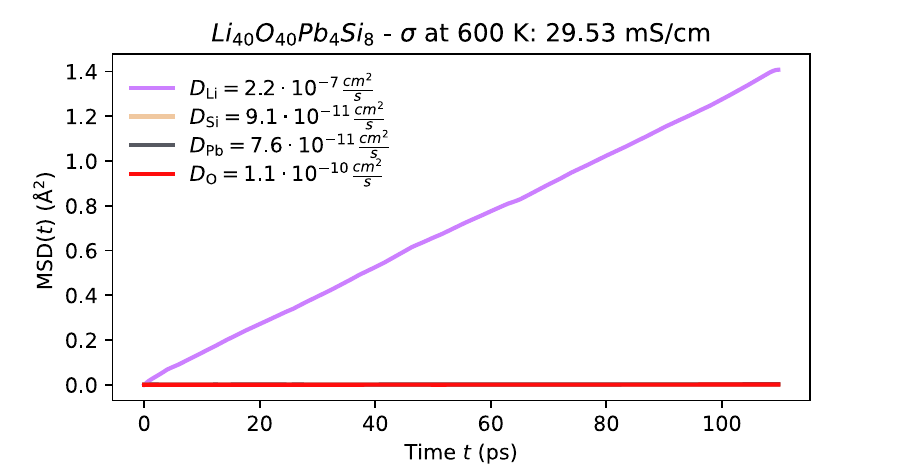}
  \includegraphics[width=\columnwidth]{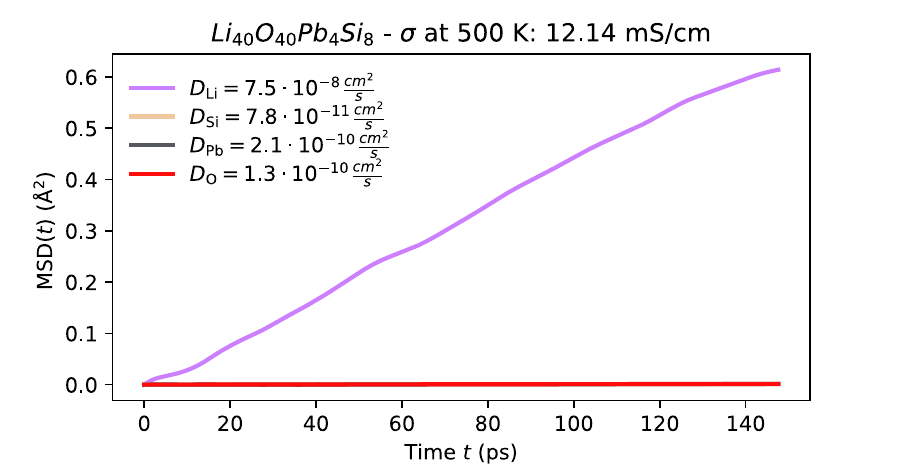}
  \caption{MSD plot of Li along with host-lattice species of $Li_{10}Si_2PbO_{10}$ at all temperatures studied with FPMD}
\end{figure}

\begin{figure}[H]
\centering10
  \includegraphics[width=\columnwidth]{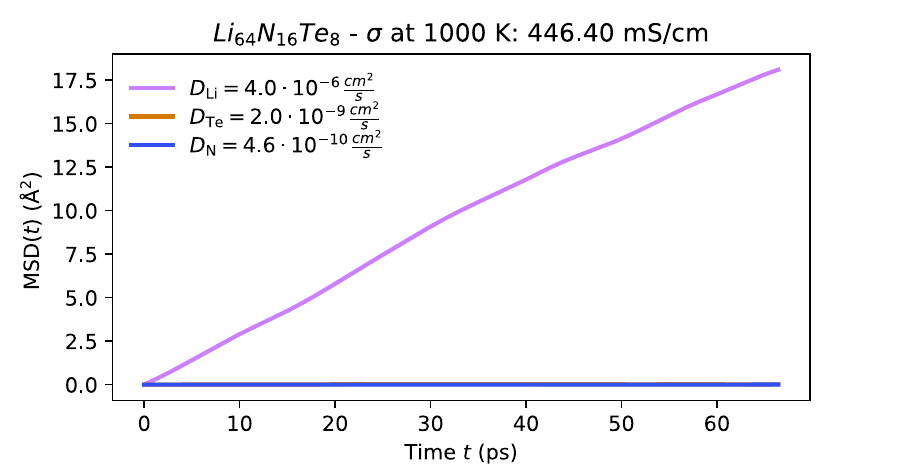}
  \includegraphics[width=\columnwidth]{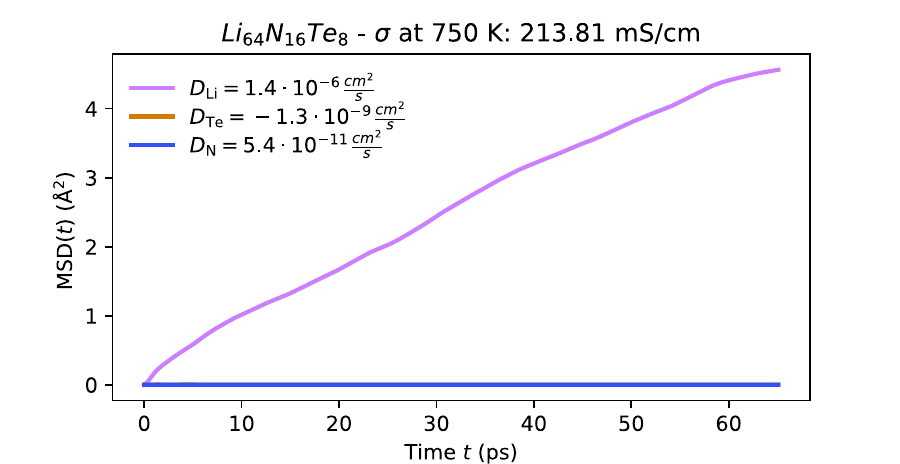}
  \includegraphics[width=\columnwidth]{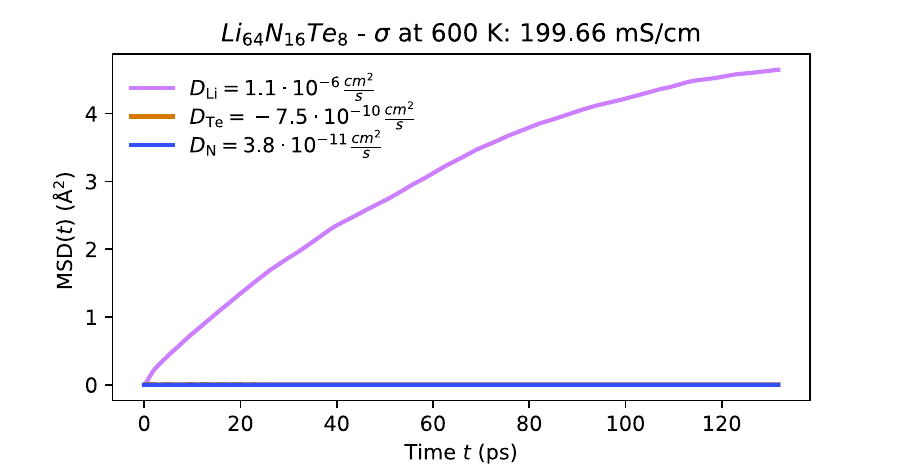}
  \includegraphics[width=\columnwidth]{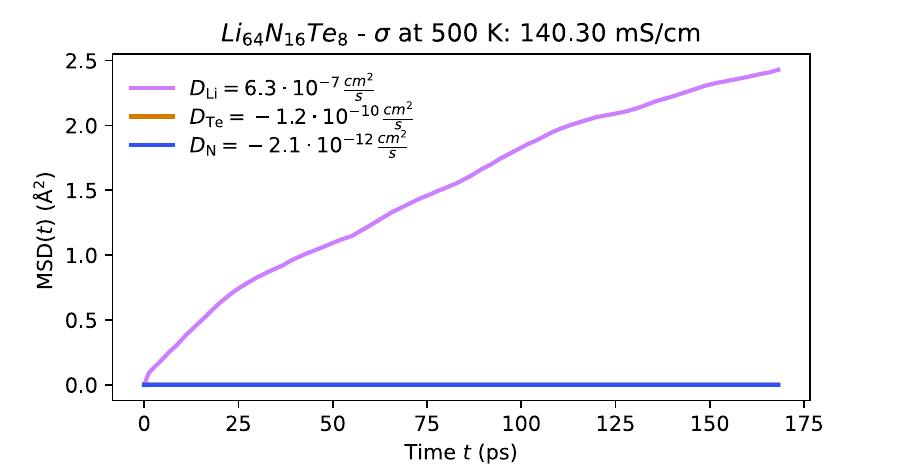}
  \caption{MSD(t) plot of Li along with host-lattice species of $Li_8TeN_2$ at all temperatures studied with FPMD.}
\end{figure}

\begin{figure}[H]
\centering
  \includegraphics[width=\columnwidth]{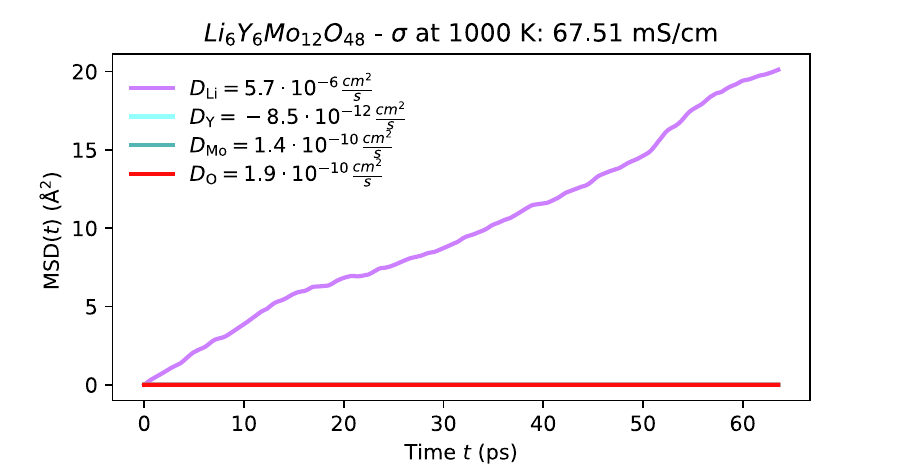}
  \includegraphics[width=\columnwidth]{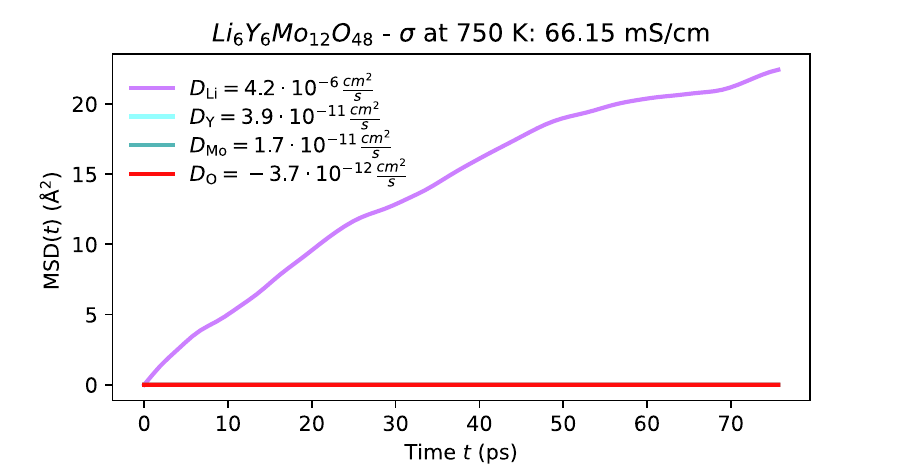}
  \includegraphics[width=\columnwidth]{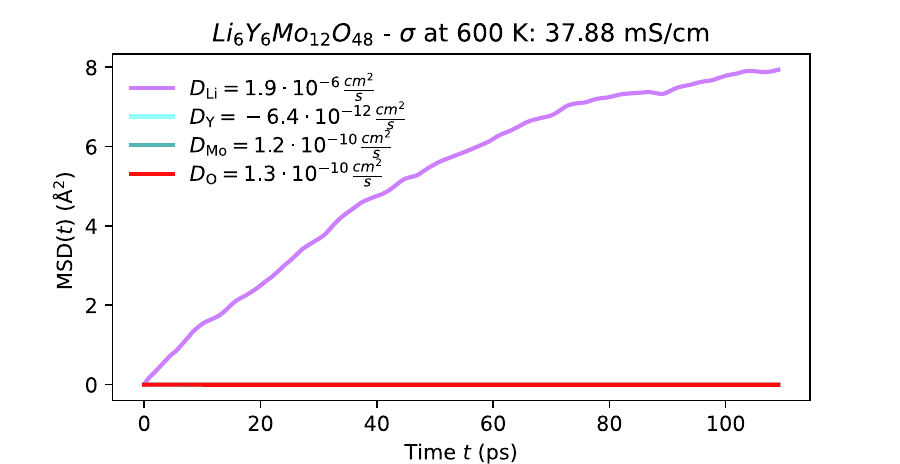}
  \includegraphics[width=\columnwidth]{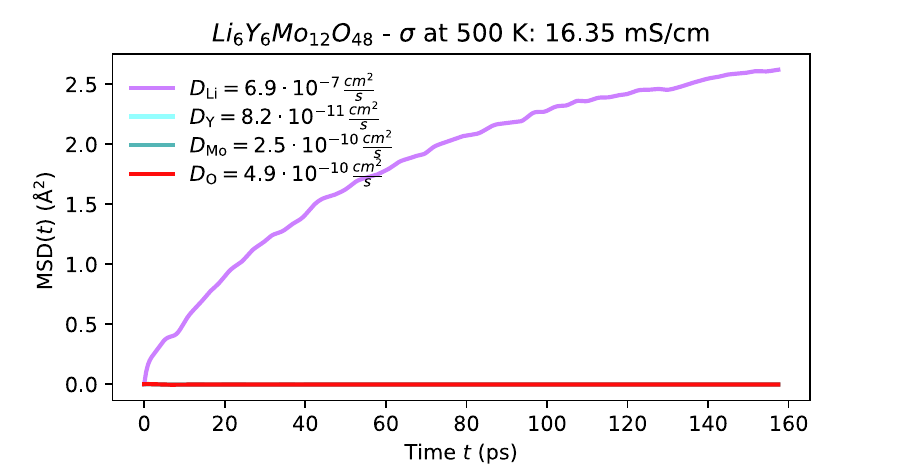}
  \caption{MSD(t) plot of Li along with host-lattice species of $LiY(MoO_4)_2$ at all temperatures studied with FPMD.}
\end{figure}

\begin{figure}[H]
  \centering
  \includegraphics[width=\columnwidth]{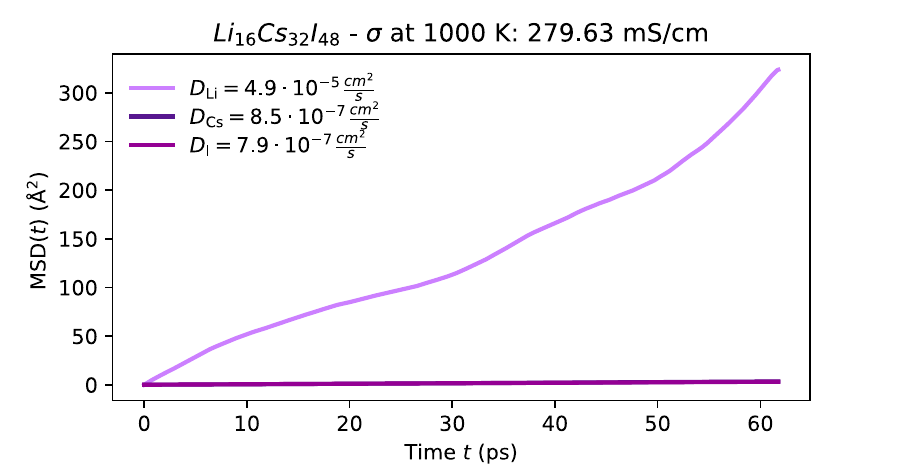}
  \includegraphics[width=\columnwidth]{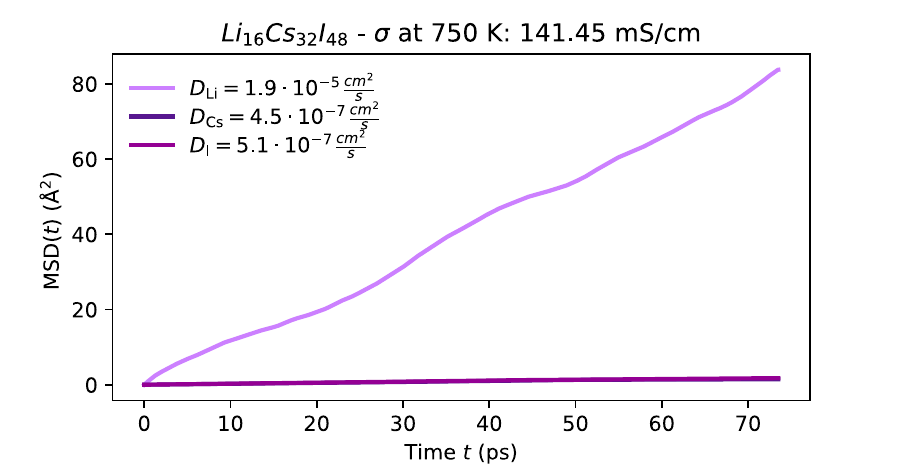}
  \includegraphics[width=\columnwidth]{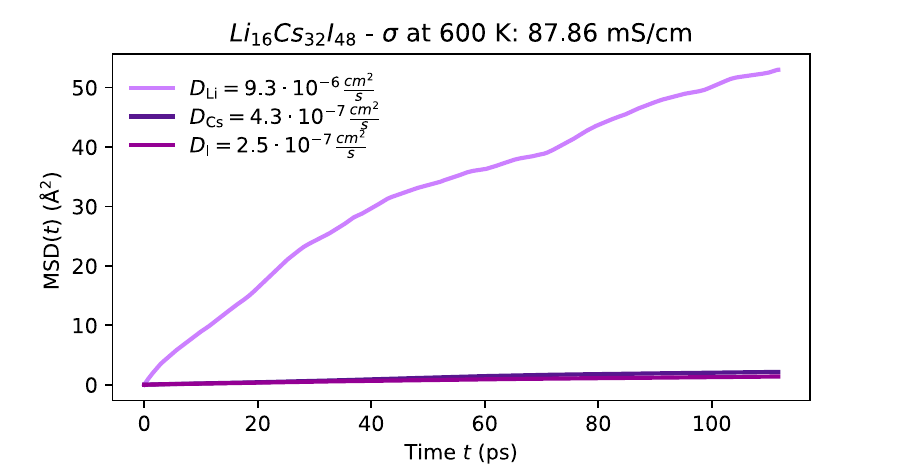}
  \includegraphics[width=\columnwidth]{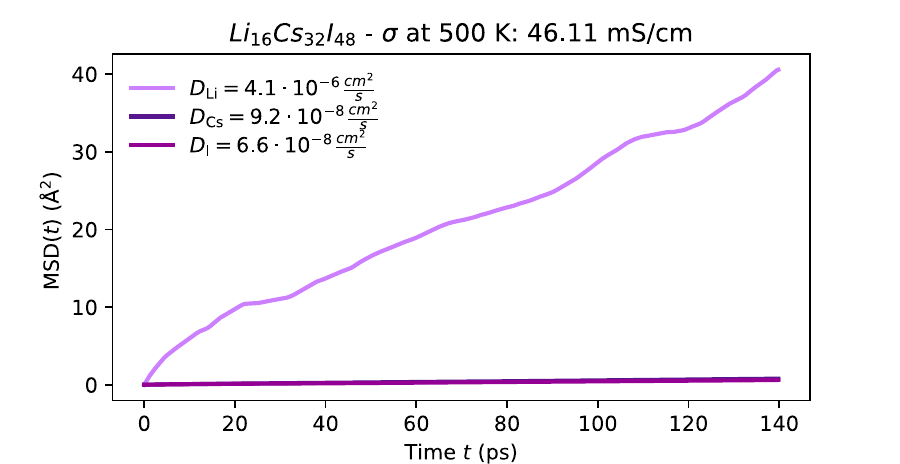}
  \caption{MSD plot of Li along with host-lattice species of $LiCs_2I_3$ at all temperatures studied with FPMD}
\end{figure}

\begin{figure}[H]
\centering
  \includegraphics[width=\columnwidth]{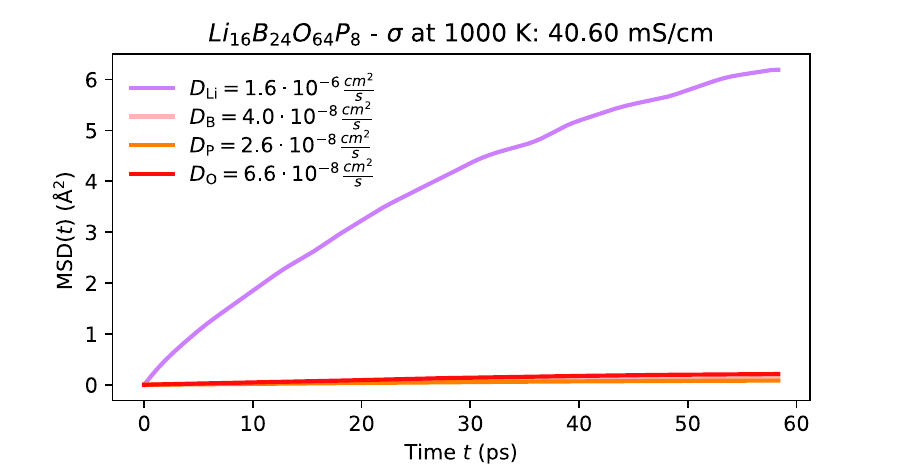}
  \includegraphics[width=\columnwidth]{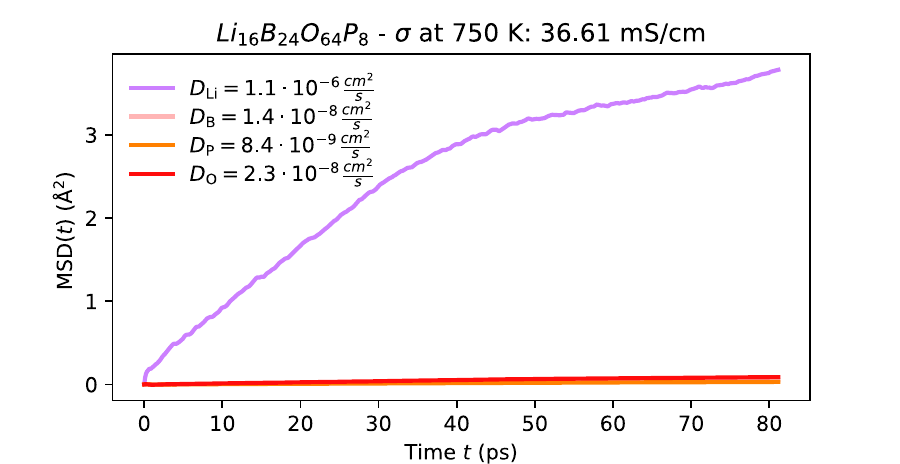}
  \includegraphics[width=\columnwidth]{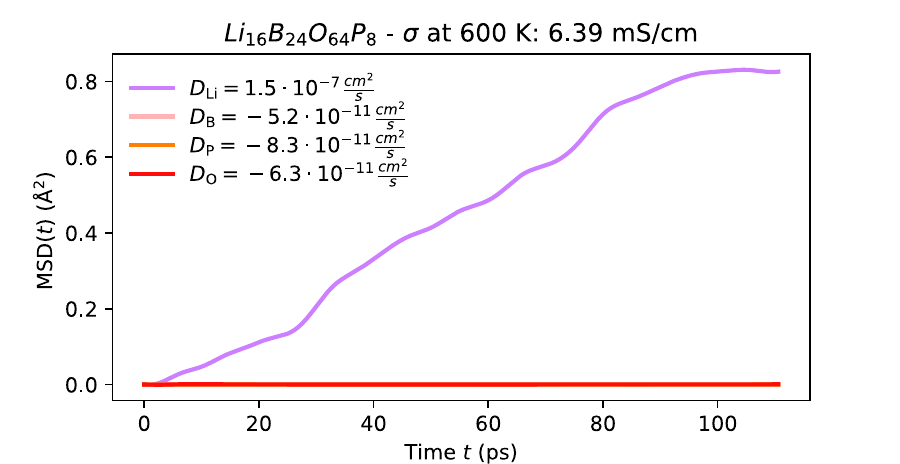}
  \includegraphics[width=\columnwidth]{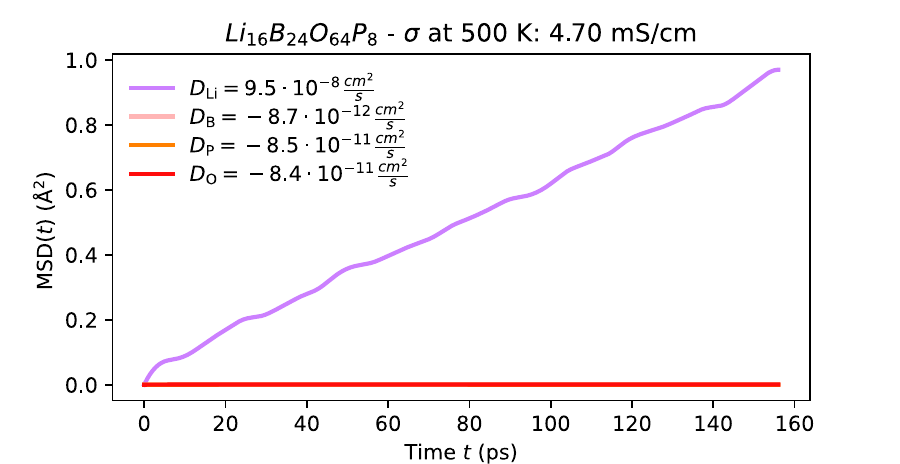}
  \caption{MSD plot of Li along with host-lattice species of $Li_2B_3PO_8$ at all temperatures studied with FPMD}
\end{figure}

\begin{figure}[H]
\centering
  \includegraphics[width=\columnwidth]{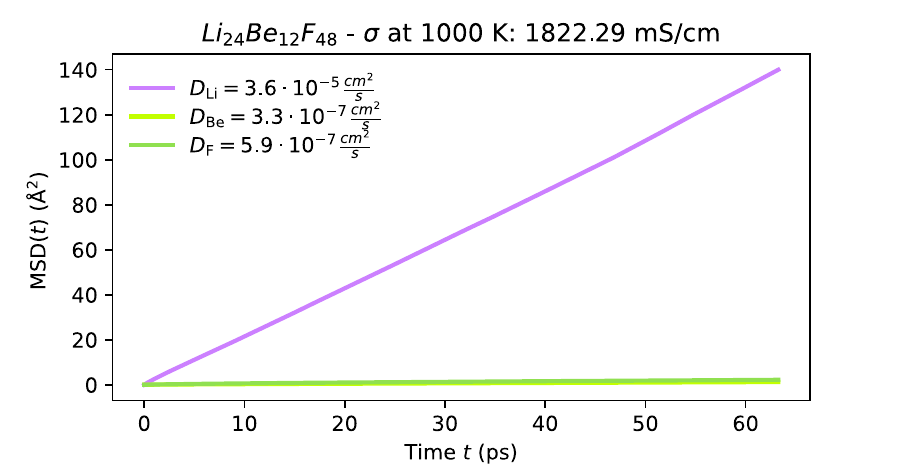}
  \includegraphics[width=\columnwidth]{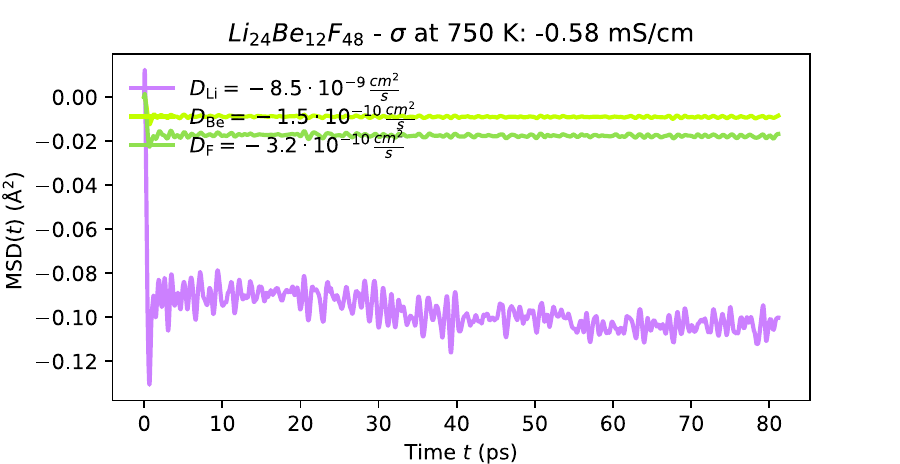}
  \includegraphics[width=\columnwidth]{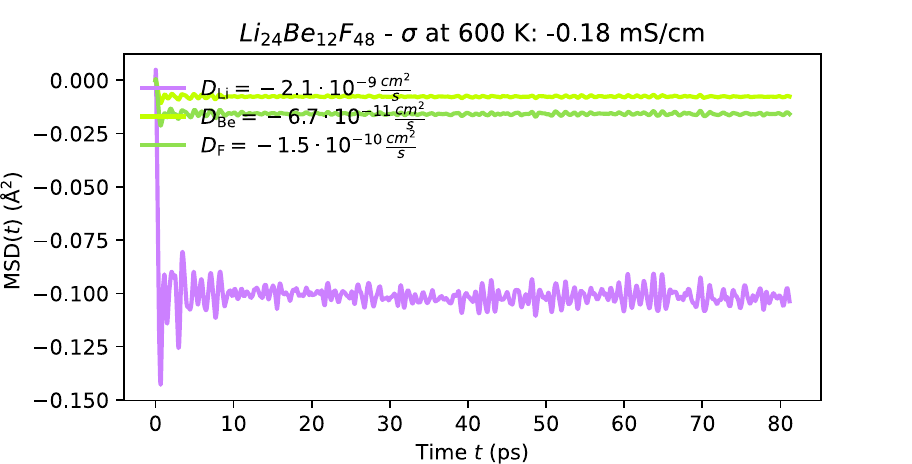}
  \includegraphics[width=\columnwidth]{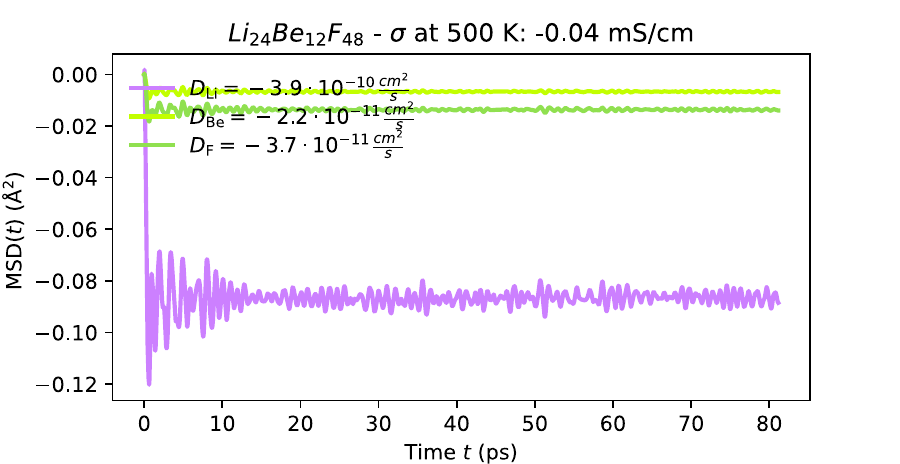}
  \caption{MSD plot of Li along with host-lattice species of $Li_2BeF_4$ at all temperatures studied with FPMD}
\end{figure}

\begin{figure}[H]
\centering
  \includegraphics[width=\columnwidth]{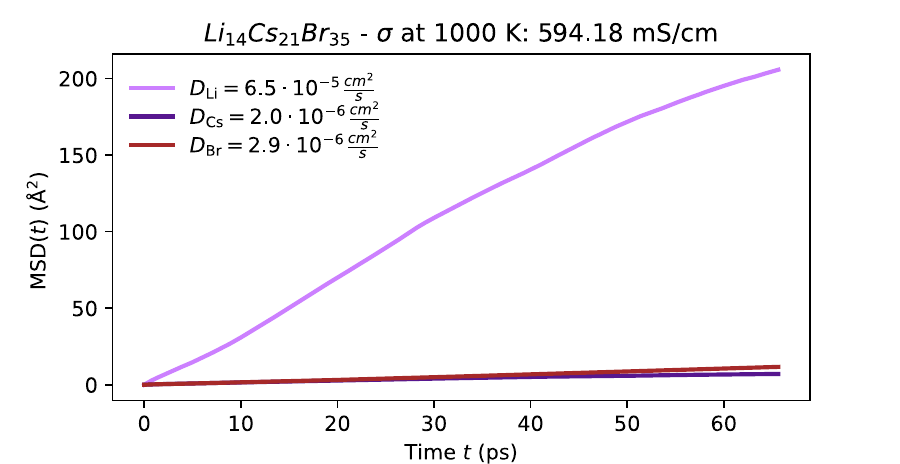}
  \includegraphics[width=\columnwidth]{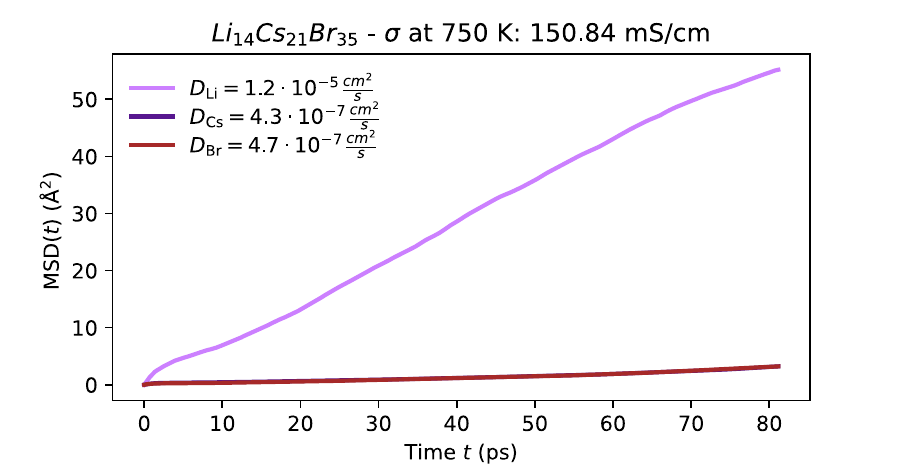}
  \includegraphics[width=\columnwidth]{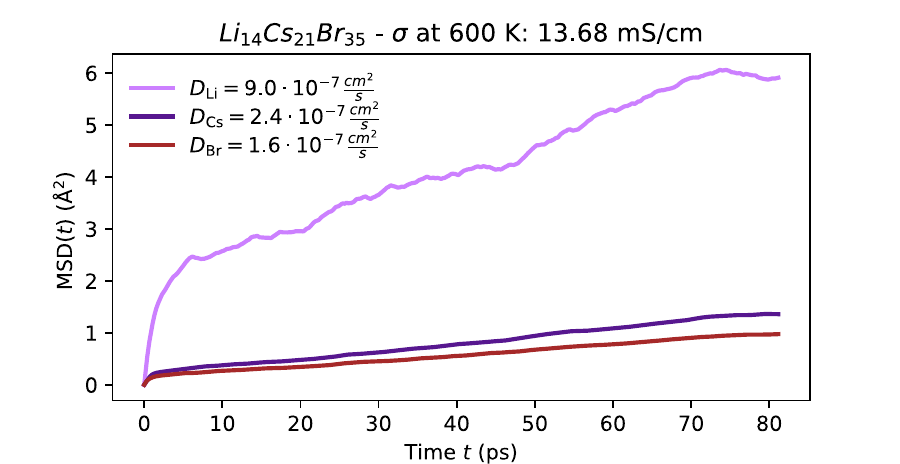}
  \includegraphics[width=\columnwidth]{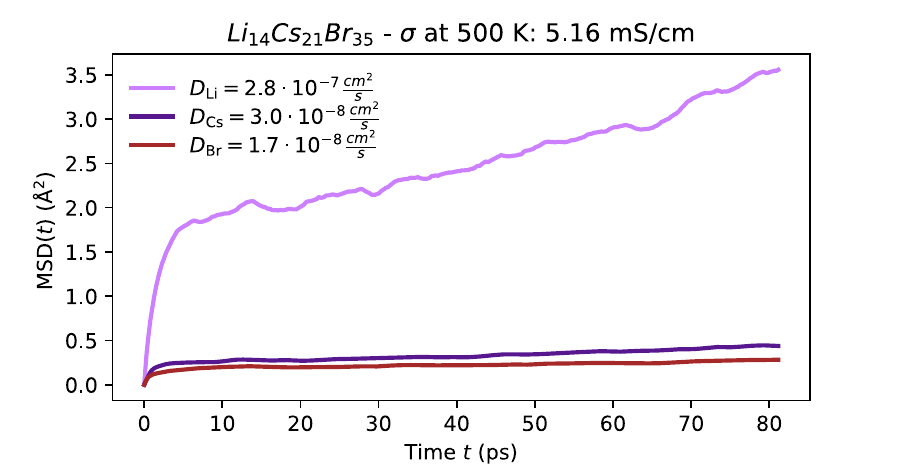}
  \caption{MSD plot of Li along with host-lattice species of $Li_2Cs_3Br_5$ at all temperatures studied with FPMD}
\end{figure}

\begin{figure}[H]
\centering
  \includegraphics[width=\columnwidth]{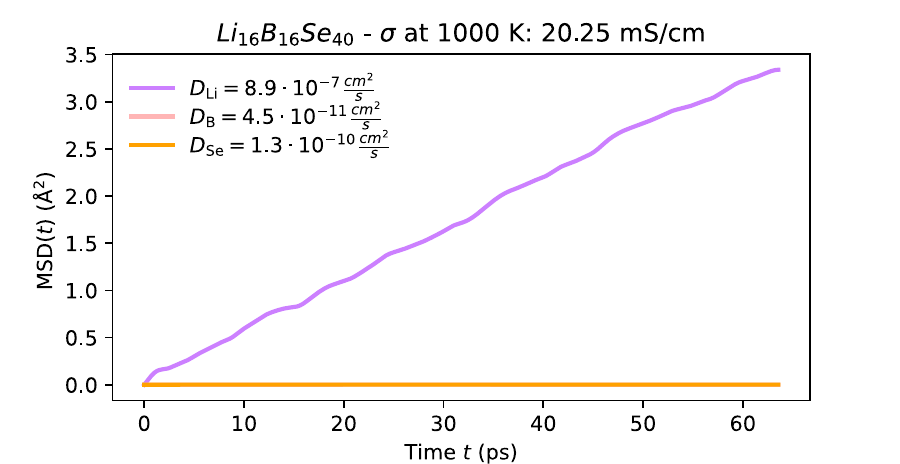}
  \includegraphics[width=\columnwidth]{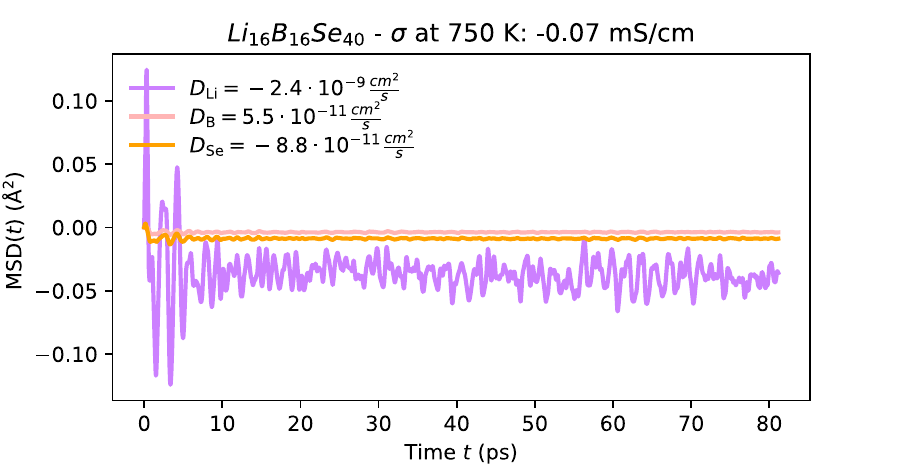}
  \includegraphics[width=\columnwidth]{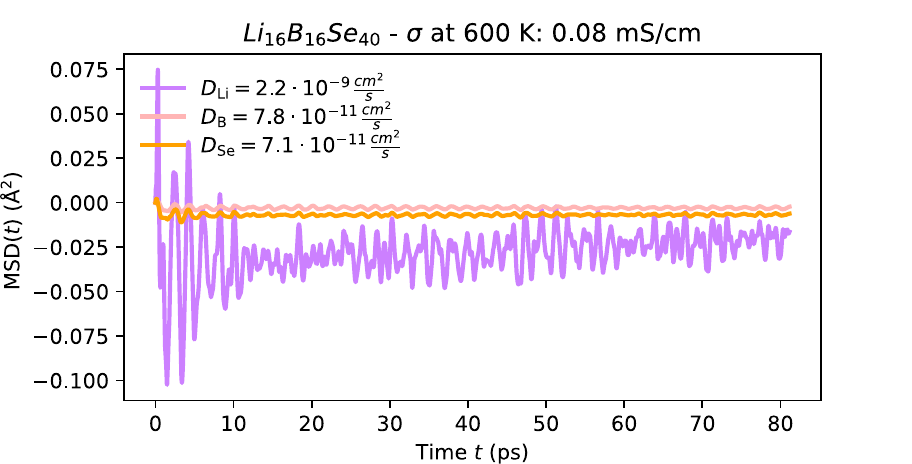}
  \includegraphics[width=\columnwidth]{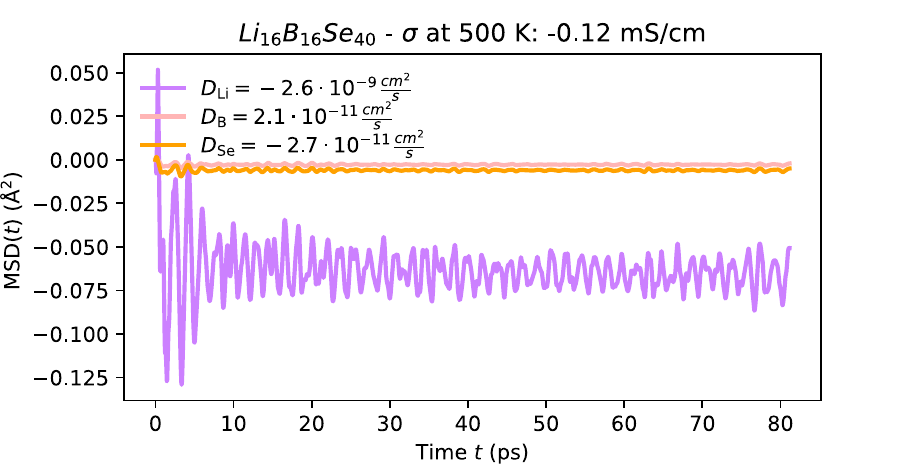}
  \caption{MSD plot of Li along with host-lattice species of $Li_2B_2Se_5$ at all temperatures studied with FPMD}
\end{figure}

\begin{figure}[H]
\centering
  \includegraphics[width=\columnwidth]{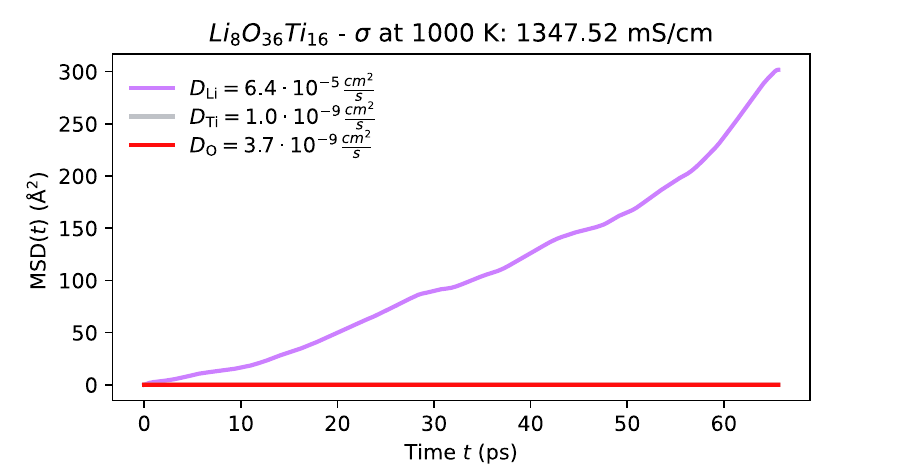}
  \includegraphics[width=\columnwidth]{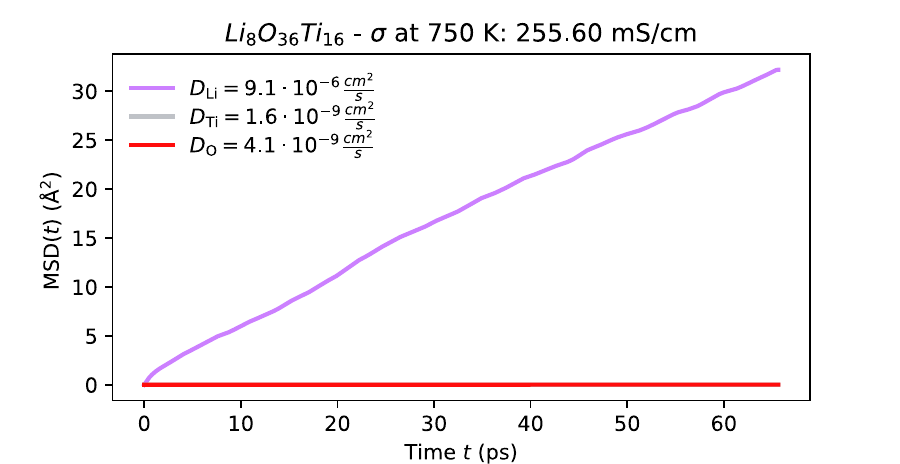}
  \includegraphics[width=\columnwidth]{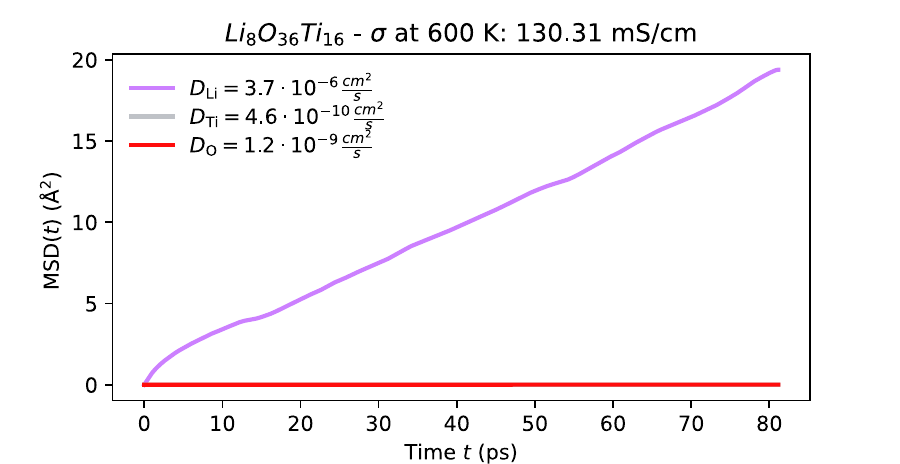}
  \includegraphics[width=\columnwidth]{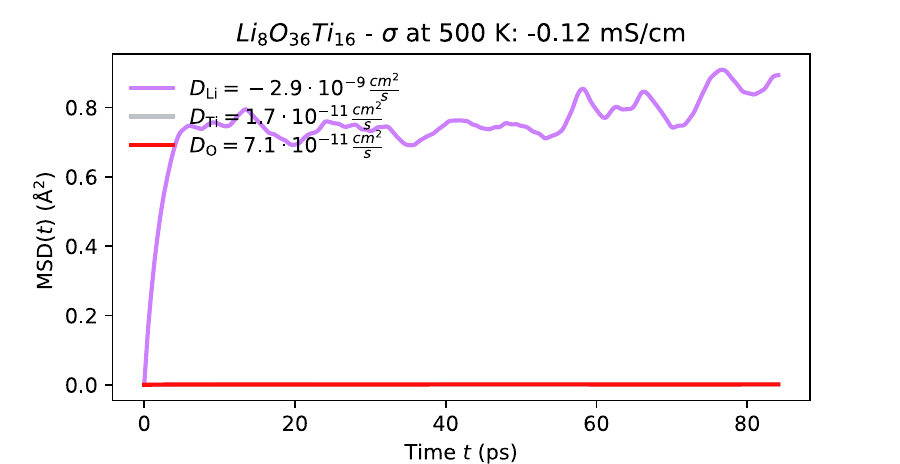}
  \caption{MSD plot of Li along with host-lattice species of $Li_2Ti_4O_9$ at all temperatures studied with FPMD}
\end{figure}

\begin{figure}[H]
\centering
  \includegraphics[width=\columnwidth]{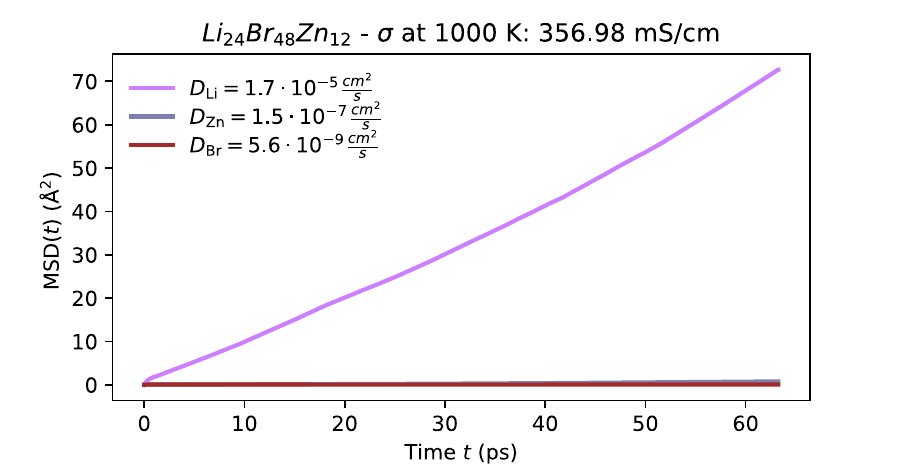}
  \includegraphics[width=\columnwidth]{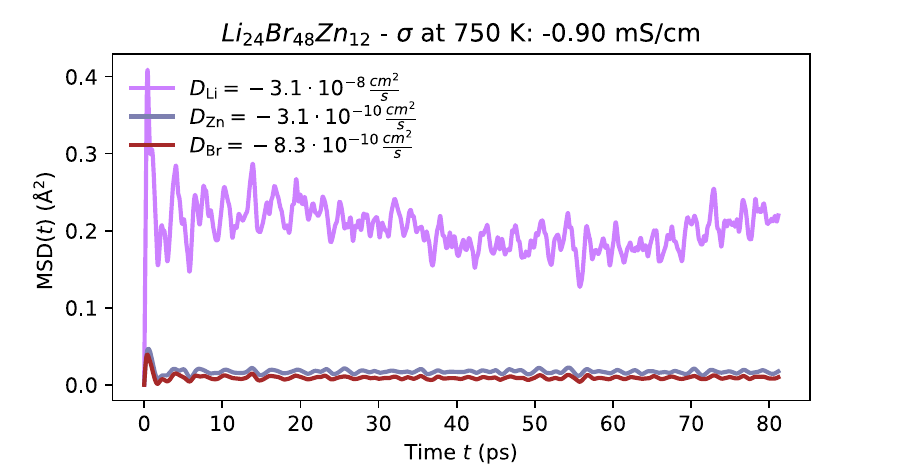}
  \includegraphics[width=\columnwidth]{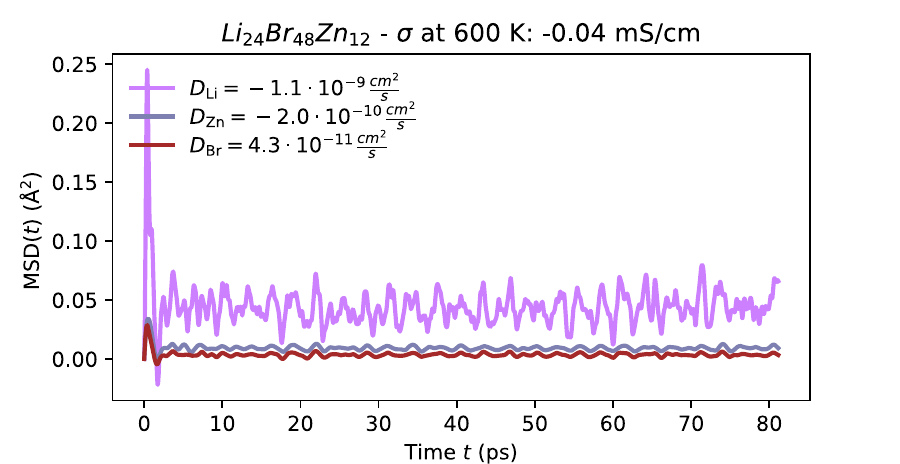}
  \includegraphics[width=\columnwidth]{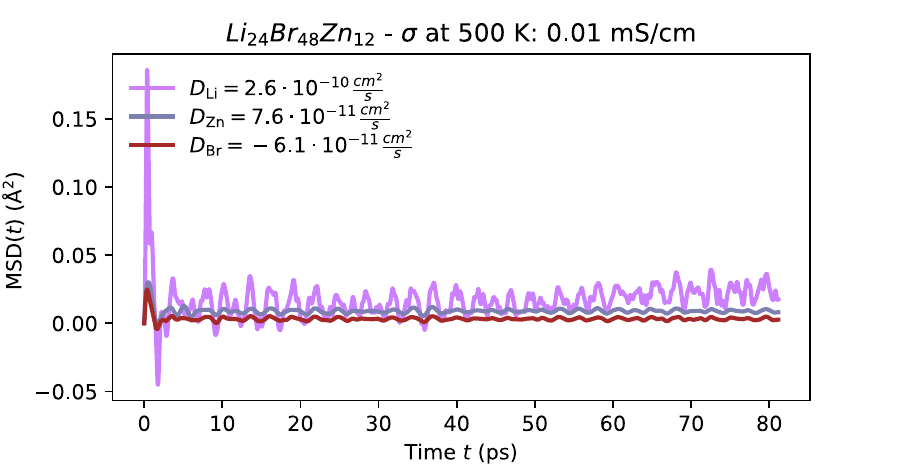}
  \caption{MSD plot of Li along with host-lattice species of $Li_2ZnBr_4$ at all temperatures studied with FPMD}
\end{figure}

\begin{figure}[H]
\centering
  \includegraphics[width=\columnwidth]{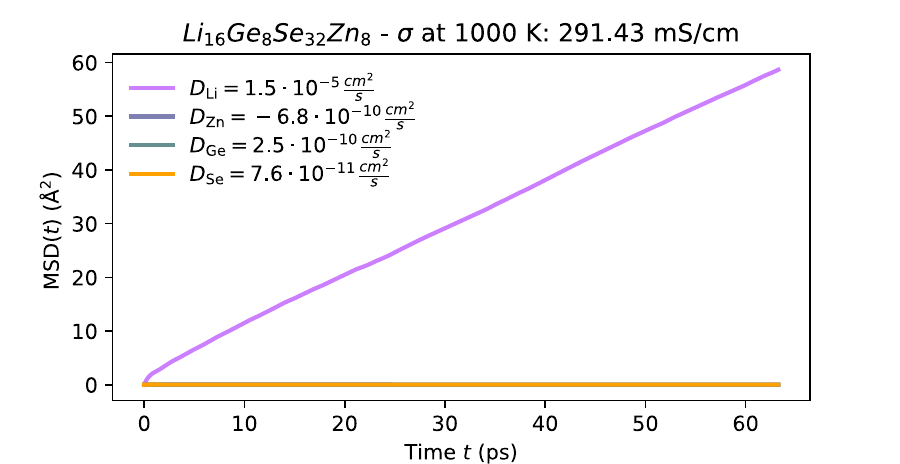}
  \includegraphics[width=\columnwidth]{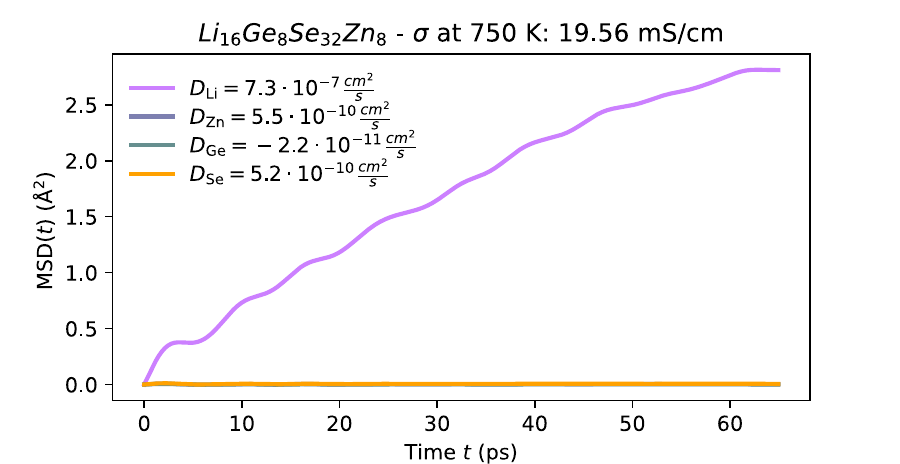}
  \includegraphics[width=\columnwidth]{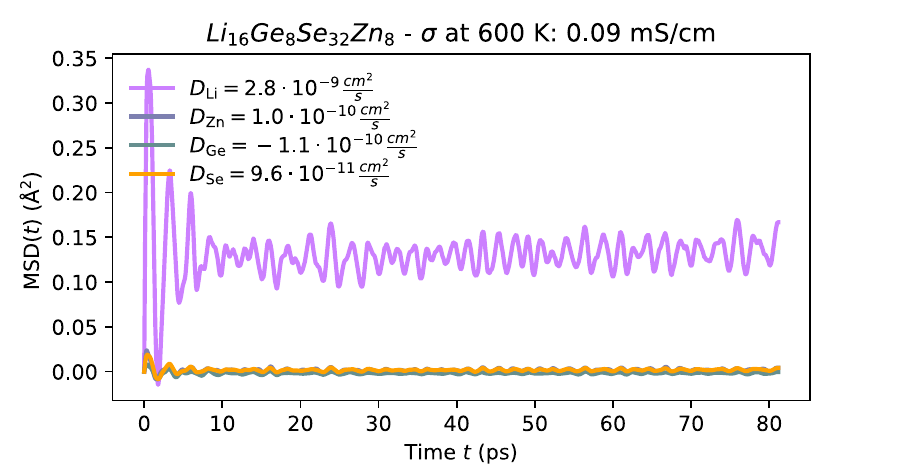}
  \includegraphics[width=\columnwidth]{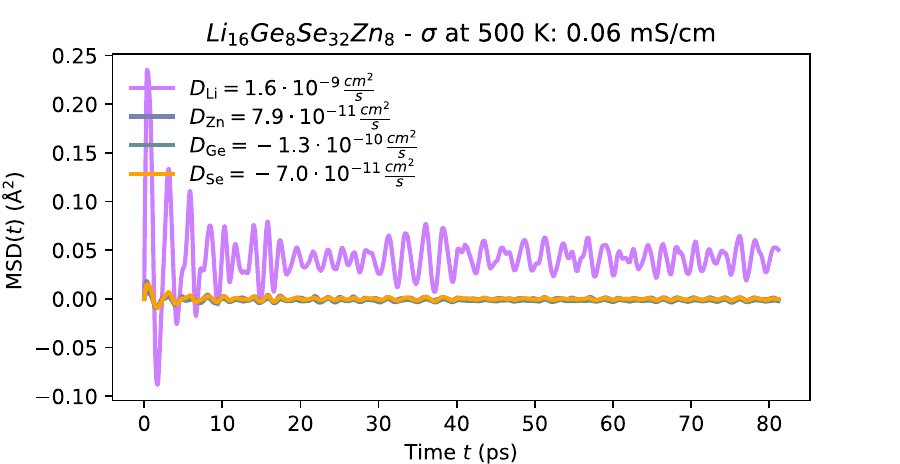}
  \caption{MSD plot of Li along with host-lattice species of $Li_2ZnGeSe_4$ at all temperatures studied with FPMD}
\end{figure}

\begin{figure}[H]
\centering
  \includegraphics[width=\columnwidth]{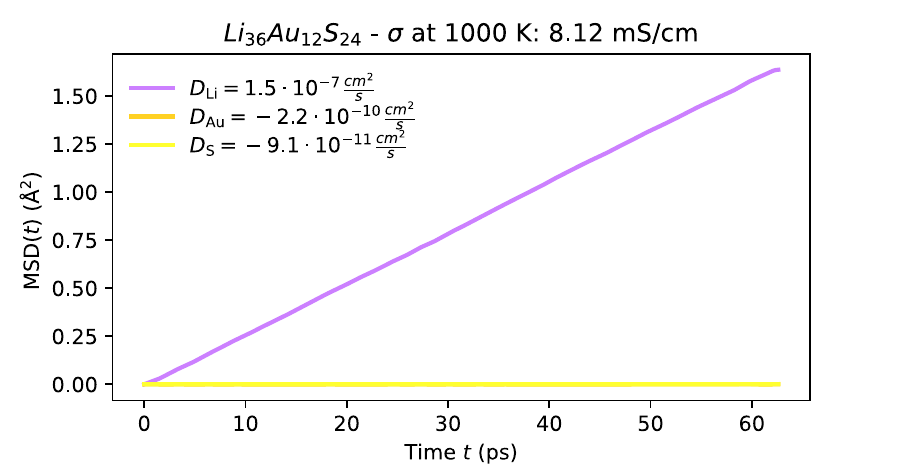}
  \includegraphics[width=\columnwidth]{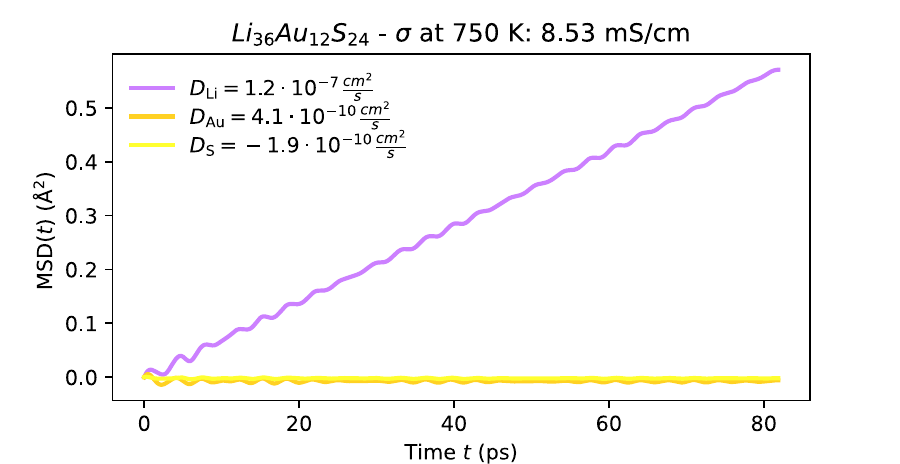}
  \includegraphics[width=\columnwidth]{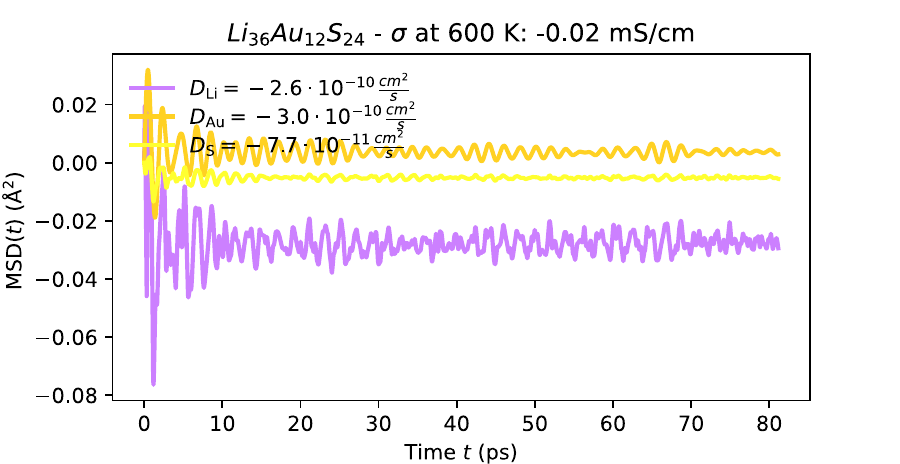}
  \includegraphics[width=\columnwidth]{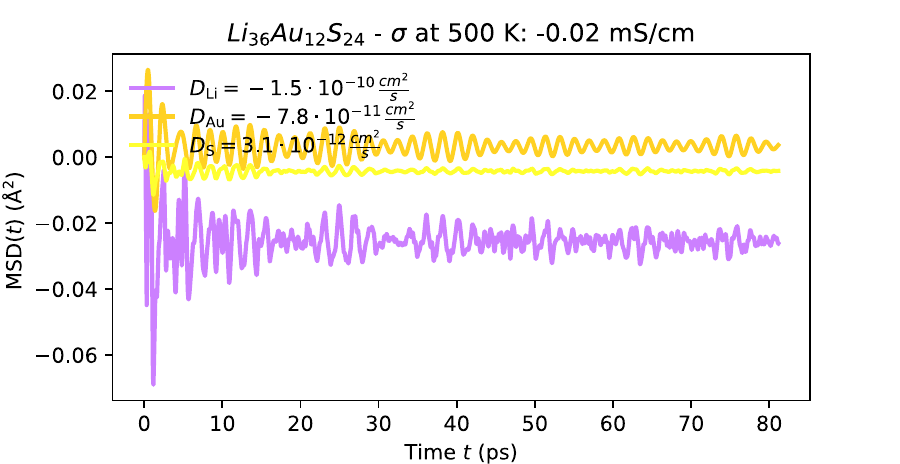}
  \caption{MSD plot of Li along with host-lattice species of $Li_3AuS_2$ at all temperatures studied with FPMD}
\end{figure}

\begin{figure}[H]
\centering
  \includegraphics[width=\columnwidth]{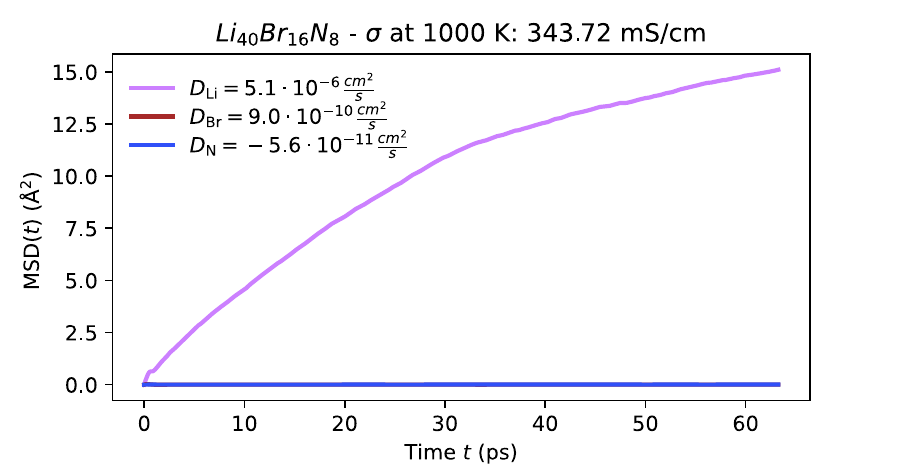}
  \includegraphics[width=\columnwidth]{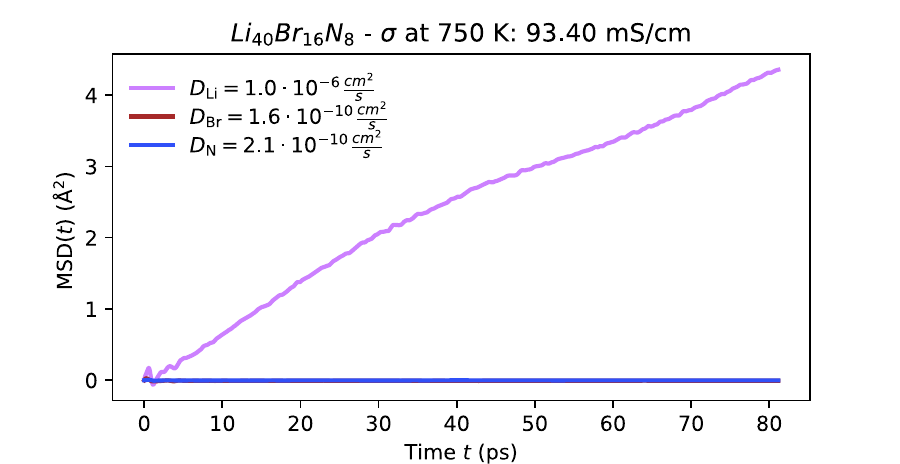}
  \includegraphics[width=\columnwidth]{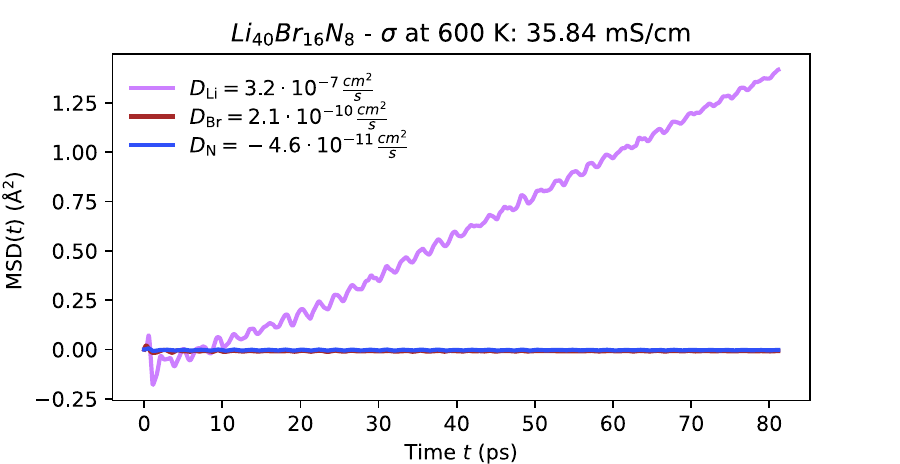}
  \includegraphics[width=\columnwidth]{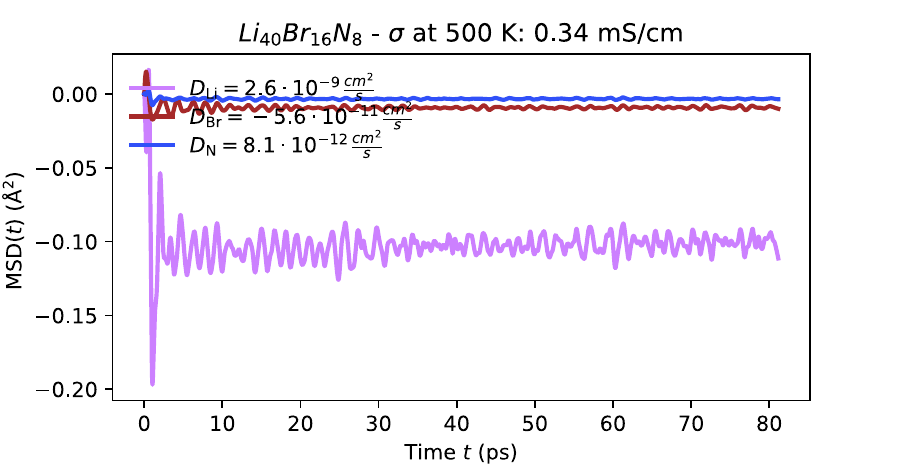}
  \caption{MSD plot of Li along with host-lattice species of $Li_5Br_2N$ at all temperatures studied with FPMD}
\end{figure}

\begin{figure}[H]
\centering
  \includegraphics[width=\columnwidth]{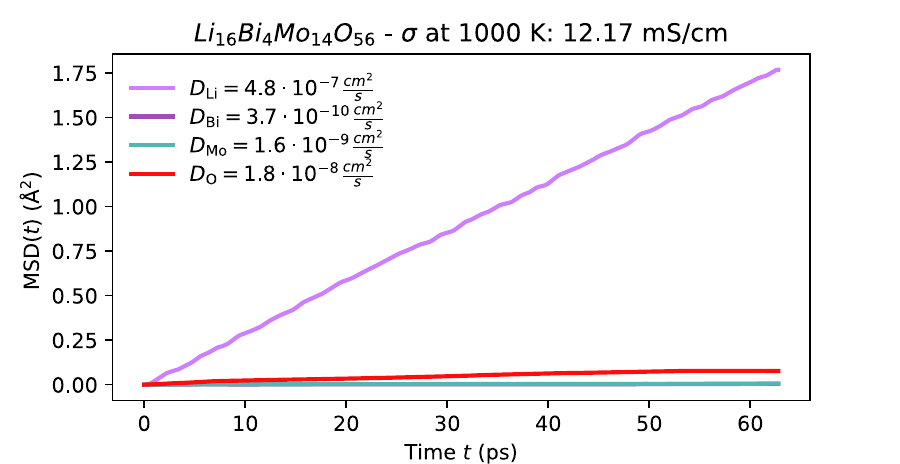}
  \includegraphics[width=\columnwidth]{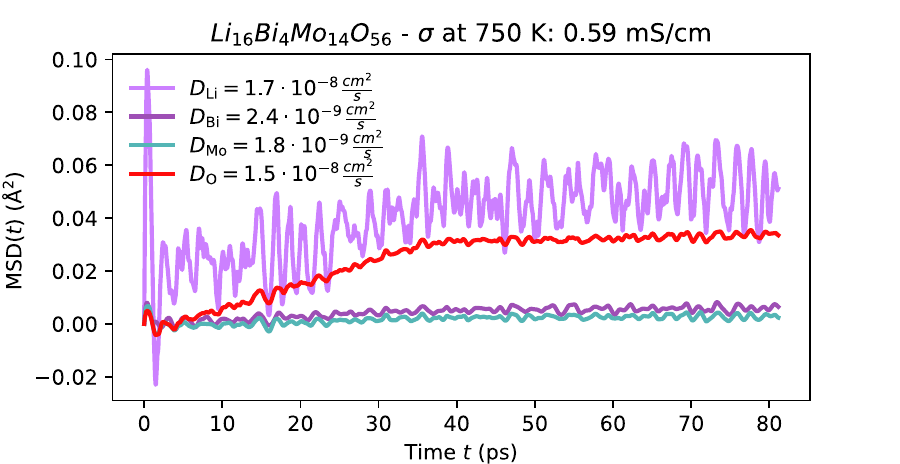}
  \includegraphics[width=\columnwidth]{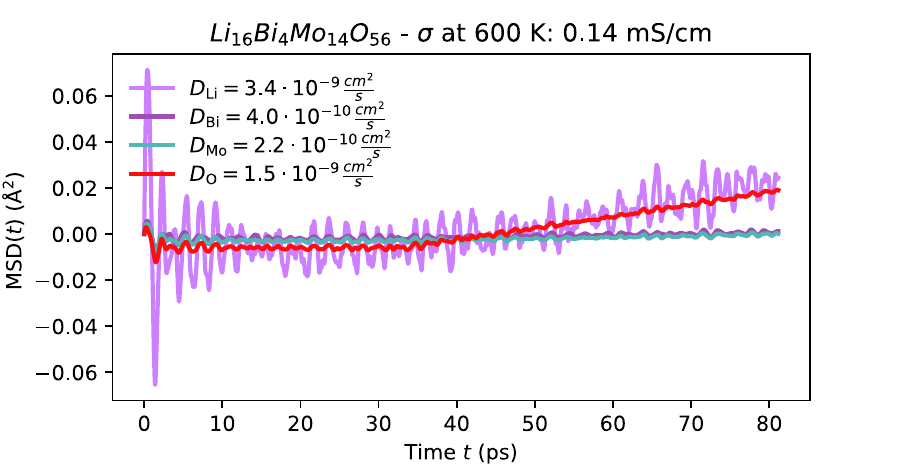}
  \includegraphics[width=\columnwidth]{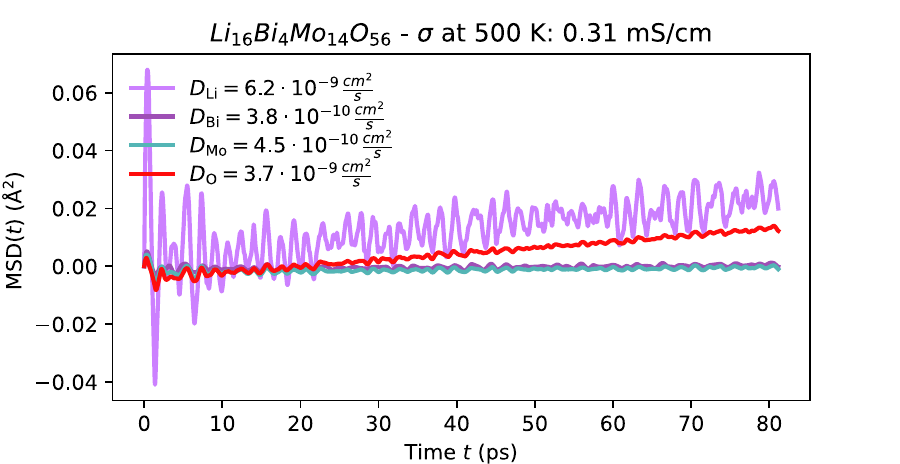}
  \caption{MSD plot of Li along with host-lattice species of $Li_8Bi_2(MoO_4)_7$ at all temperatures studied with FPMD}
\end{figure}

\begin{figure}[H]
\centering
  \includegraphics[width=\columnwidth]{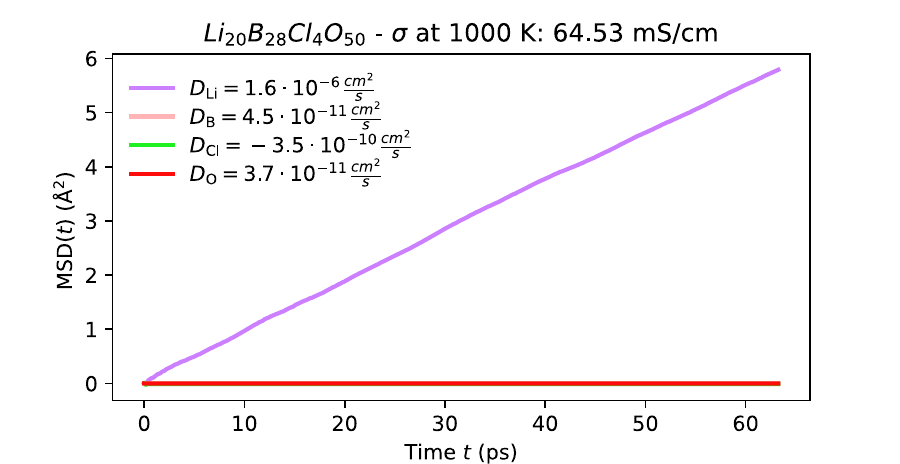}
  \includegraphics[width=\columnwidth]{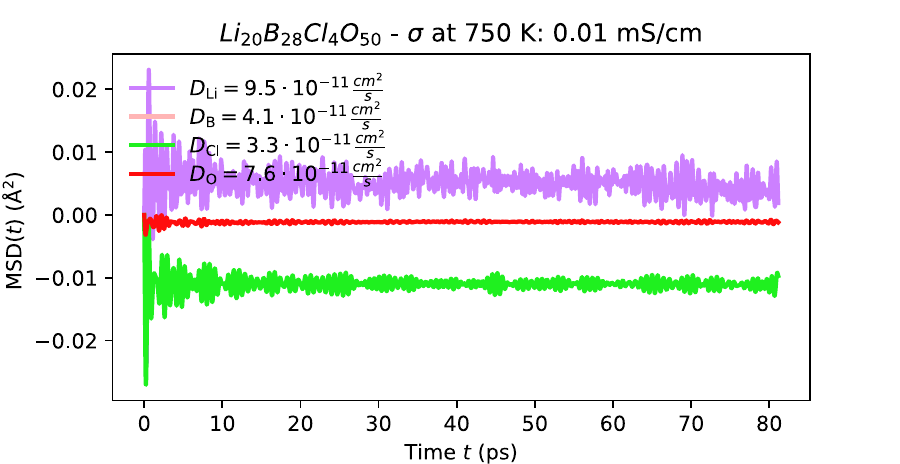}
  \includegraphics[width=\columnwidth]{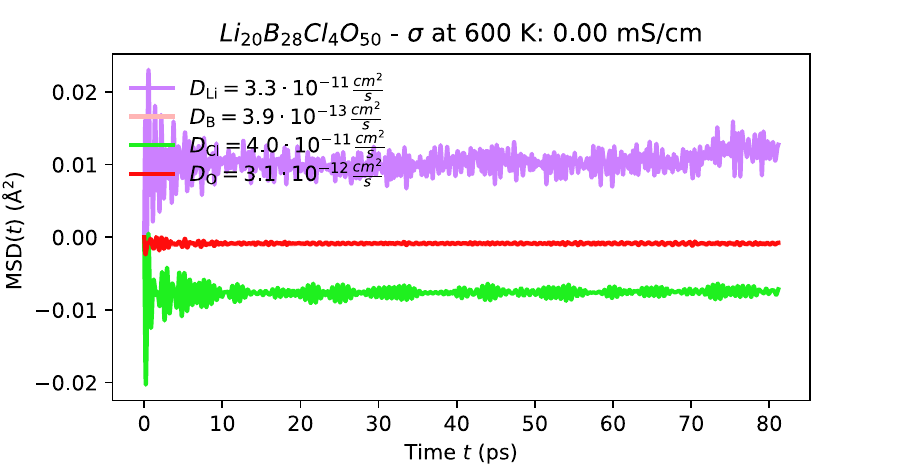}
  \includegraphics[width=\columnwidth]{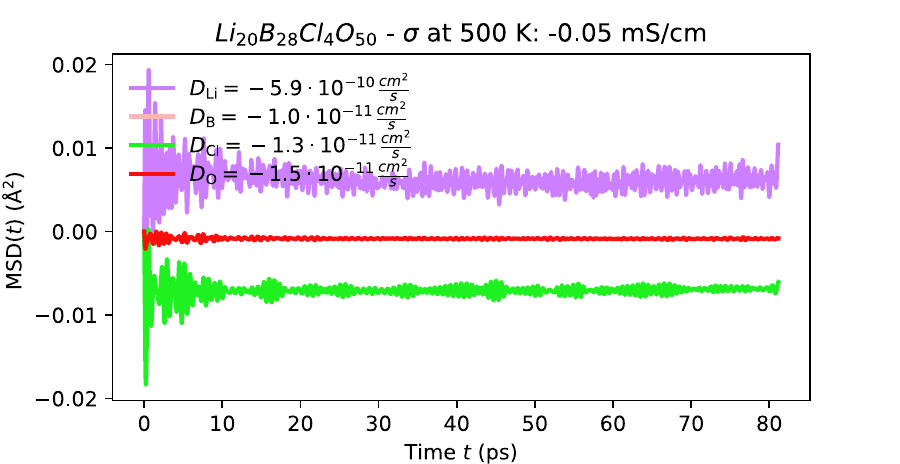}
  \caption{MSD plot of Li along with host-lattice species of $Li_{10}B_{14}Cl_2O_{25}$ at all temperatures studied with FPMD}
\end{figure}

\begin{figure}[H]
\centering
  \includegraphics[width=\columnwidth]{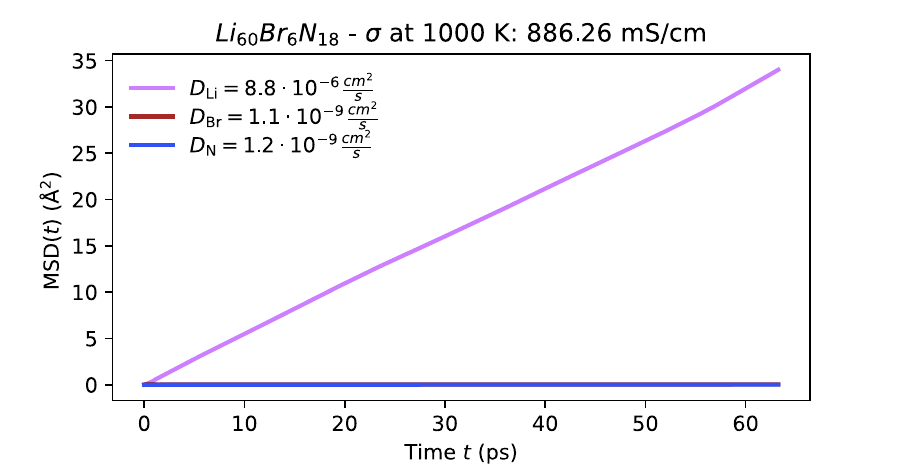}
  \includegraphics[width=\columnwidth]{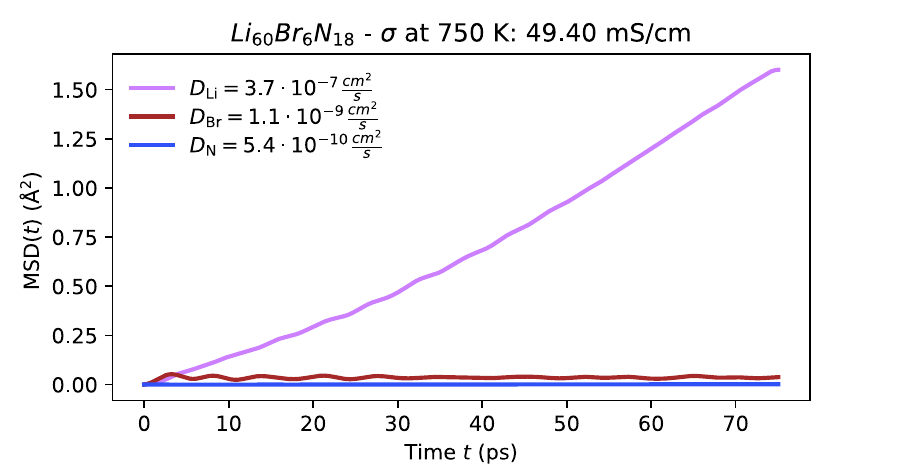}
  \includegraphics[width=\columnwidth]{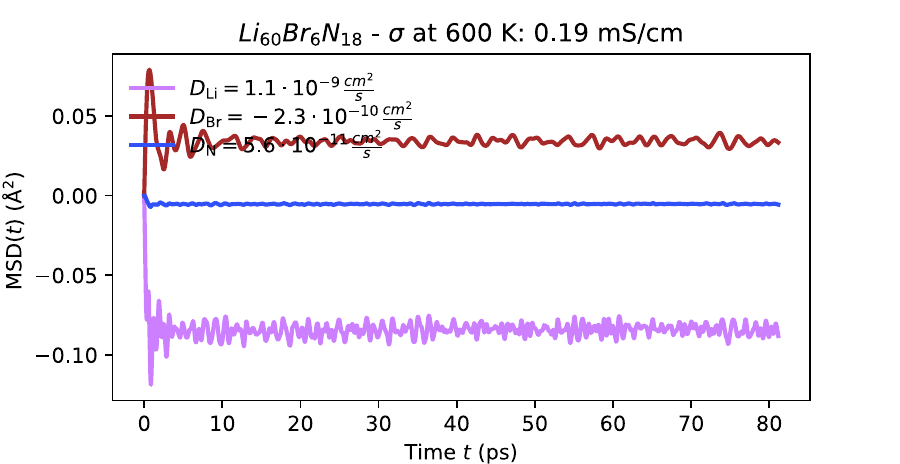}
  \includegraphics[width=\columnwidth]{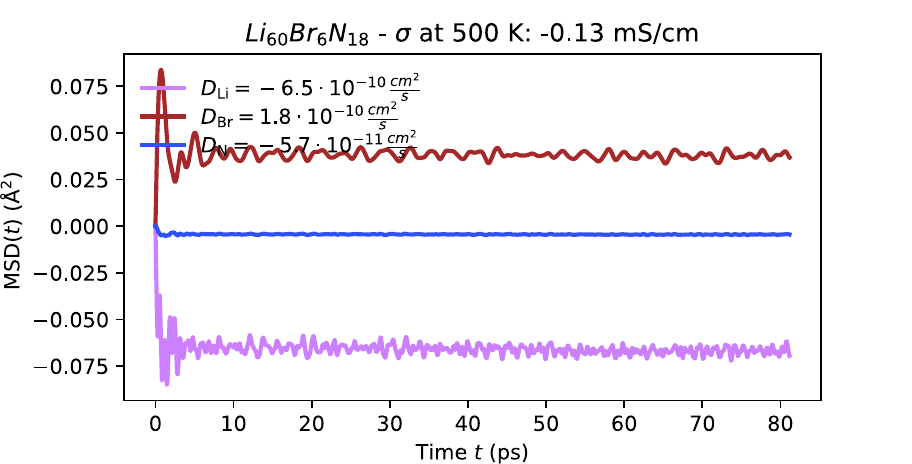}
  \caption{MSD plot of Li along with host-lattice species of $Li_10BrN_3$ at all temperatures studied with FPMD}
\end{figure}

\begin{figure}[H]
\centering
  \includegraphics[width=\columnwidth]{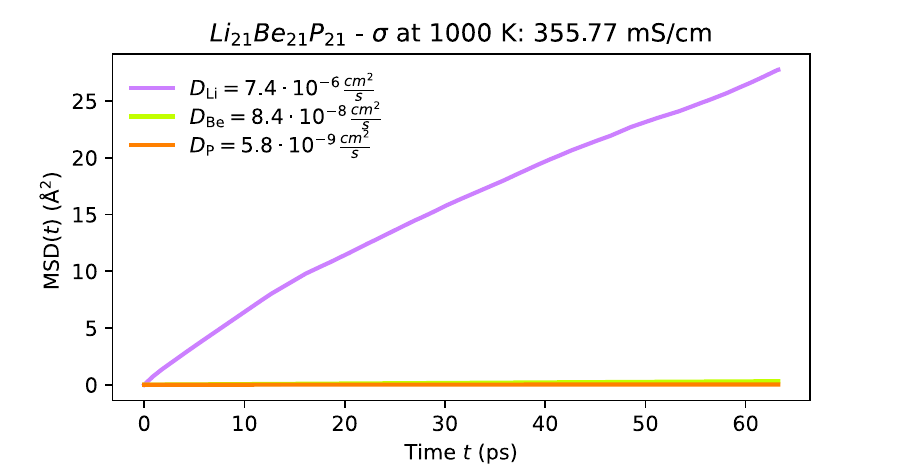}
  \includegraphics[width=\columnwidth]{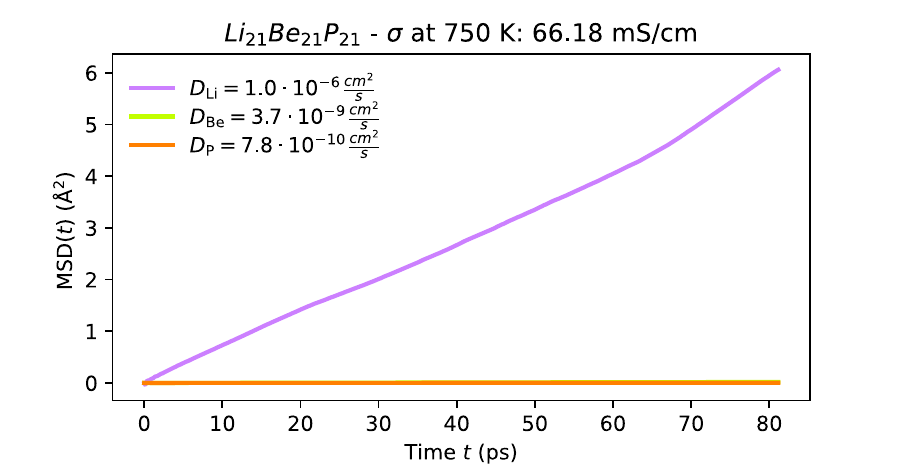}
  \includegraphics[width=\columnwidth]{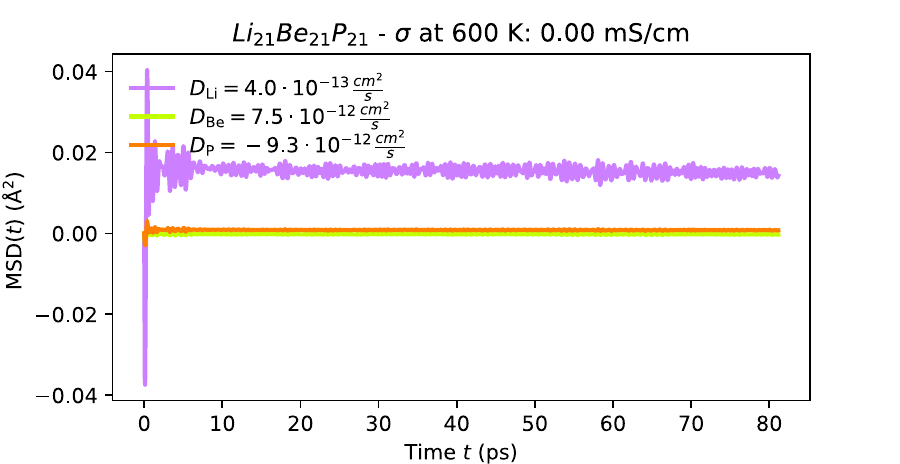}
  \includegraphics[width=\columnwidth]{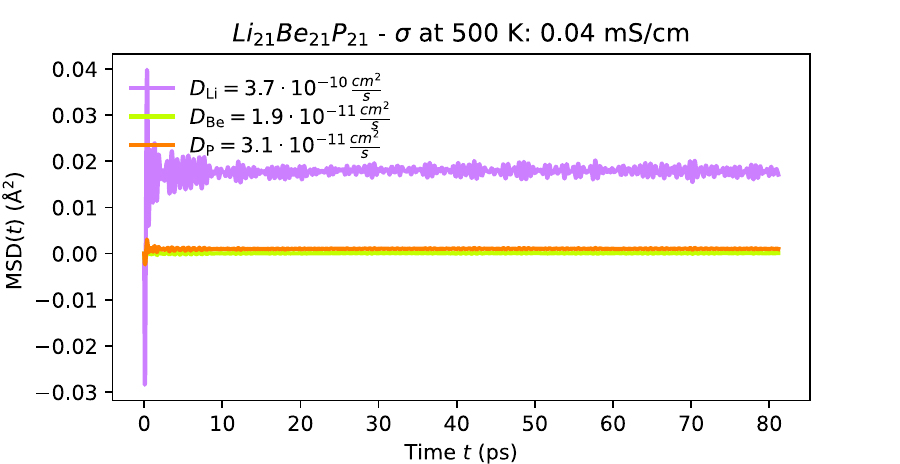}
  \caption{MSD plot of Li along with host-lattice species of $LiBeP$ at all temperatures studied with FPMD}
\end{figure}

\begin{figure}[H]
\centering
  \includegraphics[width=\columnwidth]{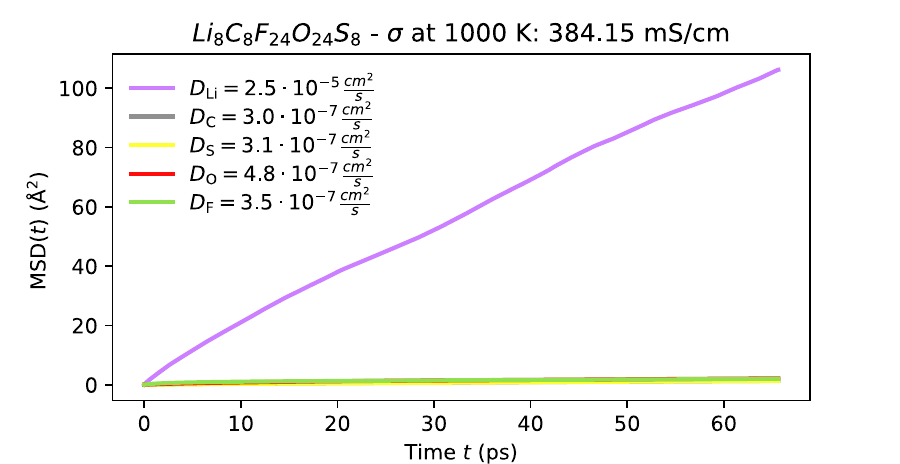}
  \includegraphics[width=\columnwidth]{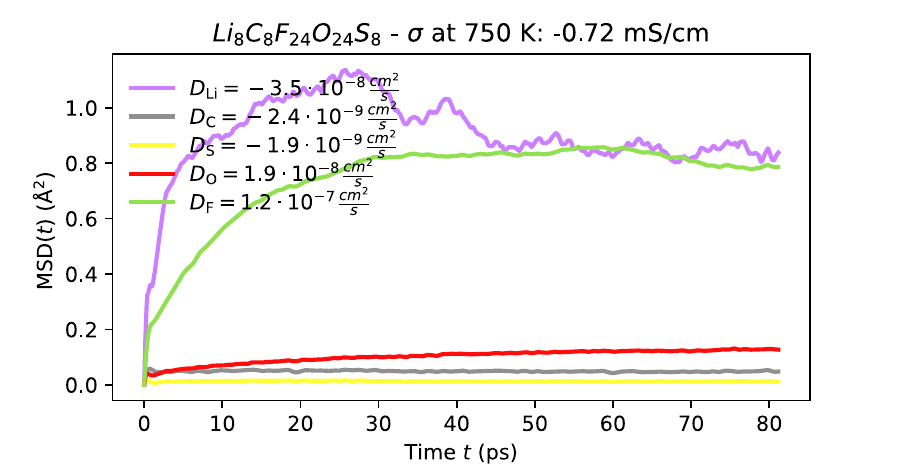}
  \includegraphics[width=\columnwidth]{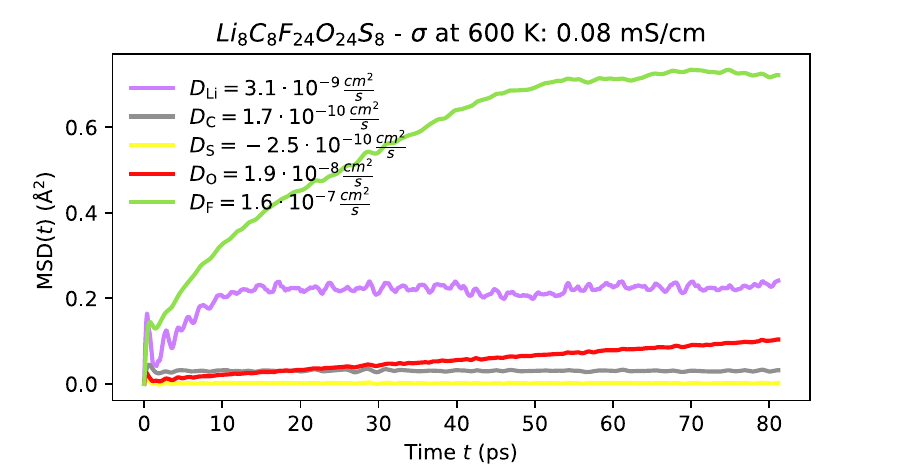}
  \includegraphics[width=\columnwidth]{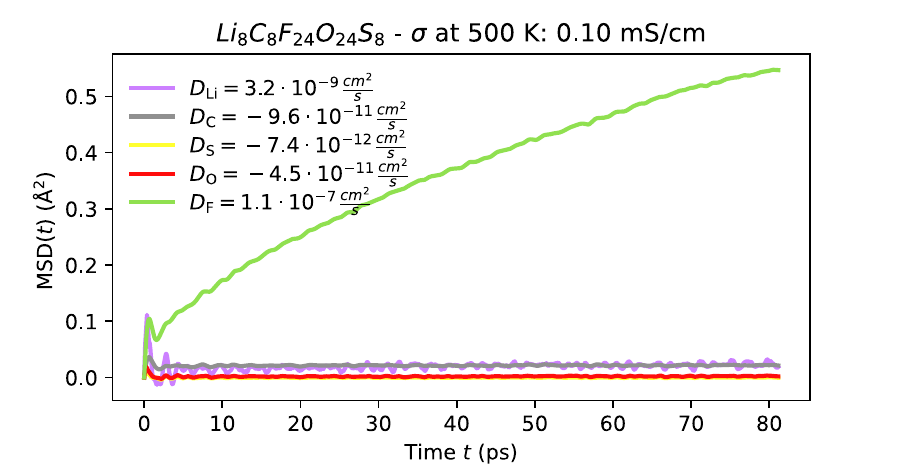}
  \caption{MSD plot of Li along with host-lattice species of $LiCS(OF)_3$ at all temperatures studied with FPMD}
\end{figure}

\begin{figure}[H]
\centering
  \includegraphics[width=\columnwidth]{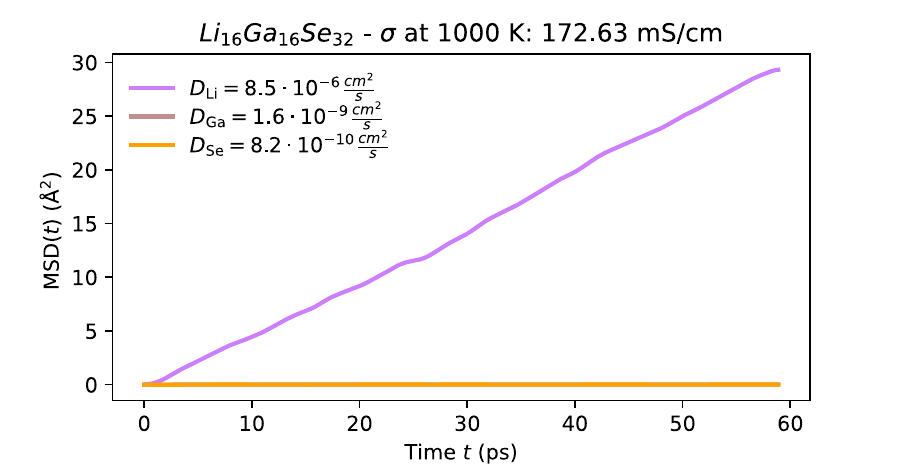}
  \includegraphics[width=\columnwidth]{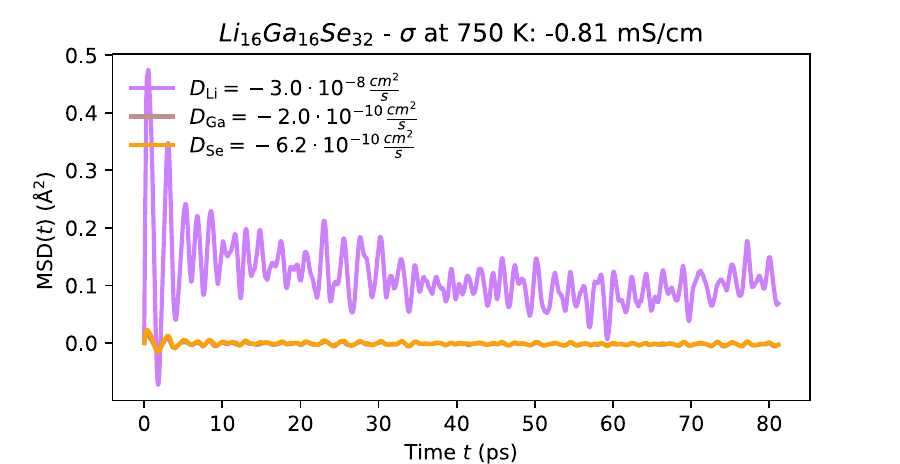}
  \includegraphics[width=\columnwidth]{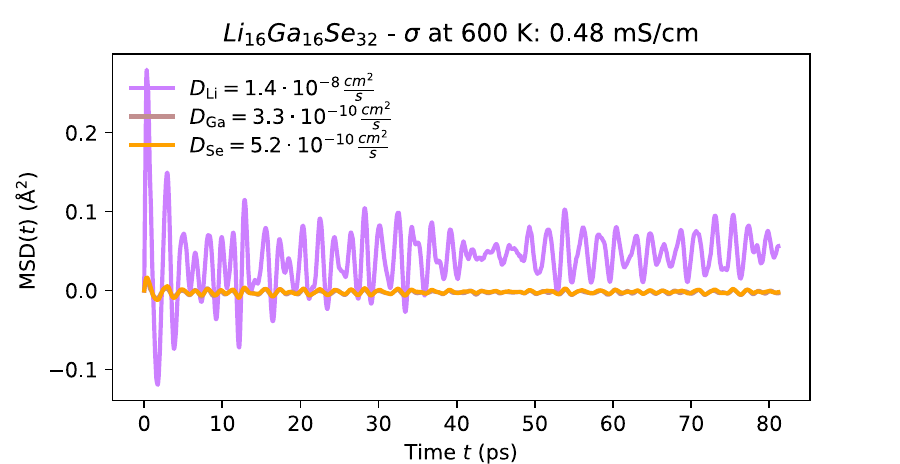}
  \includegraphics[width=\columnwidth]{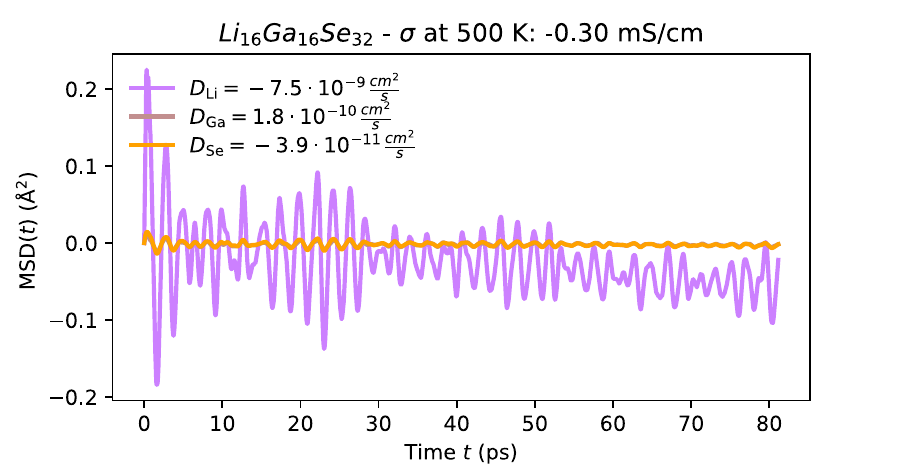}
  \caption{MSD plot of Li along with host-lattice species of $LiGaSe_2$ at all temperatures studied with FPMD}
\end{figure}

\begin{figure}[H]
\centering
  \includegraphics[width=\columnwidth]{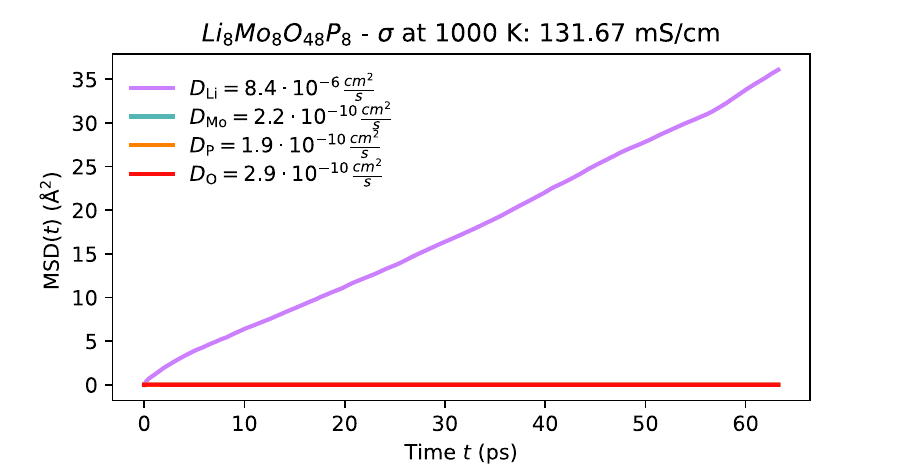}
  \includegraphics[width=\columnwidth]{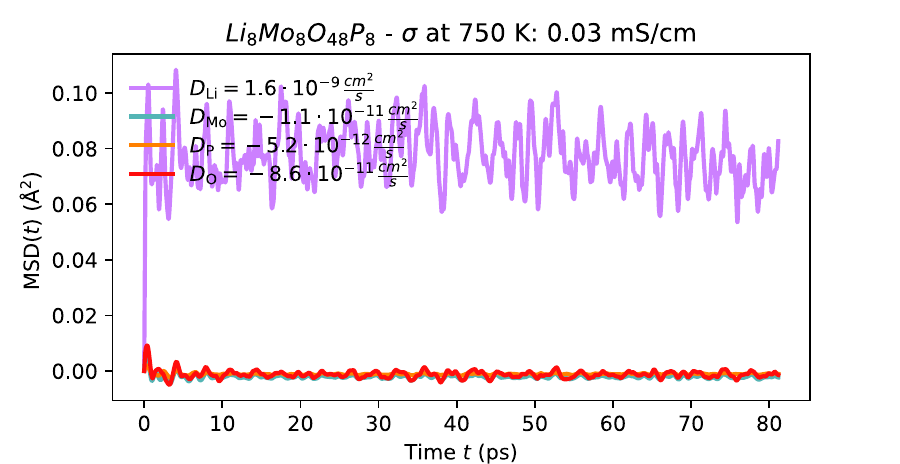}
  \includegraphics[width=\columnwidth]{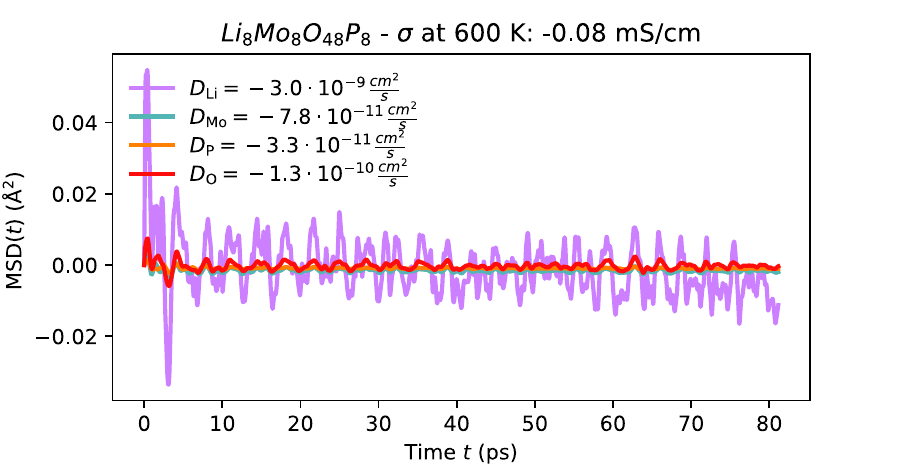}
  \includegraphics[width=\columnwidth]{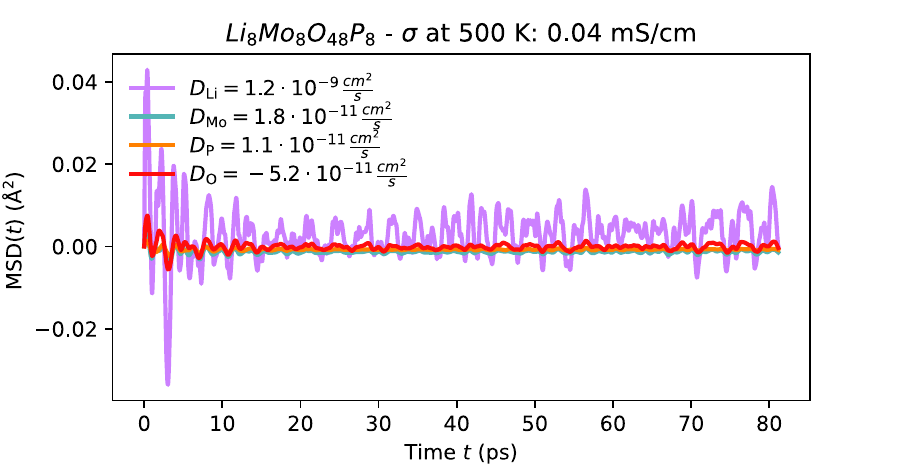}
  \caption{MSD plot of Li along with host-lattice species of $LiMoPO_6$ at all temperatures studied with FPMD}
\end{figure}

\begin{figure}[H]
\centering
  \includegraphics[width=\columnwidth]{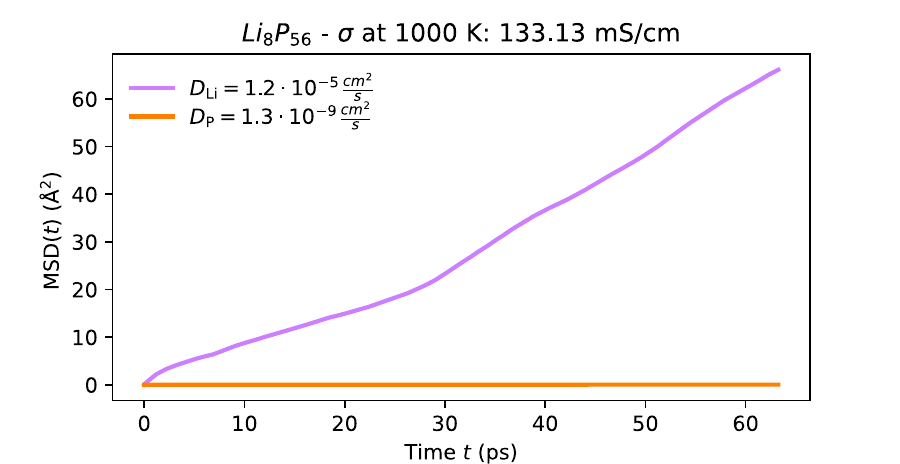}
  \includegraphics[width=\columnwidth]{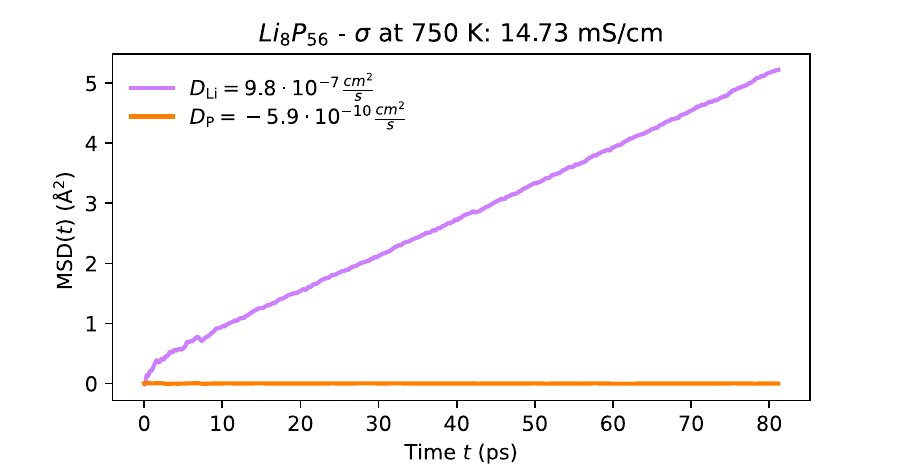}
  \includegraphics[width=\columnwidth]{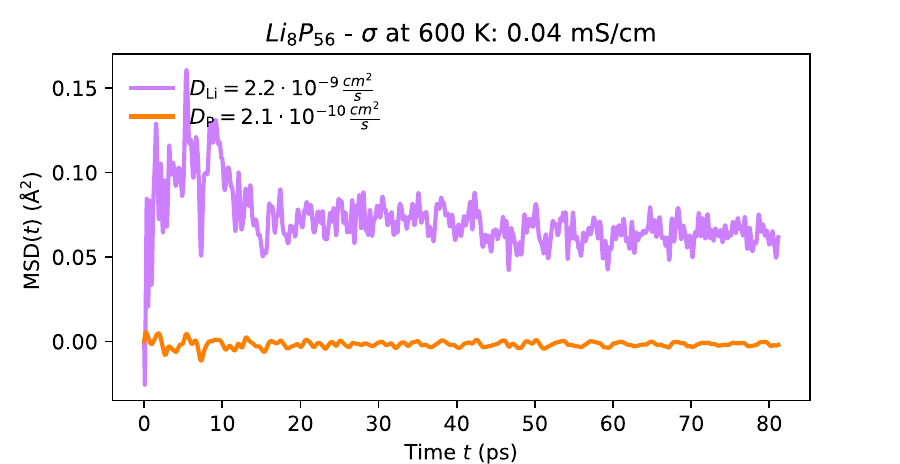}
  \includegraphics[width=\columnwidth]{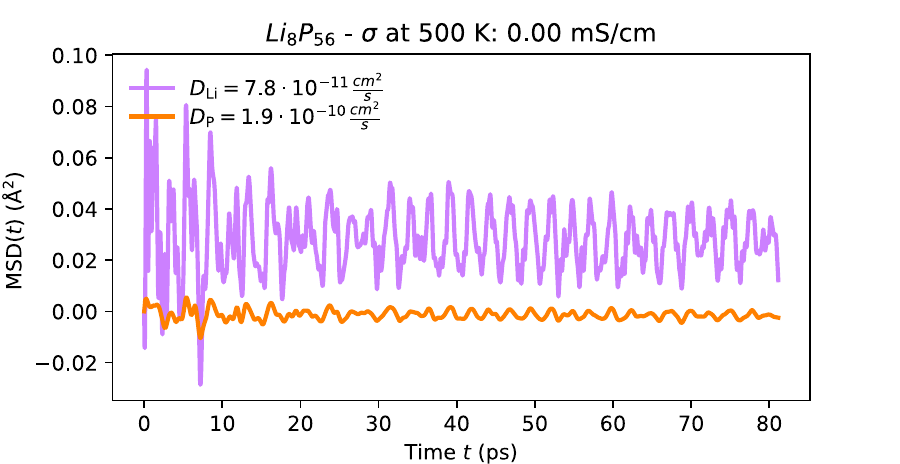}
  \caption{MSD plot of Li along with host-lattice species of $LiP_7$ at all temperatures studied with FPMD}
\end{figure}

\begin{figure}[H]
\centering
  \includegraphics[width=\columnwidth]{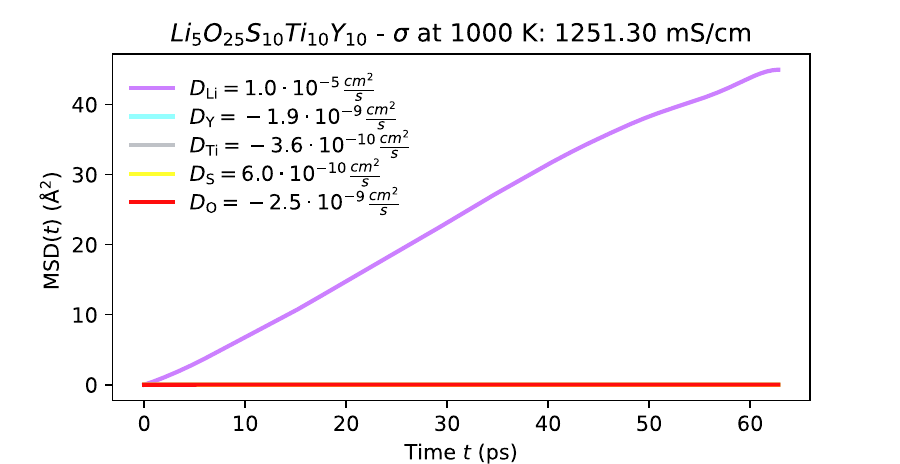}
  \includegraphics[width=\columnwidth]{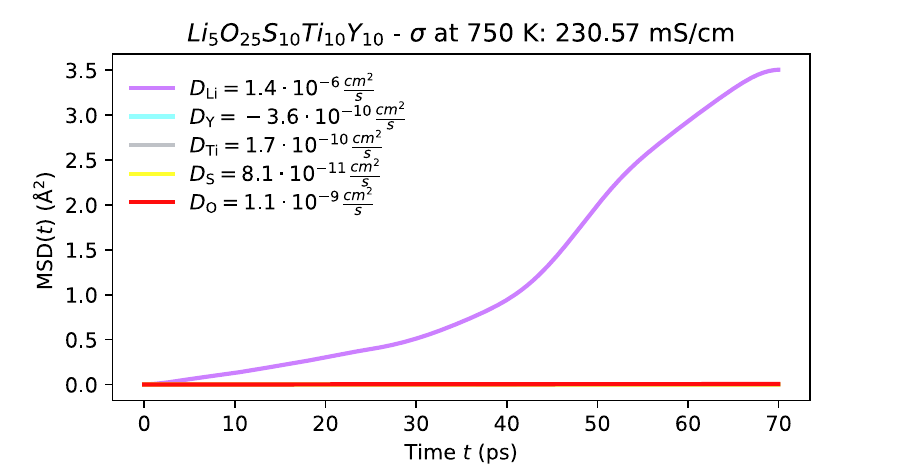}
  \includegraphics[width=\columnwidth]{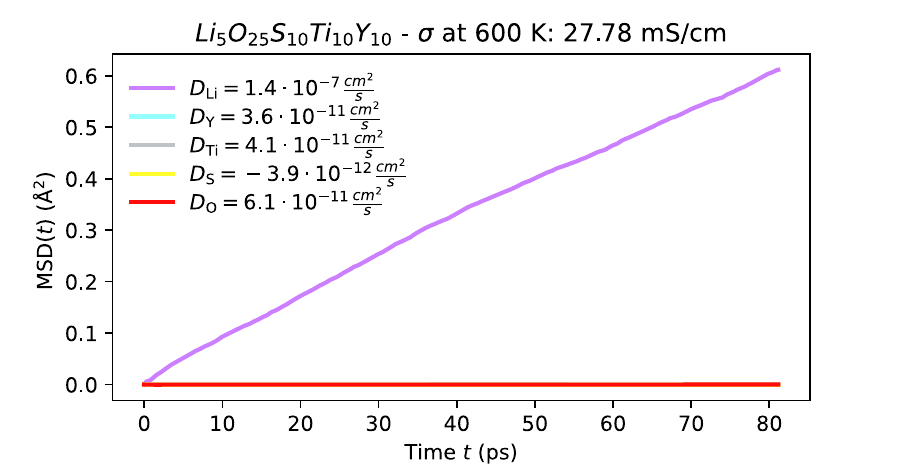}
  \includegraphics[width=\columnwidth]{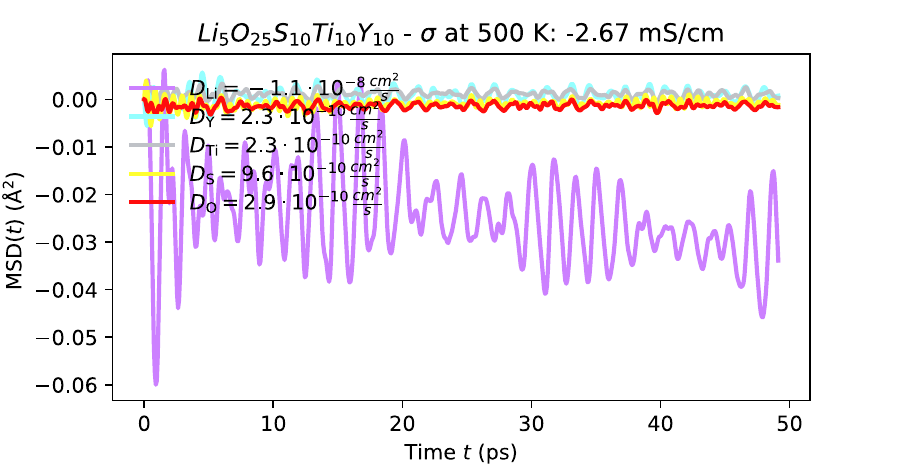}
  \caption{MSD plot of Li along with host-lattice species of $LiY_2Ti_2S_2O_5$ at all temperatures studied with FPMD}
\end{figure}

\begin{figure}[H]
\centering
  \includegraphics[width=\columnwidth]{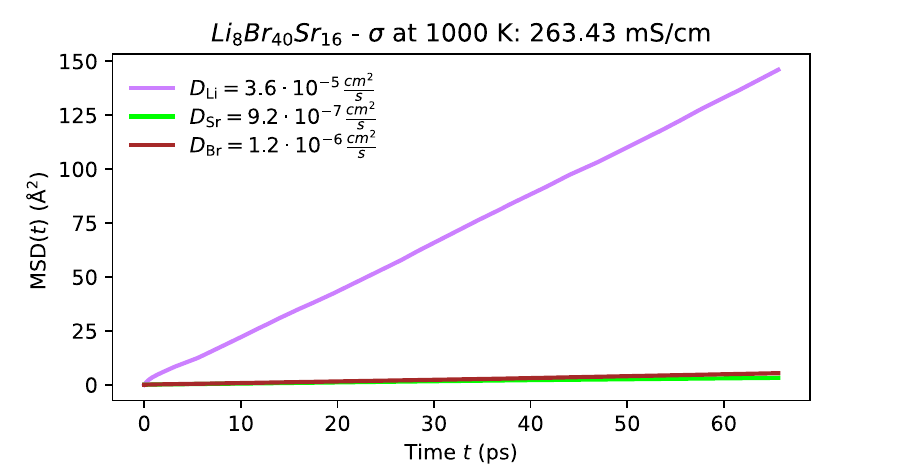}
  \includegraphics[width=\columnwidth]{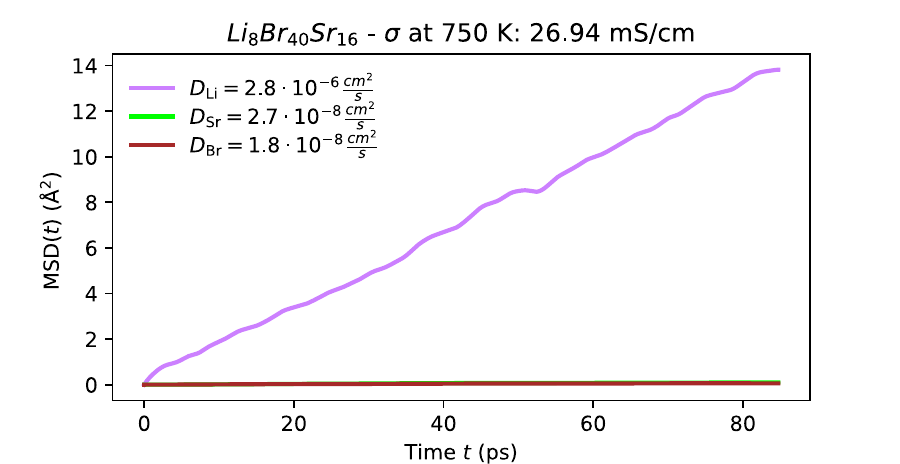}
  \includegraphics[width=\columnwidth]{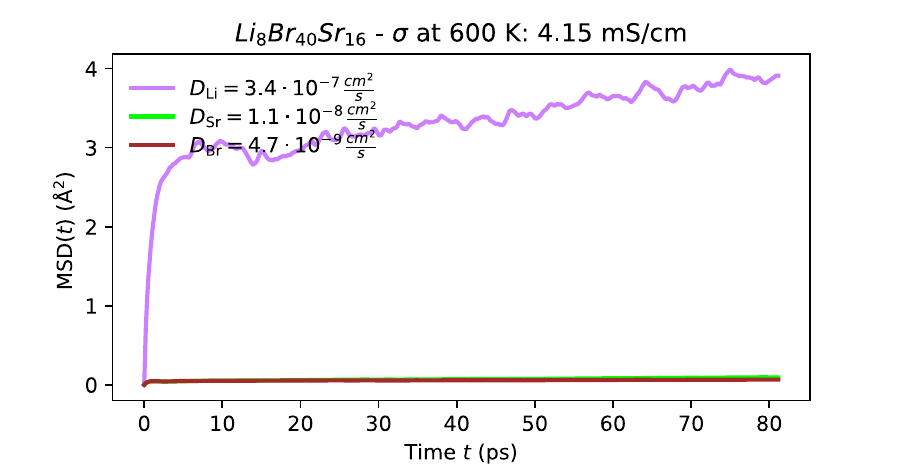}
  \includegraphics[width=\columnwidth]{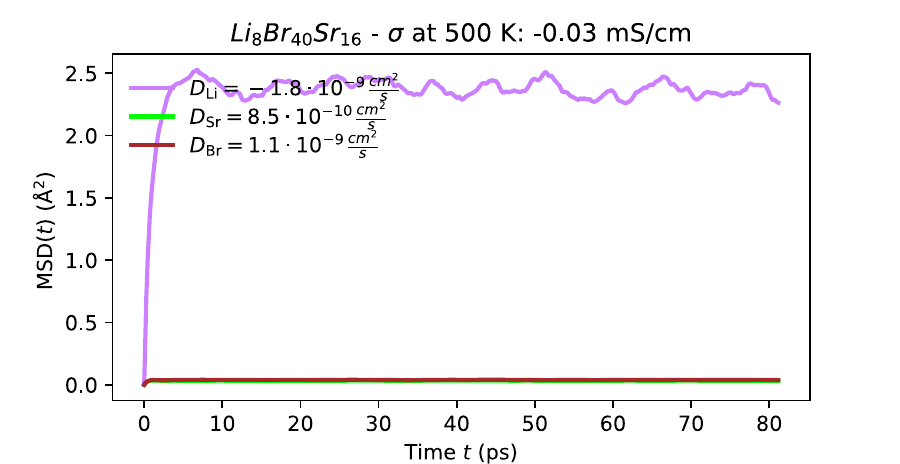}
  \caption{MSD plot of Li along with host-lattice species of $LiSr_2Br_5$ at all temperatures studied with FPMD}
\end{figure}

\FloatBarrier

\phantomsection
\section{Non diffusive structures} \label{S1_slow}

We find 18 materials that do not exhibit Li-ion diffusion in our FPMD simulations at 1000 K. 
We show the MSD plots only at 1000 K as we did not run FPMD at other temperatures on account of not being able to resolve diffusion at 1000 K. 

\begin{figure}[H]
\centering
  \includegraphics[width=\columnwidth]{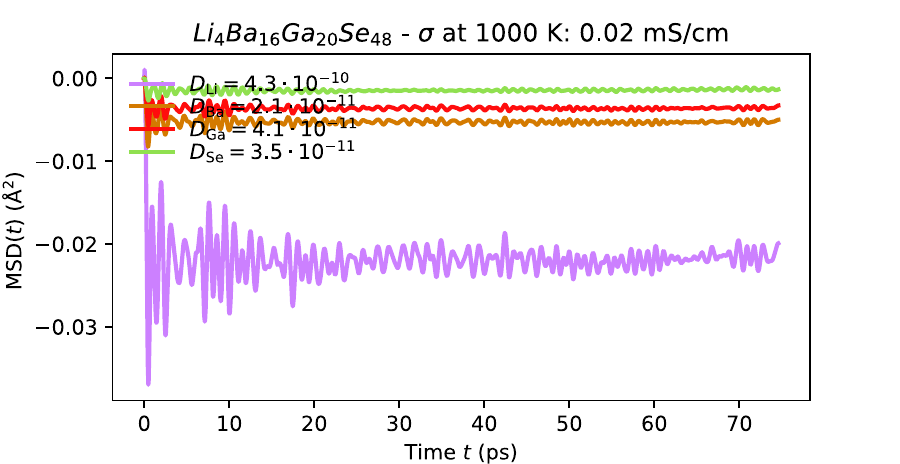}
  \caption{MSD plot of Li along with host-lattice species of $LiBa_4Ga_5Se_12$ at 1000 K studied with FPMD}
\end{figure}

\begin{figure}[H]
\centering
  \includegraphics[width=\columnwidth]{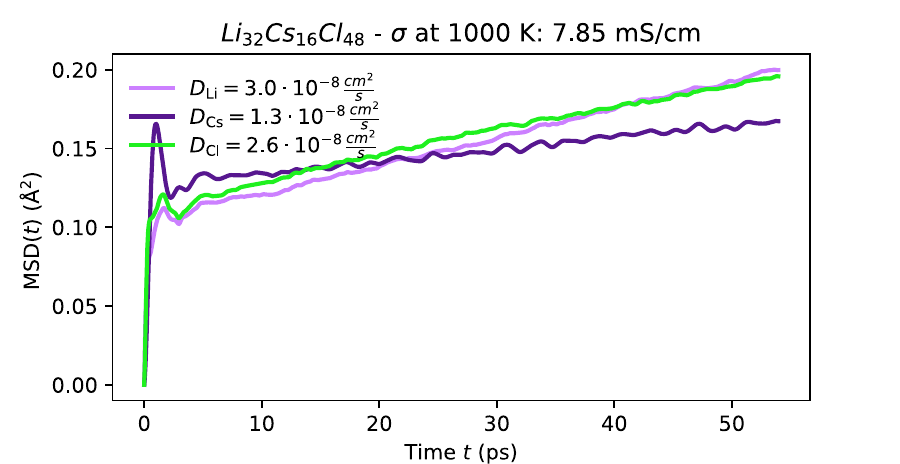}
  \caption{MSD plot of Li along with host-lattice species of $Li_2CsCl_3$ at 1000 K studied with FPMD}
\end{figure}

\begin{figure}[H]
\centering
  \includegraphics[width=\columnwidth]{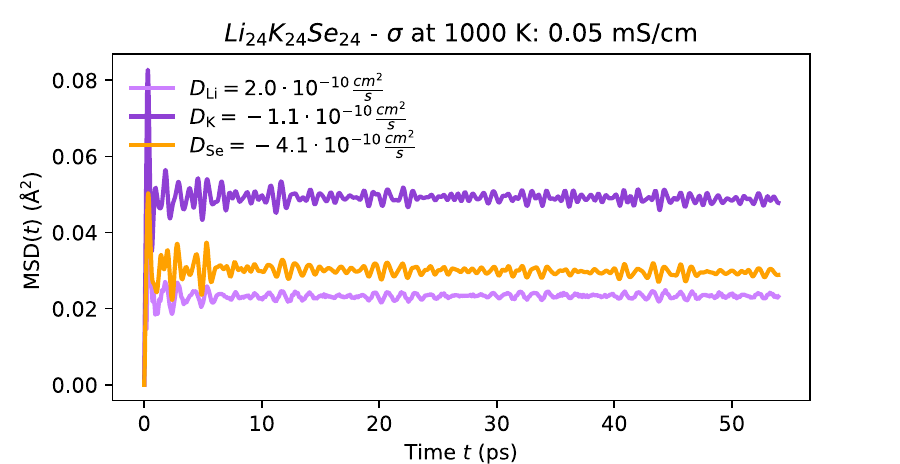}
  \caption{MSD plot of Li along with host-lattice species of $LiKSe$ at 1000 K studied with FPMD}
\end{figure}

\pagebreak
\FloatBarrier

\begin{figure}[H]
\centering
  \includegraphics[width=\columnwidth]{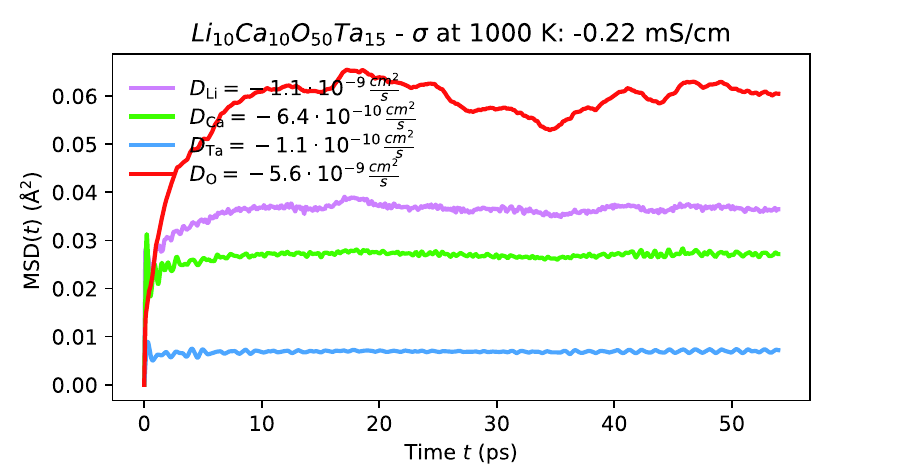}
  \caption{MSD plot of Li along with host-lattice species of $Li_2Ca_2Ta_3O_{10}$ at 1000 K studied with FPMD}
\end{figure}

\begin{figure}[H]
\centering
  \includegraphics[width=\columnwidth]{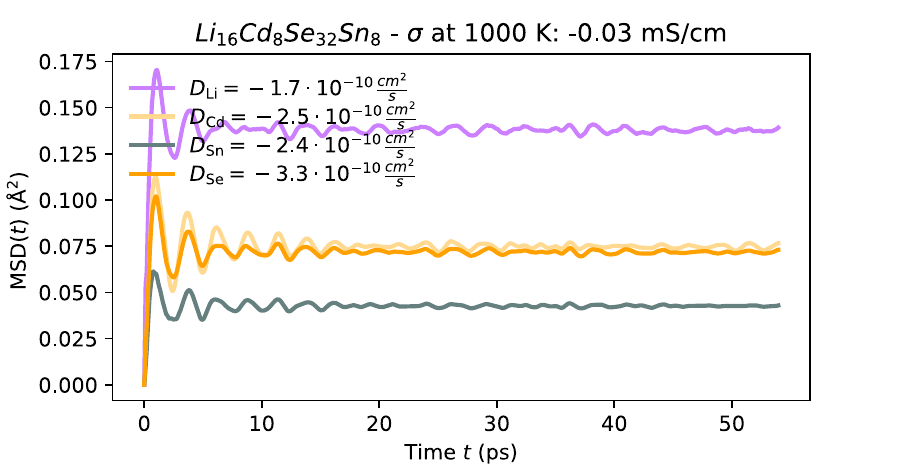}
  \caption{MSD plot of Li along with host-lattice species of $Li_2CdSnSe_4$ at 1000 K studied with FPMD}
\end{figure}

\begin{figure}[H]
\centering
  \includegraphics[width=\columnwidth]{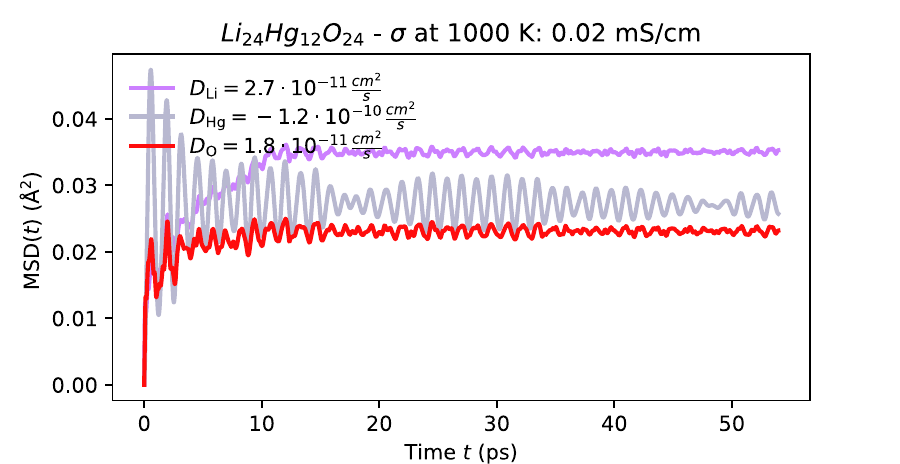}
  \caption{MSD plot of Li along with host-lattice species of $Li_2HgO_2$ at 1000 K studied with FPMD}
\end{figure}

\begin{figure}[H]
\centering
  \includegraphics[width=\columnwidth]{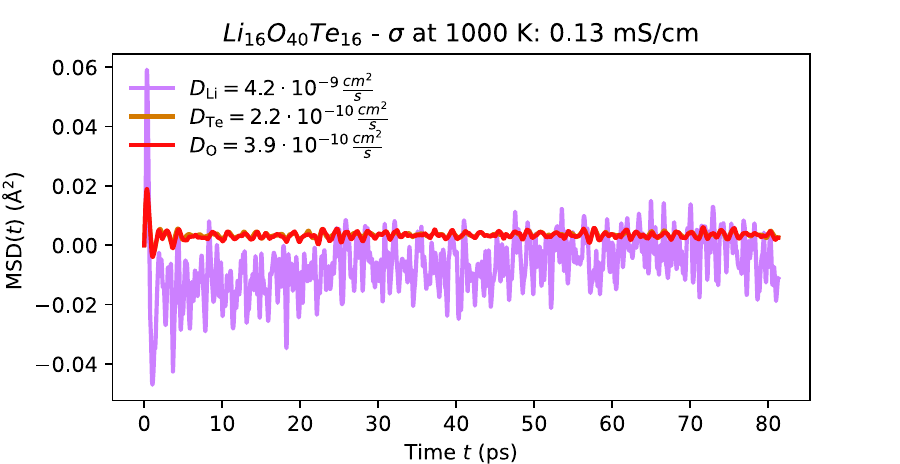}
  \caption{MSD plot of Li along with host-lattice species of $Li_2Te_2O_5$ at 1000 K studied with FPMD}
\end{figure}

\begin{figure}[H]
\centering
  \includegraphics[width=\columnwidth]{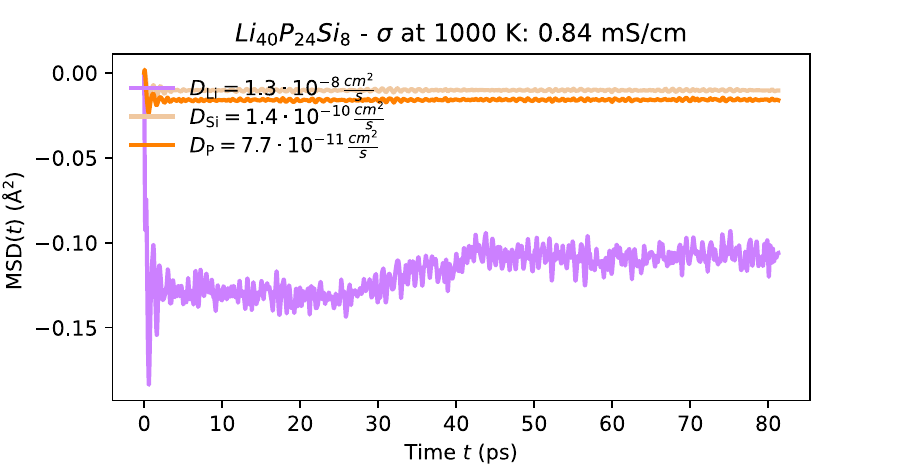}
  \caption{MSD plot of Li along with host-lattice species of $Li_5SiP_3$ at 1000 K studied with FPMD}
\end{figure}

\begin{figure}[H]
\centering
  \includegraphics[width=\columnwidth]{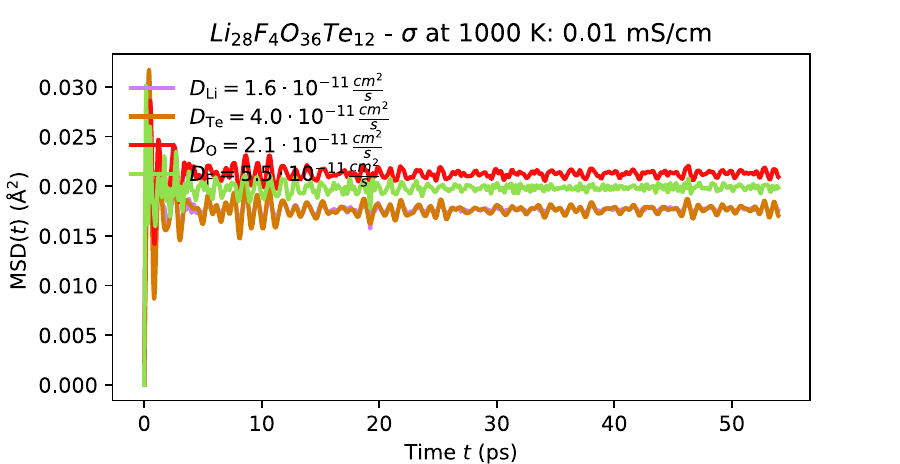}
  \caption{MSD plot of Li along with host-lattice species of $Li_7Te_3O_9F$ at 1000 K studied with FPMD}
\end{figure}

\pagebreak
\FloatBarrier

\begin{figure}[H]
\centering
  \includegraphics[width=\columnwidth]{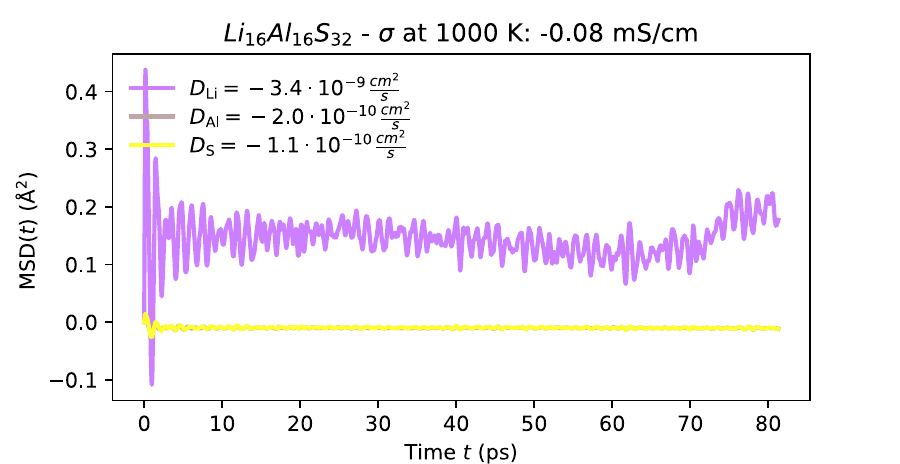}
  \caption{MSD plot of Li along with host-lattice species of $LiAlS_2$ at 1000 K studied with FPMD}
\end{figure}

\begin{figure}[H]
\centering
  \includegraphics[width=\columnwidth]{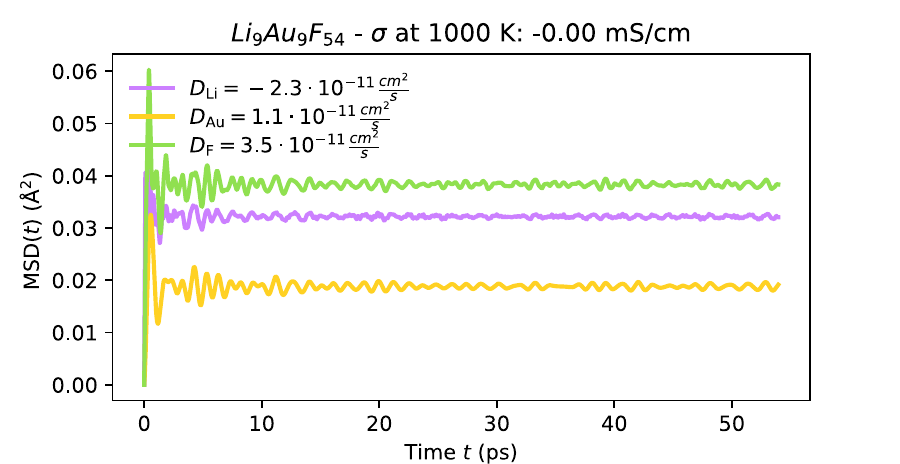}
  \caption{MSD plot of Li along with host-lattice species of $LiAuF_6$ at 1000 K studied with FPMD}
\end{figure}

\begin{figure}[H]
\centering
  \includegraphics[width=\columnwidth]{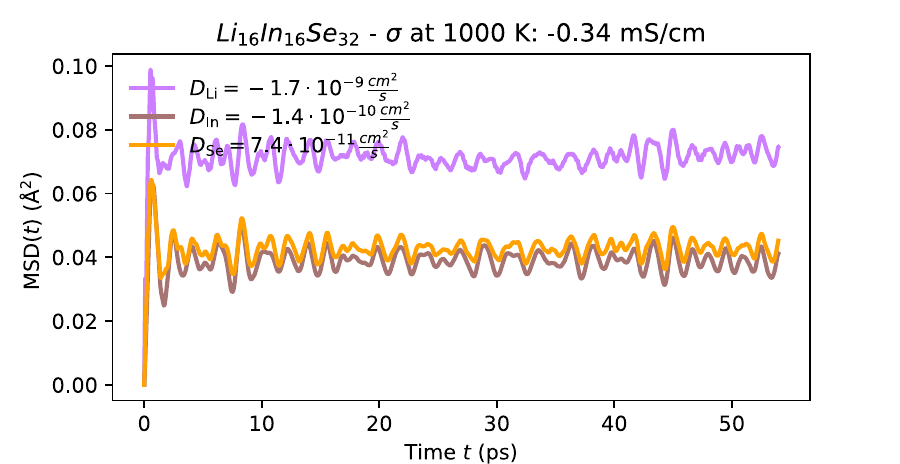}
  \caption{MSD plot of Li along with host-lattice species of $LiInSe_2$ at 1000 K studied with FPMD}
\end{figure}

\begin{figure}[H]
\centering
  \includegraphics[width=\columnwidth]{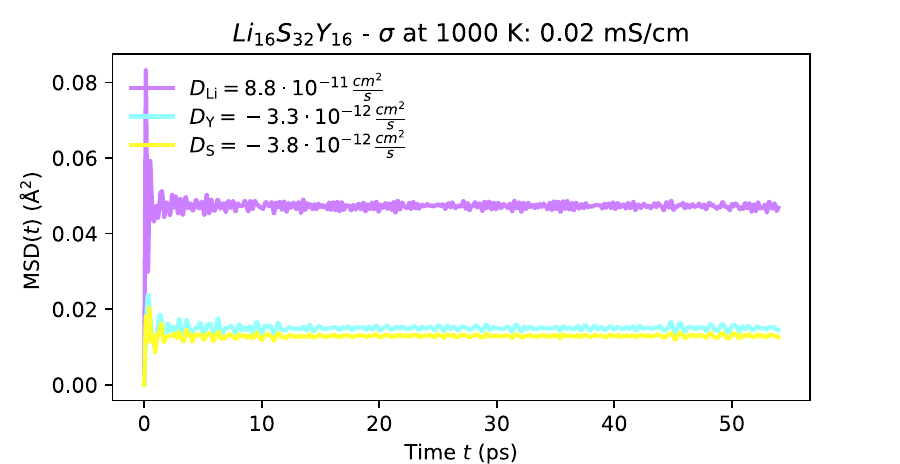}
  \caption{MSD plot of Li along with host-lattice species of $LiYS_2$ at 1000 K studied with FPMD}
\end{figure}

\begin{figure}[H]
\centering
  \includegraphics[width=\columnwidth]{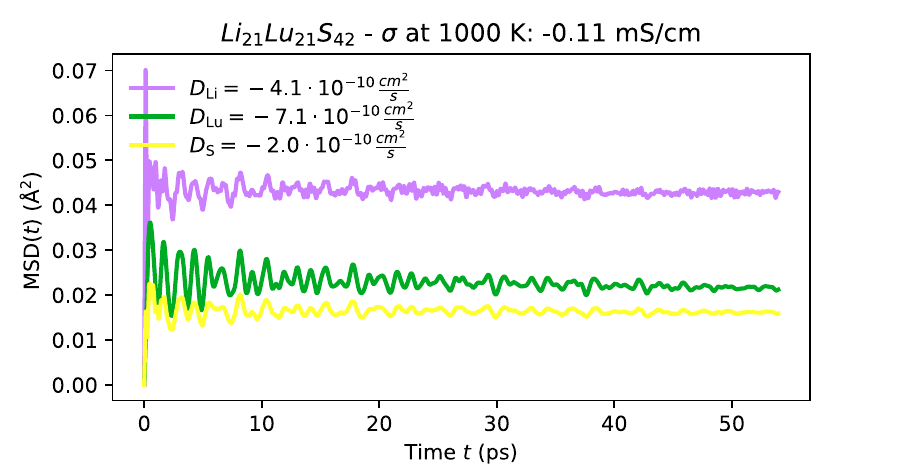}
  \caption{MSD plot of Li along with host-lattice species of $LiLuS_2$ at 1000 K studied with FPMD}
\end{figure}

\begin{figure}[H]
\centering
  \includegraphics[width=\columnwidth]{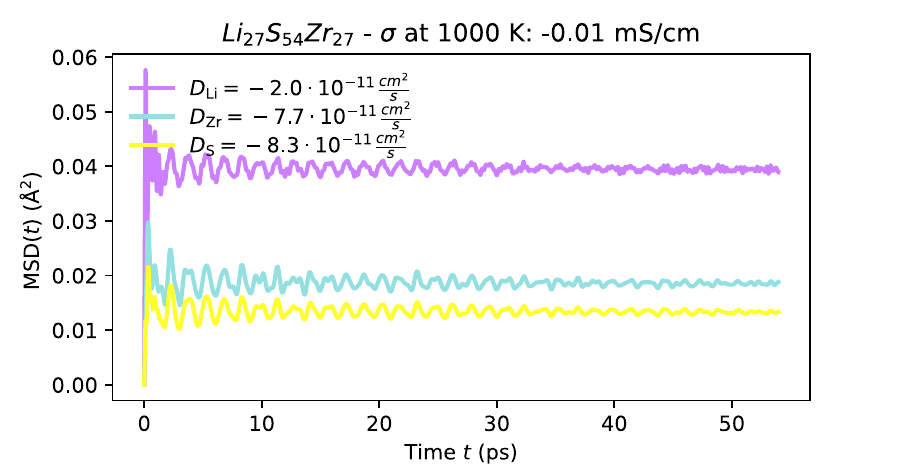}
  \caption{MSD plot of Li along with host-lattice species of $LiZrS_2$ at 1000 K studied with FPMD}
\end{figure}

\pagebreak
\FloatBarrier

\begin{figure}[H]
\centering
  \includegraphics[width=\columnwidth]{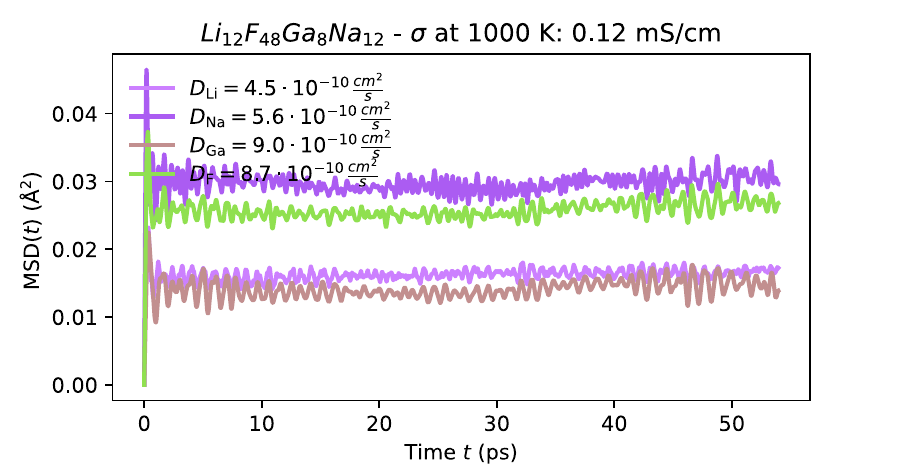}
  \caption{MSD plot of Li along with host-lattice species of $Li_3Na_3Ga_2F_{12}$ at 1000 K studied with FPMD}
\end{figure}

\begin{figure}[H]
\centering
  \includegraphics[width=\columnwidth]{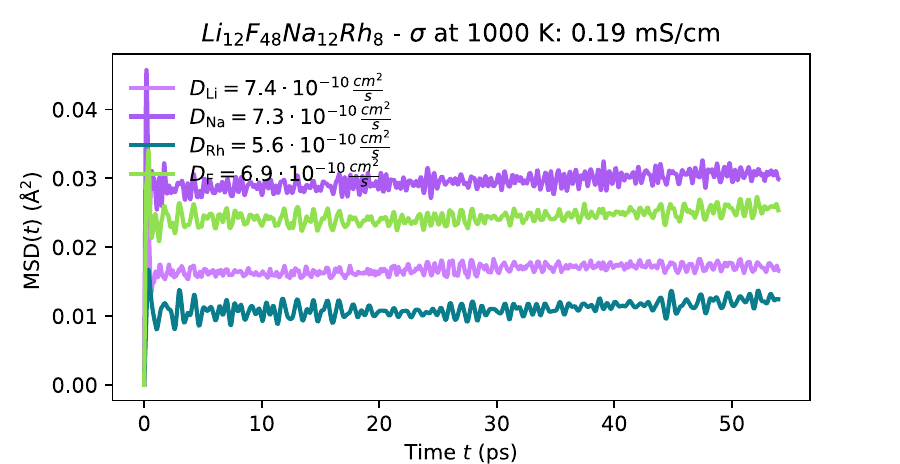}
  \caption{MSD plot of Li along with host-lattice species of $Li_3Na_3Rh_2F_{12}$ at 1000 K studied with FPMD}
\end{figure}

\begin{figure}[H]
\centering
  \includegraphics[width=\columnwidth]{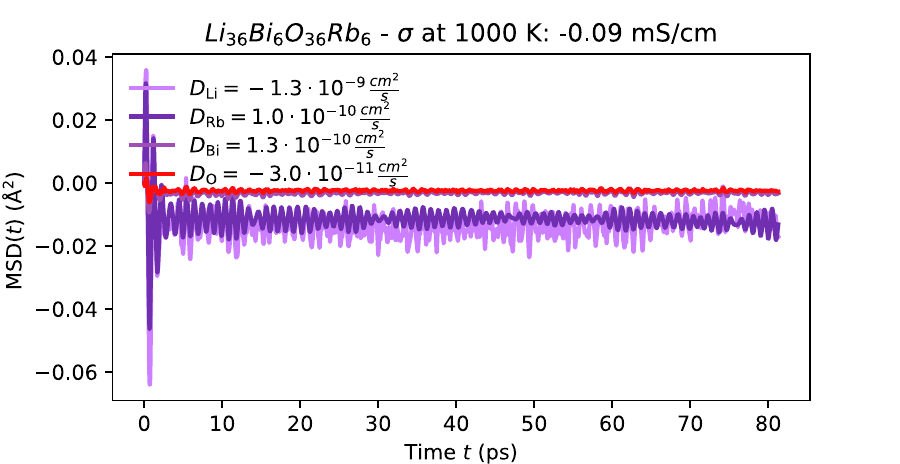}
  \caption{MSD plot of Li along with host-lattice species of $Li_6RbBiO_6$ at 1000 K studied with FPMD}Fig. 5
\end{figure}


\end{document}